\documentclass[preprint,showpacs,showkeys,preprintnumbers,amsmath,amssymb,aps]{revtex4}
\usepackage{epsfig}
\usepackage{graphicx}
\usepackage{dcolumn}
\usepackage{bm}
\usepackage{tabularx}
\usepackage{multirow}
\usepackage{color}
\def\math{\textfont1=\testfont \skewchar\testfont=\skewtrial
 \mathchardef\Gamma="100 \mathchardef\Delta="101
 \mathchardef\Theta="102 \mathchardef\Lambda="103 \mathchardef\Xi="104
 \mathchardef\Pi="105 \mathchardef\Sigma="106 \mathchardef\Upsilon="107
 \mathchardef\Phi="108 \mathchardef\Psi="109 \mathchardef\Omega="10A
 \def\ii{i} \def\jj{j}
 \def\\##1{|##1|+}\mathtrial
 \def\\##1{##1_2+}\mathtrial
 \def\\##1{##1^2+}\mathtrial
 \def\\##1{##1/2+}\mathtrial
 \def\\##1{2/##1+}\mathtrial
 \def\\##1{##1,{}+}\mathtrial
 \def\\##1{d##1+}\mathtrial
 \let\ii=\imath \let\jj=\jmath \def\\##1{\hat##1+}\mathtrial}
\newcount\skewtrial \skewtrial='177
\def\mathtrial{$\\A \\B \\C \\D \\E \\F \\G \\H \\I \\J \\K \\L \\M \\N \\O
 \\P \\Q \\R \\S \\T \\U \\V \\W \\X \\Y \\Z \\a \\b \\c \\d \\e \\f \\g
 \\h \\\ii \\\jj \\k \\l \\m \\n \\o \\p \\q \\r \\s \\t \\u \\v \\w \\x \\y
 \\z \\\alpha \\\beta \\\gamma \\\delta \\\epsilon \\\zeta \\\eta \\\theta
 \\\iota \\\kappa \\\lambda \\\mu \\\nu \\\xi \\\pi \\\rho \\\sigma \\\tau
 \\\upsilon \\\phi \\\chi \\\psi \\\omega \\\vartheta \\\varpi \\\varphi
 \\\Gamma \\\Delta \\\Theta \\\Lambda \\\Xi \\\Pi \\\Sigma \\\Upsilon
 \\\Phi \\\Psi \\\Omega \\\partial \\\ell \\\wp$\par}

\DeclareFontFamily{U}{mathb}{\hyphenchar\font45}
\DeclareFontShape{U}{mathb}{m}{n}{
      <5> <6> <7> <8> <9> <10> gen * mathb
      <10.95> mathb10 <12> <14.4> <17.28> <20.74> <24.88> mathb12
      }{}
\DeclareSymbolFont{mathb}{U}{mathb}{m}{n}
\DeclareFontSubstitution{U}{mathb}{m}{n}
\DeclareMathSymbol{\drsh}                  {3}{mathb}{"EB}

\newcommand{\gsim}{ \mathop{}_{\textstyle \sim}^{\textstyle >} }
\newcommand{\lsim}{ \mathop{}_{\textstyle \sim}^{\textstyle <} }

\newcommand{\MEV}{ ~{\rm MeV} }
\newcommand{\GEV}{ ~{\rm GeV} }
\newcommand{\TEV}{ ~{\rm TeV} }
\newcommand{\ifb}{ ~{\rm fb}^{-1}_{} }

\newcommand{\e}{{\rm e}}

\newcommand{\az}{c_{1ZZ}^{}}
\newcommand{\azp}{c'_{1ZZ}}
\newcommand{\azsq}{c_{1ZZ}^{2}}
\newcommand{\bz}{c_{2ZZ}^{}}
\newcommand{\bzp}{c'_{2ZZ}}
\newcommand{\cz}{c_{3ZZ}^{}}
\newcommand{\czp}{c'_{3ZZ}}

\newcommand{\aw}{c_{1WW}^{}}
\newcommand{\awp}{c'_{1WW}}
\newcommand{\awsq}{c_{1WW}^2}
\newcommand{\bw}{c_{2WW}^{}}
\newcommand{\dw}{c_{3WW}^{}}

\newcommand{\bg}{c_{2Z\gamma}^{}}
\newcommand{\bgSM}{c_{2Z\gamma}^{\rm SM}}
\newcommand{\cg}{c_{3Z\gamma}^{}}

\newcommand{\bgg}{c_{2\gamma\gamma}^{}}

\newcommand{\bggSM}{c_{2\gamma\gamma}^{\rm SM}}

\newcommand{\fWW}{f_{\rm WW}^{}}
\newcommand{\fBB}{f_{\rm BB}^{}}
\newcommand{\fBW}{f_{\rm BW}^{}}
\newcommand{\fB}{f_{\rm B}^{}}
\newcommand{\fW}{f_{\rm W}^{}}
\newcommand{\fphione}{f_{\phi 1}^{}}
\newcommand{\fphitwo}{f_{\phi 2}^{}}
\newcommand{\fphifour}{f_{\phi 4}^{}}

\newcommand{\sw}{s_{\rm W}^{}}

\newcommand{\swsq}{s_{\rm W}^{2}}

\newcommand{\cw}{c_{\rm W}^{}}

\newcommand{\cwsq}{c_{\rm W}^{2}}

\newcommand{\gz}{g_{Z}^{}}
\newcommand{\gzsq}{g_{Z}^{2}}

\newcommand{\mz}{m_{Z}^{}}
\newcommand{\mzsq}{m_{Z}^{2}}
\newcommand{\mw}{m_{W}^{}}
\newcommand{\mwsq}{m_{W}^{2}}
\newcommand{\mh}{m_{H}^{}}
\newcommand{\mhsq}{m_{H}^{2}}

\newcommand{\mhhatsq}{\hat{m}_{H}^{2}}
\newcommand{\mt}{m_{t}^{}}

\newcommand{\pH}{p_{H}^{}}

\newcommand{\pTH}{ {p_{\rm T}}_{H} }
\newcommand{\EH}{E_{H}^{}}
\newcommand{\thetaH}{\theta_{H}^{}}
\newcommand{\yH}{y_{H}}
\newcommand{\thetaZ}{\theta_{Z}^{}}

\newcommand{\pone}{p_{1}^{}}
\newcommand{\ptwo}{p_{2}^{}}
\newcommand{\pT}{p_{\rm T}}

\def\bsub{\begin{subequations}}
\def\esub{\end{subequations}}

\begin{document}
\begin{flushright}
Pre-Print  KEK-TH-1256
\vskip - 0.5 cm
\end{flushright} 
\title{ Measuring the Higgs-Vector boson Couplings 
at Linear $e^{+} e^{-}$ Collider } 
\author{Sukanta Dutta$^{1,2}$}\email{Sukanta.Dutta@kek.jp}
\author{ Kaoru Hagiwara$^{2,3}$ }
\author{Yu Matsumoto$^{2,3}$} \email{ymatsu@post.kek.jp}
\affiliation{$^1$SGTB Khalsa College, University of
Delhi. Delhi-110007. India.}
\affiliation{$^2$KEK Theory Division, 1-1 Oho Tsukuba, 305-0801, Japan.}
\affiliation{$^3$Graduate University for Advanced Studies(Soken-dai), 
1-1 Oho Tsukuba, 305-0801, Japan.}
\begin{abstract}
\baselineskip 0.6 cm
 We estimate the accuracy with which the  coefficient of the CP even
 dimension six operators involving Higgs and two vector bosons ($HVV$)
 can be measured at linear $e^+ e^-$ colliders. Using the optimal
 observables method for the kinematic distributions, our analysis is
 based on the five different processes. First is the $WW$ fusion process
 in the $t$-channel ($e^+e^- \rightarrow \bar{\nu}_e \nu_e H$), where we
 use the rapidity $y$ and the transverse momentum $\pT$ of the Higgs
 boson as observables. Second is the $ZH$ pair production process in the
 $s$-channel, where we use the scattering angle of the $Z$ and the $Z$ decay
 angular distributions, reproducing the results of the previous
 studies. Third is the $t$-channel $ZZ$, fusion processes ($e^+e^-
 \rightarrow e^+e^-H$), where we use the energy and angular
 distributions of the tagged $e^+$ and $e^-$. In the fourth, we consider
 the rapidity distribution of the untagged  $e^+e^-H$ events, which can
 be approximated well as the $\gamma \gamma$ fusion of the
 bremsstrahlung photons from $e^+$ and $e^-$ beams. As the last process,
 we consider the single tagged $e^+e^-H$ events, which probe the $\gamma
 e^{\pm} \rightarrow He^{\pm}$ process. All the results are presented in
 such a way that statistical errors of the constraints on the effective
 couplings and their correlations are read off when all of them are
 allowed to vary simultaneously, for each of the above processes, for
 $m_H=120 $ GeV, at $\sqrt{s}=250\GEV$,  $350\GEV$  $500\GEV$ and
 $1\TEV$, with and without $e^-$ beam polarization of 80\%. We find for
 instance that the  $HZZ$ and $HWW$ couplings can be measured with 0.6\%
 and 0.9\% accuracy, respectively, for the integrated luminosity of
 $L=100\ifb$ at $\sqrt{s}=250\GEV$, $350\GEV$ and $L=500\ifb$  at
 $\sqrt{s}=500\GEV$, $1\TEV$, for the luminosity uncertainty of 1\% at
 each energy. We find that the luminosity uncertainty affects only one
 combination of the non-standard couplings which are proportional to the
 standard $HWW$ and $HZZ$ couplings, while it does not affect the errors
 of the other independent combinations of the couplings. As a
 consequence, we observe that a few combinations of the eight dimension
 six operators can be constrained as accurately as the two operators
 which have been constrained by the precision measurements of the $Z$
 and $W$ boson properties.
\end{abstract}

\maketitle
\baselineskip 0.7cm
\section{INTRODUCTION}
\label{INTRO}
The Standard Model (SM) of the elementary particles  based on the SU(3) 
$\times$ SU(2) $\times$ U(1) gauge symmetry has proved to be a successful 
theory to interpret all the precision data available to date. SM predicts a 
light Higgs boson whose discovery  is one of the prime tasks of the upcoming 
future colliders.  

In fact, the present electroweak precision measurements indicate the
existence of a light Higgs boson \cite{lepbound,pdg2006}. Experiments at
the CERN Large Electron Positron collider (LEP) set the lower bound on
its mass of $114.1\GEV$ at the $95 \%$ confidence level (CL)
\cite{lepbound}. The Fermilab  Tevatron, which collides proton and
anti-proton at $\sqrt{s}=2\TEV$, is currently the only collider  which
can produce low mass Higgs bosons. Analysis with Run IIb data samples by
the CDF and D$\slash\! \! \! 0$ detectors indicates that the Tevatron
experiments can observe the Higgs boson with about 10 $\ifb$ total
integrated luminosity for the mass of around $120\GEV$
\cite{tevreport}. The Large Hadron Collider (LHC) at CERN will start
colliding two protons at $\sqrt{s}=14\TEV$ in the year 2008, and is
geared to detect the Higgs boson in gluon-gluon and vector-boson fusion
processes. It will measure ratios of various Higgs boson couplings
through variety of decay channels at accuracies of order 10 to 15\% with
100 $\ifb$ luminosity \cite{zeppkinPRD62013009}.

Despite the success, SM presents the naturalness problem due to the
quadratic sensitivity of the Higgs boson mass to the new physics scale
at high energies, which implies that there is a need of subtle fine tuning
to keep the electroweak symmetry breaking theory below the\TEV scale. To
put it in another way, this may suggest an existence of a new physics
scale $\Lambda $ not far above the TeV scale. The key to probe the new
physics beyond the SM theory is to clarify the origin of the electroweak
symmetry breaking, the Higgs mechanism. Therefore, it is necessary to
measure the Higgs boson properties as precisely as possible, especially
the $HVV$ couplings, because they are expected to be sensitive to  the
symmetry breaking physics that gives rise to the weak boson masses.

With this motivation, we  re-examine the potential of the future
$e^+e^-$ linear collider, the International Linear Collider  (ILC) in
the precise measurement of the $HVV$ couplings. Clean experimental
environment, well defined initial state, tunable energy, and beam
polarization renders ILC to be  the best machine to study the Higgs
boson properties with  high precision.  In this paper, we study the
sensitivity of the  ILC measurements on all the $HVV$ ($HWW$, $HZZ$,
$HZ\gamma$ and $H\gamma\gamma$) couplings comprehensively and
semi-quantitatively by using all the available processes with a light
Higgs boson ($m_H \lsim$ 120 GeV); $e^+\,e^-\to \nu_e\,\bar{\nu_e}\,H$
with $t$-channel $W$ exchange, $e^+\,e^-\to Z\,H$ with $s$-channel $Z$
exchange, $e^+\,e^-\to e^+\,e^-\,H$ with $t$-channel $Z$ exchange,
no-tag $e^+\,e^-\to (e^+\,e^-)\,H$ process from $\gamma\gamma$ fusion,
and single-tagged $e^+\,e^-\to (e^\pm)\,e^\mp\,H$ process that probes 
$\gamma\, e^\mp \to H\,\ e^\mp$ via $t$-channel $\gamma $ and $Z$ exchange.

In order to quantify the ILC sensitivity to measure various $HVV$
couplings simultaneously, we adopt the powerful technique of the optimal
observables method \cite{atwood, davier, diehl, gunion}. It allows us to
measure several couplings simultaneously as long as the non-standard
couplings give rise to different observable kinematic distributions. The
results can be summarized in terms  of the covariance matrix of the
measurement errors, from each process at each energy, that scales
inversely as the integrated luminosity.

In order to combine results from different processes and at different
energies, we adopt the effective Lagrangian of the SM particles with
operators of mass dimension six to parametrize all the $HVV$ couplings
\cite{linear,Grinstein,STpram}. This allows us not only to compare the
significance  of the measurements of various $HVV$ couplings at
different energies and at different colliders, but also to study  what
ILC can add to the precision measurements of the $Z$ and $W$ boson
properties in the search for new physics via quantum effects. We
therefore parametrize the $HVV$ couplings as  linear combination of all
the  dimension six operators that are  allowed by the electroweak gauge
symmetry and $CP$ invariance.

Some of the previous studies based on the optimal observables method
are found for $CP$-violating effects in $e^+\,e^-\to Z\,H$ via $HZZ$ and
$HZ\gamma$ couplings~\cite{gunion,ma}, and also in 
$e^+\,e^-\to t\,\bar t\,H$ \cite{gunion}. CP conserving and CP violating
effects in $e^+\,e^-\to Z\,H$ process has been studied in
ref.~\cite{kniehl,StongHagi}. In refs.~\cite{gunion,kniehl,StongHagi}
all the relevant couplings are varied simultaneously, and their
correlations are studied. More recently, the ILC sensitivity to the
$HZZ$ and $HWW$ couplings has been studied in
refs.~\cite{debchou,mamta,TaoHan}. Bounds on the coefficients of the
Higgs-vector boson dimension-6 operators have been found in
refs.~\cite{tevatron, photonhiggs} based on non-observation of the
Higgs boson signal at the Tevatron. Whenever relevant, we compare our
results with the previous observations.

This paper is organized as follows. In section~\ref{Couplings}, we 
describe the low energy effective interactions among the Higgs boson and
the electroweak gauge bosons arising from new physics that is
parametrized in terms of the effective Lagrangian of the SM particles
with operators up to mass dimension six. In section \ref{OptPhaseSpace},
we introduce the optimal observables method and explain how we perform
the phase space integration when some of the kinematic distributions are
unobservable, such as neutrino momenta and a distinction between quark
and anti-quark jets. Although we present numerical results for
unpolarized beams  and for 80\% polarized $e^-$ beam only, all the
formulas are presented for an arbitrary polarization of $e^-$ and $e^+$
beams. After introducing final state cuts, such as those for the $e^\pm$
tagging and those for selecting or excluding $Z\to f\,\bar f$ events, we
present the total cross sections for all the five processes at
$\sqrt{s}$=200 GeV-1 TeV for $\mh$ =120 GeV, and at $\sqrt{s}$=250 GeV,
500 GeV, 1 TeV for $\mh$ = 100-200 GeV. Then in section \ref{WWFusion} we
compute the statistical errors of the non-standard $HWW$ couplings
extracted from measurements of the $WW$-fusion process, 
$e^+e^- \to \nu_e\,\bar{\nu}_e  H$. In section \ref{ZHProd}, we study
the constraints on the $HZZ$ and $HZ\gamma$ couplings extracted from
$ZH$ production. In section \ref{ZZ}, not only the $HZZ$ and $HZ\gamma$
couplings but also the $H\gamma\gamma$ coupling are studied in the
double-tag $e^+e^-\to e^+e^-\, H$ process via $t$-channel $Z$ and
$\gamma$ exchange. In section \ref{EPF}, we obtain the constraints on
the $H\gamma\gamma$ coupling from the $\gamma\gamma$ fusion, in no-tag
$e^+e^-\to (e^+e^-) H$ events, using the equivalent real photon
approximation. In section \ref{SingleTag}, we consider the single-tag
$e^+e^-\to (e^\pm) \, e^\mp \,H$ process to constrain the
$H\gamma\gamma$ and $H\gamma Z$ couplings. In section \ref{luminos}, we
address the implication of luminosity uncertainty on the measurement of
these couplings. In section \ref{conclu}, we summarize all our results,
compare them with previous studies, and present our estimates for the
ILC constraints on the dimension six operators, which are then compared
with the constraints from the precision electroweak measurements of the
$W$ and $Z$ boson properties. In Appendices we present our
parameterizations of the 3-body phase space (Appendix
\ref{app:phasespace}), and the explicit forms the $t$-channel and
$s$-channel currents and their contractions that appear in the helicity
amplitudes (Appendix \ref{app:current}).


\section{Generalized $HVV$ vertex with dimension six operators}
\label{Couplings}

In our study, we adopt the effective Lagrangian of the Higgs and
the gauge bosons with operators up to mass dimension six,
\begin{eqnarray}
 L_{\rm eff} = L_{\rm SM} +\sum_{i} \frac{f_{i}^{(6)}}{\Lambda^{2}} 
  {\cal O}_{i}^{(6)},
  \label{fulllagrangianop6}
\end{eqnarray}
\noindent 
where $L_{\rm SM}$ denotes the renormalizable SM Lagrangian and 
${\cal O}_i^{(6)}$'s are the gauge-invariant operators of mass dimension 
6. The index $i$ runs over all operators of the given mass dimension. 
The mass scale is set by $\Lambda$, and the coefficients $f_i^{(6)}$ are 
dimensionless parameters, which are determined once the full theory is
known. Excluding the dimension 5 operators for the neutrino Majorana
masses, and the dimension 6 operators with quark and lepton fields, we
are left with the following eight CP even operators that affect the
$HVV$ couplings. Notation of the operators are taken from the
reference~\cite{LowEnegyEff}.
\bsub
 \begin{eqnarray}
  {\cal O}_{WW} &=& 
   \Phi^{\dagger} \overline{W}^{\mu \nu} \overline{W}_{\mu \nu} \Phi, 
   \label{opstart} \label{Operatorsf}\\
  {\cal O}_{BB} &=& 
   \Phi^{\dagger} \overline{B}^{\mu \nu} \overline{B}_{\mu \nu} \Phi, \\
  {\cal O}_{BW} &=& 
   \Phi^{\dagger} \overline{B}^{\mu \nu} \overline{W}_{\mu \nu} \Phi, \\
  {\cal O}_{W} &=& 
   (D^{\mu} \Phi)^{\dagger} \overline{W}_{\mu \nu} (D^{\nu} \Phi), \\
  {\cal O}_{B} &=& 
   (D^{\mu} \Phi)^{\dagger} \overline{B}_{\mu \nu} (D^{\nu} \Phi), \\
  {\cal O}_{\phi 1} &=& 
   \Big[(D_{\mu} \Phi)^{\dagger} \Phi \Big]\,\,
   \Big[\Phi^{\dagger} (D^{\mu} \Phi)\Big], \\
  {\cal O}_{\phi 4} &=& 
   (\Phi^{\dagger} \Phi) (D_{\mu} \Phi)^{\dagger} (D^{\mu} \Phi), \\
  {\cal O}_{\phi 2} &=& 
   \frac{1}{2} \partial_{\mu}(\Phi^{\dagger} \Phi) 
   \partial^{\mu} (\Phi^{\dagger} \Phi). \label{Operatorsl}
 \end{eqnarray}
 \label{Operators}
\esub
\noindent 
\hskip -1.4mm
Here $\Phi$ denotes the Higgs doublet field with the hyper-charge
$Y=\frac{1}{2}$, and the covariant derivative is
$D_{\mu} = \partial_{\mu} + i \hat{g}_W T^a \hat{W}^a_{\mu} 
  + i \hat{g_Y} Y \hat{B}_{\mu}$,
where the gauge couplings and the gauge fields with a caret represent
those of the SM, in the absence of higher dimensional operators. The
gauge-covariant and invariant tensors $\overline{W}_{\mu \nu}$ and
$\overline{B}_{\mu \nu}$, respectively, are 
$\overline{W}_{\mu \nu} = i \hat{g}_W T^a \hat{W}^{a}_{\mu \nu}$, and
$\overline{B}_{\mu \nu} = \frac{i}{2} \hat{g}_Y \hat{B}_{\mu \nu}$. The
coefficients of the operators (\ref{Operatorsf})-(\ref{Operatorsl}),
which are denoted as $f_i^{(6)}/\Lambda^2$ in the effective Lagrangian of
eq.(\ref{fulllagrangianop6}), should give us information about
physics beyond the SM. So far, the precision measurements of the weak
boson properties~\cite{pdg2006} constrained the operaotrs 
${\cal O}_{\rm BW}$ and ${\cal O}_{\phi 1}$, which have been useful in
testing some models of the electroweak symmetry
breakdown~\cite{Grinstein,Oblique1}. In this report, we explore the
accuracy with which the ILC experiments can measure the coefficients of
all these eight operators when a light Higgs boson exists.

When the Higgs field acquires the vacuum expectation value
$<\Phi> = \frac{1}{\sqrt{2}} (0,v)^T$, the bilinear part of the
effective Lagrangian of eq.(\ref{fulllagrangianop6}) is expressed as
\begin{eqnarray}
 L_{\rm eff}
  &=& - \frac{1}{4}\,\, \left( 1 + \frac{\hat{g}_W^2 v^2}{2\,\, \Lambda^2}\,\,
                         \fWW \right)
  \,\, \hat{W}^{a}_{\mu \nu}\,\, \hat{W}^{a \mu \nu} 
  - \frac{1}{4} \,\, \left( 1 + \frac{\hat{g}_Y^2 v^2}{2\,\, \Lambda^2} \,\,
                      \fBB \right)
  \,\, \hat{B}_{\mu \nu}\,\, \hat{B}^{\mu \nu} \nonumber \\
 & & + \,\frac{\hat{g}_W \hat{g}_Y v^2}{8\,\, \Lambda^2}\,\, 
  \fBW\,\, \hat{B}_{\mu \nu}\,\, \hat{W}^{3 \mu \nu} 
  + \frac{\hat{g}_W^2 \,\, v^2}{8} 
  \left( 1 + \frac{v^2}{2\,\, \Lambda^2}\,\, \fphifour \right) \,\,
  ( \hat{W}^1_{\mu} \,\, \hat{W^1}^{\mu} + \hat{W}^2_{\mu} \,\,
  \hat{W^2}^{\mu}) \nonumber \\
 & & + \frac{v^2}{8} \,\,
  \left( 1 + \frac{v^2}{2\,\, \Lambda^2} \,\, (\fphione + \fphifour) \right)
  \,\, (\hat{g}_W \,\, \hat{W^3}_{\mu} - \hat{g}_Y \,\, \hat{B}_{\mu} ) 
  \,\, (\hat{g}_W \,\, \hat{W^3}^{\mu} - \hat{g}_Y \,\, \hat{B}^{\mu} ) 
  \nonumber \\
 & & + \frac{1}{2}\,\, \left( 1 + \frac{v^{2}}{2\,\, \Lambda^{2}} 
                        \,\,  \left(\fphione + \fphifour + 2\,\,\fphitwo
                                 \right) \right)
 \,\,\, \left(\partial_{\mu} \hat{H}_0 \right)\,\,  
 \left( \partial^{\mu} \hat{H}_0 \right) 
 - \frac{1}{2}\,\, \mhhatsq \,\, \hat{H}_0^2 + \cdots.
 \label{Eq.ModLag1}
\end{eqnarray}
After renormalization of gauge fields and their couplings,
\bsub
\begin{eqnarray}
 W_\mu^a = \left( 1+\frac{\hat g_W^2 \,\, v^2}{4\,\,\Lambda^2} \,\,\,
            f_{WW} \right)\,\, \hat W_\mu^a\, ;
 &\hskip 1 cm&
 g_W\,\,\, W_\mu^a = \hat g_W\,\,\, \hat W_\mu^a\label{def1}\, ;\\
 B_\mu = \left(1+\frac{\hat g_Y^2\,\, v^2}{4\,\,\Lambda^2}\,\,\, 
          f_{BB}\right)\,\, \hat B_\mu\, ;
 &\hskip 1 cm&
 g_Y\,\,\, B_\mu = \hat g_Y\,\,\, \hat B_\mu\, ; \label{def2}
\end{eqnarray}
\label{defboth12}
\esub
\noindent 
\hskip -1.4mm 
and after diagonalization of the mass squared matrices,
the effective Lagrangian reads
\begin{eqnarray}
 L_{\rm eff} &=& -\,\,\frac{1}{2} W^+_{\mu\nu}\,\,{W^-}^{\mu\nu}
  -\,\,\frac{1}{4} Z_{\mu\nu}\,\,Z^{\mu\nu}
  -\,\,\frac{1}{4} A_{\mu\nu}\,\,A^{\mu\nu}
  +\frac{g_W\,\,g_Y\,\, v^2}{8\,\,\Lambda^2}\,\, \fBW \,\, 
  B_{\mu\nu}\,\,{W^3}^{\mu\nu} \nonumber\\
 & & + m_W^2\,\, W^+_{\mu}\,\,{W^-}^{\mu}
  +\frac{m_Z^2}{2}\,\,Z_{\mu}\,\,Z^{\mu}
  -\frac{1}{2}\,\, \big(\partial_\mu H \big)\,\, 
  \big( \partial^\mu H \big)
  -\frac{1}{2}\,\, m_H^2\,\, H^2+\cdots ,\label{efflag2}
\end{eqnarray}
where 
\bsub
\begin{eqnarray}
 m_W^2 &=& \frac{g_W^2\,\, v^2}{4}\,\,\,
  \left[1+\frac{v^2}{2\,\,\Lambda^2}\,\,\, \fphifour\right],\\
 m_Z^2 &=& \frac{g_Z^2\,\, v^2}{4}\,\,\,
  \left[1+\frac{v^2}{2\,\,\Lambda^2}\,\,\, 
   \big( \fphione+\fphitwo \big) \right],\\
 m_H^2 &=& \hat m_H^2\,\,\, 
  \left[1-\frac{v^2}{2\,\,\Lambda^2}\,\,\, 
   \big( \fphione+\fphifour+2\,\, \fphitwo \big) \right], \\
 H &=& \left[ 1+\frac{v^2}{4\,\,\Lambda^2}\,\,\, 
      \big( \fphione+\fphifour+2\,\, \fphitwo\big) \right]^{1/2}
 \,\,\, \hat H_0 \, .\label{def6}
\end{eqnarray}
\label{def5}
\esub
\noindent
\hskip -1.4mm 
All the remaining terms in the effective Lagrangian, denoted by dots in
eq.(\ref{efflag2}), are expressed in terms of the renormalized fields,
couplings and masses, as defined in eqs.(\ref{defboth12}) and
eqs.(\ref{def5}). The standard gauge interactions are dictated by the
covariant derivative
\begin{eqnarray}
 D_\mu =\partial_\mu +i\,\,\frac{g_W}{\sqrt{2}}\,\, \left( T^+\,\,
  W^+_\mu+T^-\,\, W^-_\mu\right) +i\,\, g_Z \,\, \left(T^3-s_W^2\,\,
  Q\right) \,\, Z_\mu+i\,\, e\,\, Q\,\, A_\mu,\label{stdgaugeint1}
\end{eqnarray}
where $T^\pm=\big(T^1\pm i\,\, T^2\big)/\,\, 2, \,\,\,\, Q=T^3+Y$, and 
$e=g_W\,\,s_W=g_Y\,\, c_W=g_Z\,\,c_W\,\,s_W\,$.

Before expressing the $HVV$ interactions of $L_{\rm eff}$, let us
briefly review the observable consequence of new physics in the gauge
boson two point functions in eq.(\ref{efflag2}). First, the ratio of
the neutral current and the charged current interactions at low energies
deviate~\cite{Oblique0, Oblique1} from unity,
\begin{eqnarray}
 \rho \left(=\frac{1}{1-\alpha\, T} \right)
  = \frac{G_{\rm N.\,C.}}{G_{\rm C.\,C.}} 
  = \frac{g_Z^2/\,\mzsq}{g_W^2/\, \mwsq}
  = 1 - \frac{\fphione}{2\,\,\Lambda^2}\,.
\end{eqnarray}
\noindent Second, the extra kinetic mixing between $B_{\mu\nu}$ and
$W^3_{\mu\nu}$ modifies the $\gamma$ and $Z$ boson propagators
\begin{eqnarray}
\Delta\Pi_T^{QQ}(q^2) = 2\,\,\Delta\Pi_T^{3Q}(q^2) =
 \left(-\,\, \frac{v^2}{2\,\,\Lambda^2}\,\, f_{BW}\right)\,\,q^2
\end{eqnarray}
\noindent 
in the notation of ref.~\cite{Hagiwara:1998}, which contributes to the
$S$ parameter~\cite{Oblique1}
\begin{eqnarray}
 S= 16\,\,\pi\,\,
  \left[ \frac{\overline{s}^2(m_Z^2)\,\,\overline{c}^2(m_Z^2)}
   {\overline{e}^2(m_Z^2)}-\frac{1}{\overline{g}_Z^2(0)} \right]
  = -\,\,\frac{4\,\,\pi\,\,v^2}{\Lambda^2}\,\,\,f_{BW}\, .
\end{eqnarray}
\noindent 
Here the over-lined couplings $\overline{e}^2(q^2)$,
$\overline{s}^2(q^2)=1-\overline{c}^2(q^2)$ and $\overline{g}^2_Z(q^2)$
are the effective couplings that contain the gauge-boson propagator
corrections at the momentum transfer $q^2$ \cite{Hagiwara:1998}. We will
examine the constraints on $\fphione$ and $f_{BW}$ from the precision
measurements of the weak boson properties in the last section of this
report.

The terms describing the $HVV$ couplings in the effective Lagrangian are
now expressed as
\begin{eqnarray}
 L_{\rm eff}^{HVV} &=& 
 (1+\aw) \, g \,\mw H W_{\mu}^{+} W^{- \mu} + (1+\az) \,\, \frac{\gz \,\mz}{2} \,\, H \, Z_{\mu}\,  Z^{\mu} \nonumber \\
 & + & \frac{\gz}{\mz} \left[ 
  \bw \,\, H \,W_{\mu \nu}^{+}\,  W^{- \mu \nu} + \frac{\dw}{2} 
  \Bigg\{ \bigg(\left(\partial_{\mu} H\right) W_{\nu}^{-} 
           - \left(\partial_{\nu} H\right)\, W_{\mu}^{-}\bigg)\,\, W^{+ \mu \nu} 
  + {\rm h.c.} \bigg\} \right] \nonumber \\
 & + &  \frac{\gz}{\mz} \left[ \frac{\bz}{2} \,\, H \,Z_{\mu \nu} \, Z^{\mu \nu}
  + \frac{\cz}{2} \Big( (\partial_{\mu} H) \, Z_{\nu} - (\partial_{\nu} H) \, Z_{\mu}\Big)\,
  Z^{\mu\nu}  \right]+ \frac{\gz}{\mz}\,\, \Big[ \frac{\bgg}{2} H A_{\mu \nu} A^{\mu \nu} \Big]  \nonumber \\
 & + &  \frac{\gz}{\mz}\,\, \left[ \bg H \,Z_{\mu \nu} \, A^{\mu \nu}
  + \cg \bigg((\partial_{\mu} H) \, Z_{\nu} - (\partial_{\nu} H) \, Z_{\mu}\bigg)\,\, 
  A^{\mu\nu} \right]   \label{EL}
\end{eqnarray}
\noindent
where the 9 dimensionless couplings, $c_i$, parametrize all the
non-standard $HVV$ interactions: 
\bsub
\begin{eqnarray}
 \label{aZ}
  \az &=& \frac{v^{2}}{4 \Lambda^{2}} 
  \,\bigg(3\,\fphione + 3\,\fphifour - 2\,\fphitwo\bigg)\,,  \label{ninecoupstart} \\
 \aw &=& \frac{v^{2}}{4 \Lambda^{2}} 
  \,\bigg(-\,\fphione + 3\,\fphifour - 2\,\fphitwo\bigg)\,, \label{aWeq}\\
 && \nonumber \\
 \bz &=& \frac{\mzsq}{\Lambda^{2}}
  \,\bigg(-s_{\rm W}^4 \,\fBB - \swsq \,\, \cwsq \,\fBW - c_{\rm W}^4 \,\fWW
  \bigg)\,, \label{bzeq}\\ 
 \bg &=& \frac{\mzsq}{\Lambda^{2}} \,\bigg(\swsq \,\fBB 
  + \frac{1}{2} (\cwsq - \swsq) \,\fBW - \cwsq \,\fWW\bigg)\, 
  \sw \cw, \label{bgeq} \\
 \bgg &=& \frac{m_{Z}^{2}}{\Lambda^{2}} 
  \,\bigg(-\,\fBB + \,\fBW - \,\fWW\bigg)\, \cwsq \,\, \swsq, \label{bggeq}\\
 \bw &=& \frac{\mzsq \cwsq}{\Lambda^{2}} \,\bigg(-\,\fWW\bigg)\,, \label{bweq}\\
 && \nonumber \\
 \cz &=& \frac{\mzsq}{2 \Lambda^{2}} \,\bigg(-\swsq \,\fB - \cwsq
  \,\fW\bigg)\,, \label{czeq}\\ 
 \cg &=& \frac{\mzsq}{4 \Lambda^{2}} \,\bigg(\,\fB-\,\fW\bigg)\, \sw\,\,
  \cw, \label{cgeq}\\ 
 \dw &=& \frac{\mzsq \cwsq}{2 \Lambda^{2}} \,\bigg(-\,\fW\bigg).
  \label{ninecoupend}
\end{eqnarray}
\label{effcoup_Operator}
\esub
\noindent 

From the effective Lagrangian of eq.(\ref{EL}), we obtain the Feynman rule 
for $V_1^{\mu}(\pone)-V_2^{\nu}(\ptwo)-H(\pH)$ vertex as
\begin{eqnarray}
 \Gamma^{H V_{1} V_{2}}_{\mu \nu} (p_H,\,p_1,\,p_2) &=& \gz \,\,\mz\,\, 
  \left[ h_{1}^{V_{1} V_{2}} \,\, g_{\mu \nu} 
   + \frac{h_{2}^{V_{1} V_{2}}}{m_{Z}^{2}} \,\, p_{2 \mu} \,\, p_{1 \nu} 
  \right],
  \label{HVVFormFactor}
\end{eqnarray}
\begin{figure}[htb!]
\begin{center}
\includegraphics[width=7cm, height=5cm]{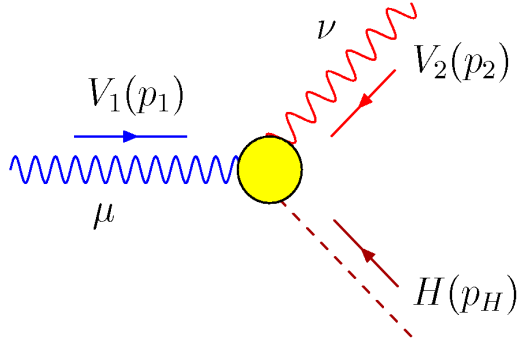} \vskip -.75 cm
\caption{\small The $HVV$ Vertex.}
\label{fig:vertex}
\end{center}
\vskip -.5 cm
\end{figure}
\noindent
\hskip -1.4mm
where all three momenta are incoming, $p_1 + p_2 + p_H = 0$, as shown in
the fig.~\ref{fig:vertex}. $V_1$ and $V_2$ can be $(V_{1} V_{2})=(ZZ)$,
$(Z\gamma)$, $(\gamma Z)$, $(\gamma \gamma)$, $(W^+W^-)$ or  $(W^-W^+)$. The coefficients
$h_{i}^{V_{1}V_{2}}(p_1,\,p_2)$  are
\bsub
\begin{eqnarray}
 h_{1}^{ZZ} (p_1,\, p_2)&=& 1 +\az 
  +\frac{p_1^2 +p_2^2 -\mhsq}{\mzsq} \, \bz
  +\frac{\mhsq}{\mzsq} \, \cz,  \label{ZZh1} \\
 h_{2}^{ZZ}  (p_1,\, p_2)&=& 2 \,\,\left(\bz-\cz\right), \label{ZZh2} 
\end{eqnarray}
\esub
for the $HZZ$ couplings,
\bsub
\begin{eqnarray}
 h_{1}^{\gamma \gamma}  (p_1,\, p_2)&=& \frac{p_1^2 +p_2^2 -\mhsq}{\mzsq} \, \bgg,
  \label{AAh1} \\
 h_{2}^{\gamma \gamma}  (p_1,\, p_2)&=& 2 \, \bgg, \label{AAh2} 
\end{eqnarray}
\esub
for the $H\gamma\gamma$ couplings, 
\bsub
\begin{eqnarray}
 h_{1}^{Z \gamma}  (p_1,\, p_2)&=& \frac{p_1^2 +p_2^2 -\mhsq}{\mzsq} \, \bg 
  -\frac{ p_1^2 -p_2^2 -\mhsq}{\mzsq} \,\cg,
  \label{ZAh1} \\
 h_{1}^{\gamma Z}  (p_1,\, p_2)&=& \frac{p_1^2 +p_2^2 -\mhsq}{\mzsq} \, \bg 
  -\frac{-p_1^2 +p_2^2 -\mhsq}{\mzsq} \, \cg,
  \label{AZh1} \\
 h_{2}^{Z \gamma} (p_1,\, p_2) &=& h_{2}^{\gamma Z} (p_1,\, p_2) = 2 \, (\bg - \cg),
  \label{ZAh2} 
\end{eqnarray}
\esub
for the $HZ\gamma$ couplings. It is to be noted  that the  $HZ\gamma$ coupling $\cg $ 
has the Feynman rule which is 
 not symmetric
under an interchange of $p_1$ and $p_2$. For the $HWW$ couplings,
\bsub
\begin{eqnarray}
h_{i}^{W^+W^-} (p_1,\, p_2)&=& h_{i}^{W^-W^+} (p_1,\, p_2)=h_{i}^{WW} (p_1,\, p_2)\hskip 0.5 cm (i=1,\,2)\\
 h_{1}^{WW} (p_1,\, p_2)&=& (1+\aw) \,\, \cos^{2} \theta_{W}
  +\frac{p_1^2 +p_2^2 -\mhsq}{\mzsq} \, \bw
  +\frac{\mhsq}{\mzsq} \, \dw, \label{WWh1} \\
 h_{2}^{WW} (p_1,\, p_2)&=& 2\,\, \left(\bw-\dw\right). \label{WWh2} 
\end{eqnarray}
\label{WWh}
\esub
\hskip -1.4mm
\noindent Although we do not consider off-shell Higgs boson contributions in this report, 
$\mhsq$ should be replaced by $(p_1+p_2)^2$ in the above Feynman rules 
when the Higgs-boson is off-shell. 

\section{ Optimal Observables and Phase Space}
\label{OptPhaseSpace}
\subsection{Optimal observables method}
The optimal observables method \cite{gunion} makes use of all the
kinematic distributions which are observable in experiments. We
therefore summarize our phase-space parameterizations for all the Higgs
boson production processes in $e^+ e^-$ collisions considered in this
study, which can be generically written as
\begin{eqnarray}
  e^{-} \left(k_{1}, \frac{\sigma_{1}}{2} \right) 
  + e^{+} \left(k_{2} ,\frac{\sigma_{2}}{2}\right) 
  &\rightarrow &  
 f \left( \pone, \frac{\lambda_{1}}{2} \right) 
  + \bar{f} \left( \ptwo, \frac{\lambda_{2}}{2} \right)+ H(\pH) \,\, .
  \label{ffbarh}
\end{eqnarray}
\begin{figure}[htb!]
 \begin{center}
  \begin{minipage}[t]{0.50\textwidth}
   \includegraphics[width=7cm,height=5cm]{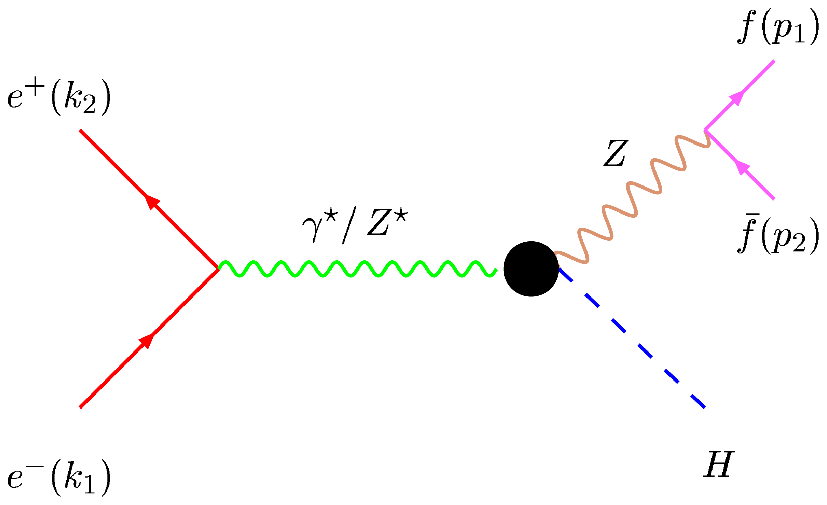}
   \par {(a) ZH production}
  \end{minipage}
  \vspace{5mm}
  \begin{minipage}[t]{0.45\textwidth}
   \includegraphics[width=7cm,height=5cm]{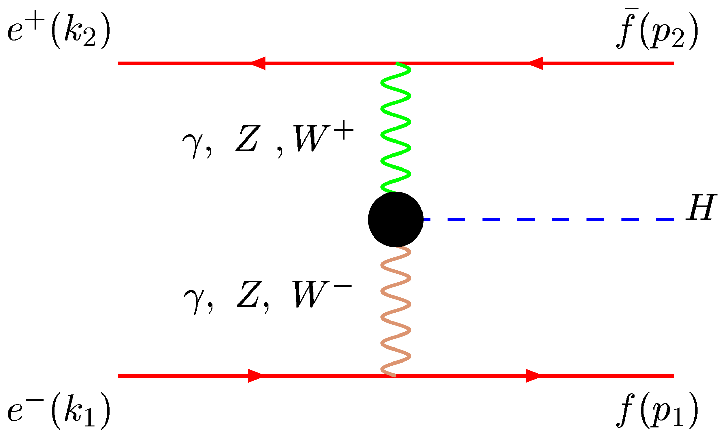}
   \par {(b) vector-boson fusion}
  \end{minipage}
  \caption{\small Feynman diagrams for 
  $e^{-} e^{+} \rightarrow f \bar{f} H$.}
  \label{diagram}
 \end{center}
\end{figure}
\noindent
\hskip -1.4 mm
Here $k_i$ and $\sigma_i/2$ are the $e^{\pm}$ four momenta and
helicities, respectively, and $p_i^{\mu}$ and $\lambda_i/2$ are the four
momenta and helicities, respectively, of the produced fermion ($i=1$) and
anti-fermion ($i=2$). For $f \neq e,\nu_e$, the processes (\ref{ffbarh})
occur only through the $ZH$ production diagram as shown in
fig.\ref{diagram}(a), whereas for $f = e$ or $\nu_e$, both the diagrams
fig.\ref{diagram}(a) and fig.\ref{diagram}(b) contribute. The effective
$HVV$ vertex is depicted by the solid circle in the Feynman
diagrams. The $ZH$ production process (a) is sensitive to the $HZZ$ and
$HZ\gamma$ couplings, while the vector-boson fusion processes (b) are
sensitive to the $HWW$ coupling for $f=\nu_e$, and the $HZZ$,
$HZ\gamma$, $H\gamma\gamma$ couplings for $f=e$. 

The matrix elements for the processes eq.(\ref{ffbarh}) can in general
be expressed as
\begin{eqnarray}
  M^{\lambda_1 \lambda_2}_{\sigma_1 \sigma_2} = 
  (M_{\rm SM})^{\lambda_1 \lambda_2}_{\sigma_1 \sigma_2} + 
  \sum_{i=1}^n c_i (M_i)^{\lambda_1 \lambda_2}_{\sigma_1 \sigma_2}
\end{eqnarray}
where $(M_{\rm SM})^{\lambda_1 \lambda_2}_{\sigma_1 \sigma_2}$ denotes
the SM helicity amplitude, and $c_i$ denotes the non-standard couplings
of eq.(\ref{effcoup_Operator}) that contribute to the process. The
matrix elements $(M_i)^{\lambda_1\lambda_2}_{\sigma_1 \sigma_2}$ give
the helicity amplitudes which are proportional to the coupling $c_i$.
If the $e^-$ and $e^+$ beam polarizations are $P$ and $\bar{P}$
($|P|,|\bar{P}| \le 1$) respectively, the differential cross section can
be expressed as 
\begin{eqnarray}
 d\sigma(P,\bar{P}) &=& 
  \Bigg[ \,\,\Sigma_{\rm SM} (P,\bar{P};\Phi_3) + \sum_{i=1}^n \,\, 
  c_i \,\, \Sigma_i (P,\bar{P};\Phi_3) \,\, \Bigg] \,\,\, d \Phi_3
  \label{dsigma}
\end{eqnarray}
where the non-standard couplings $c_i$ are assumed to be real and small,
and hence the terms quadratic in couplings are dropped. Here $d\Phi_3$
is the 3-body phase space volume of the $f\bar{f}H$ system, and
\begin{eqnarray}
 \Sigma_{\rm SM} (P,\bar{P};\Phi_3) = \frac{1}{2\,\, s} \,\,
  \sum_{\sigma_1,\sigma_2,\lambda_1,\lambda_2} 
  \left( \frac{1+\sigma_1\, P}{2} \right) \,
  \left(\frac{1+\sigma_2 \,\bar{P}}{2} \right) \,\,
  \bigg\vert (M_{\rm SM})^{\lambda_1 \lambda_2}_{\sigma_1 \sigma_2}
  \bigg\vert ^2 
\end{eqnarray}
gives the differential cross section of the SM. The term proportional to
$c_i$,
\begin{eqnarray}
 \Sigma_i (P,\bar{P};\Phi_3) = \frac{1}{2\,\, s} \,\,
  \sum_{\sigma_1,\sigma_2,\lambda_1,\lambda_2} 
  \left( \frac{1+\sigma_1 \, P}{2} \right) \, 
  \left( \frac{1+\sigma_2 \, \bar{P}}{2} \right) \,\, 2\,\, {\rm Re} 
  \bigg[ (M_i)^{\lambda_1 \lambda_2}_{\sigma_1 \sigma_2} 
  (M_{\rm SM}^*)^{\lambda_1 \lambda_2}_{\sigma_1 \sigma_2} \bigg]
\end{eqnarray}
gives the differential distribution which is proportional to  $c_i$.

In the optimal observables method, we make full use of the distribution
$\Sigma_i (P,\bar{P};\Phi_3)$ in order to constrain $c_i$. For instance,
if all $\Sigma_i (P,\bar{P};\Phi_3)$ have different shapes from each
other, then in principle, we can constrain all the coefficients $c_i$
simultaneously. For a given integrated Luminosity $L$, the statistical
errors of the $c_i$ measurement can be obtained from a $\chi^2$ function
\bsub
\begin{eqnarray}
 \chi^2(c_i) &=& \sum_{k=1}^N \,
  \left(\, \frac{N_{\rm exp}^k(P,\bar{P}) - N_{\rm
   th}^k(P,\bar{P};c_i)}{\sqrt{N_{\rm exp}^k(P,\bar{P})}} \,
  \right)^2 \label{chidef1} \\
 &=& \sum_{k=1}^N 
  \left( \frac{ L\, \Sigma_{\rm exp}(P,\bar{P};\Phi_3^k) \,
   \Delta - L\, [ \Sigma_{\rm SM}(P,\bar{P};\Phi_3^k) 
   + \sum_{i=1}^n\,c_i \, \Sigma_i(P,\bar{P};\Phi_3^k)] \,
   \Delta} {\sqrt{ L\, \Sigma_{\rm exp}(P,\bar{P};\Phi_3^k)\,
   \Delta}}  \right)^2, \label{chidef2}
\end{eqnarray}
\label{chidef}
\esub
\hskip -1.4mm
where $N_{\rm exp}^k(P,\bar{P})$ is the number of events in the k'th
bin, and $N_{\rm th}^k(P,\bar{P};c_i)$ is the corresponding prediction
of the theory which depends on the parameters of the SM and $c_i$. In the
second line (\ref{chidef2}), $\Phi_3^k$ for $k=1$ to $N$ gives the
representative phase space point of a bin number $k$ with the bin size
$\Delta$. Now, if all the coefficients $c_i$ are tiny, the experimental
result in the k'th bin should be approximated by the SM prediction as 
\begin{eqnarray}
 \Sigma_{\rm exp}(P,\bar{P};\Phi_3^k) \approx 
  \Sigma_{\rm SM}(P,\bar{P};\Phi_3^k).
  \label{ExpisSM}
\end{eqnarray}
The $\chi^2$ function can then be expressed as
\begin{eqnarray}
  \chi^2(c_i) = \chi^2_{\rm min} + \sum_{i,j} (c_i -\bar{c_i}) 
   \left[ V^{-1}_{(P,\bar{P})} \right]_{ij} (c_j - \bar{c_j})
  \label{chisq_cicj}
\end{eqnarray}
where 
\bsub
\begin{eqnarray}
 \left[ V^{-1}_{(P,\bar{P})} \right]_{ij} &=& L_0 \sum_{k=1}^N 
  \frac{\Sigma_i(P,\bar{P};\Phi_3^k) \; \Sigma_j(P,\bar{P};\Phi_3^k)}
  {\Sigma_{\rm SM}(P,\bar{P};\Phi_3^k)} \; \Delta 
  \label{Vinv-a} \\
 &\stackrel{N\to\infty}{\longrightarrow}& L_0 \int 
  \frac{\Sigma_i(P,\bar{P};\Phi_3) \; \Sigma_j(P,\bar{P};\Phi_3)}
  {\Sigma_{\rm SM}(P,\bar{P};\Phi_3)} \; d\Phi_3,
  \label{Vinv-b}
\end{eqnarray}
\label{Vinv}
\esub
\hskip -1.4mm
where we take $L_0=100\ifb$ as a nominal integrated luminosity through
out this report. If the total number of events is sufficiently large,
the integral representation in eq.(\ref{Vinv-b}) gives a good
approximation for the $V^{-1}$ matrix. The value of $\chi^2_{\rm min}$
and the mean value $\bar{c_i}$ depend on the actual experimental
results, or the small deviation from the equality in
eq.(\ref{ExpisSM}). If the SM prediction gives a reasonably good
description of the data in most of the phase space region, then the
statistical errors of $c_i$ and their correlations are determined solely
in terms of the covariance matrices $V$, which is the inverse of the
matrix given in eq.(\ref{Vinv});
\begin{eqnarray}
 c_i - \bar{c_i} = \pm \Delta c_i = \pm \sqrt{V_{ii}}, \hskip 0.5cm
  \rho_{ij} = V_{ij}/\sqrt{V_{ii} V_{jj}}.
\end{eqnarray}

In practice, however, we should address the following subtleties:
\begin{enumerate}
  \def\theenumi{(\roman{enumi})}
\item If the statistical error becomes small, systematic errors need to be 
  considered.
\item The results depend on how we split the total Luminosity to different 
  beam polarizations.
\item Not all the 3-body phase space points are observable in experiments.
\end{enumerate}
As for the first issue, we assume that the energy and angular
resolutions of ILC detectors are good enough to justify our integral
approximation of eq.(\ref{Vinv}), and consider only the impacts of the
luminosity uncertainty as a source of the systematic error which is
discussed in section \ref{luminos}. We leave the difficult problem of
background contaminations and the spectrum distribution due to
bremsstrahlung and beamstrahlung photon emissions to future studies. In
short, our results should be regarded as an ultimate accuracy of the
$HVV$ couplings measurement for a perfect detector in a background-free
environment, when the SM predictions are accurately known.

On the second point, we provide numerical results for the two very simple
cases only: 
\begin{enumerate}
 \item Unpolarized $e^{\pm}$ beam : The total integrated luminosity $L$
       is given for collisions with $(P,\bar{P})=(0,0)$ at each collider
       energy $\sqrt{s}$. However, in order to save the length of this
       article, we provide the unpolarized results specifically only for
       $s$-channel $ZH$ production at $\sqrt{s}=250\GEV$. They are
       calculated for all the processes at all energy choices and are
       used to evaluate the significance of the beam polarization after
       all the channels and energies are combined in section \ref{conclu}. 
 \item 80\% polarized $e^-$ beam : Exactly half of the total luminosity
       $L$ is given for collisions with $(P,\bar{P})=(0.8,0)$, and the
       remaining half with $(P,\bar{P})=(-0.8,0)$.
\end{enumerate}
In general, the covariance matrix depends on the partition of the total
luminosity into experiments with different sets of $e^-$ and $e^+$ beam
polarizations. If the $e^-$ and $e^+$ beams with polarizations
$P=\pm\big\vert P\big\vert$ and $\bar{P}=\pm \big\vert\bar{P}\big\vert$,
respectively, are available and if the integrated luminosity of
$L_{(P,\bar{P})}$ is distributed for each case, then by using the
definition of the inverse of the covariance matrix for the beam
polarization $(P,\bar{P})$ in eq.(\ref{Vinv}), the inverse of the total
covariance matrix is obtained as 
\begin{eqnarray}
&& V^{-1} = \hphantom{\frac{1}{1}} \nonumber \\
 & & \hskip -1.0cm 
  \frac{L_{(|P|,\bar{|P|})}}{L_0} V^{-1}_{(|P|,\bar{|P|})}
 + \frac{L_{(-|P|,\bar{|P|})}}{L_0} V^{-1}_{(-|P|,\bar{|P|})}
 + \frac{L_{(|P|,-\bar{|P|})}}{L_0} V^{-1}_{(|P|,-\bar{|P|})}
 + \frac{L_{(-|P|,-\bar{|P|})}}{L_0} V^{-1}_{(-|P|,-\bar{|P|})},
\end{eqnarray}
Our first case is simply $L(0,0)=L_{\rm total}$ and the second case
stands for
\begin{eqnarray}
 L_{(P=0.8,\bar{P}=0)} = L_{(P=-0.8,\bar{P}=0)} = \frac{L_{\rm total}}{2}.\label{polequipart}
\end{eqnarray}
It should be noted that an equal partition  of the total luminosity
as above is advantageous for the asymmetry measurements, and hence for
discriminating among different couplings, while
$L_{(P=-|P|,\bar{P}=|\bar{P}|)}=L_{\rm total}$ maximizes the $WW$-fusion
cross section.

We address the third point of limited observable kinematic distributions
in the following subsection.

\subsection{Observable phase space for each process}
\label{phasespacesubsec}
The observability of the three-body phase space point depends on 
species of $f\bar{f}$ pair, and also on  their energy and the
scattering angle in the laboratory frame. Let us discuss the following 4
cases one by one :
\begin{enumerate}
 \item $f = \mu, \tau$  
 \item $f = e$  
 \item $f = \nu_e, \nu_{\mu}, \nu_{\tau}$  
 \item $f = u, d, s, c, b$  
\end{enumerate}

First, for $f=\mu$ and $\tau$, all the four momenta $p_1,p_2,p_H$ are
measured and used in evaluating $V^{-1}$ to estimate the experimental
sensitivity. Since only one diagram fig.\ref{diagram}(a) contributes to
the cross section, a very tiny fraction of $\mu^{\pm}$ and $\tau^{\pm}$
escape detection along the beam pipe. We therefore use the whole phase
space region to evaluate the inverse of the covariance matrix
eq.(\ref{Vinv}). We also assume that the majority of $\tau^+ \tau^- H$
events can be made background free by selecting the events, in which the
$\tau^+ \tau^-$ invariant mass is reconstructed and required to match
the $Z$-boson mass. Although this reconstruction is not possible when
the Higgs boson decay has significant missing momentum, such as 
$H\to \tau^+\,\tau^-$ and $H\rightarrow W^+W^-$, we do not take account
of resulting reduction of the number of events, since it can be
considered as part of detection efficiency.

We do not consider the $\tau$ polarization in our analysis,
because it was found in ref.~\cite{kniehl} that its impact is not
significant once the $e^-$ beam polarization is available.

Second, for $f=e$, we can also measure  both $p_1$ and $p_2$
uniquely. For this process both the diagrams fig.\ref{diagram}(a) and
fig.\ref{diagram}(b) contribute to the cross section because of the
possible escape of $e^+$ or $e^-$ (or both) along the beam pipe, the
events can be divided into the following four classes :
\begin{enumerate}
 \item $ZH$ events are selected by requiring
       \begin{eqnarray}
        |\cos\theta_{e^{\pm}}| < 0.995, \hskip 0.5cm 
         |m_{e^+e^-}-\mz| < 5\,\,\Gamma_Z
         \label{ZH_cond}
       \end{eqnarray}
 \item Both outgoing $e^{\pm}$ are detected (double-tag events) when
       \begin{eqnarray}
        \pT (e^{\pm}) > 1\GEV, \hskip 0.5cm 
         |\cos\theta_{e^{\pm}}| < 0.995, \hskip 0.5cm 
         |m_{e^+e^-} - \mz| > 5\,\,\Gamma_Z \,\, .
         \label{Dtag_cond}
       \end{eqnarray}
\item When a photon is exchanged either from electron or positron in the
       $t$-channel, the corresponding outgoing $e^{\pm}$ tends to escape
       detection (single-tag events). Those events are selected by
       requiring 
       \bsub
       \begin{eqnarray}
         \pT (e^-) > 1\GEV, \,\, | \cos \theta_{e^-} | < 0.995 
          < | \cos \theta_{e^+} |,
       \end{eqnarray}
\noindent or
\begin{eqnarray}
         \pT (e^+) > 1\GEV, \,\, | \cos \theta_{e^+} | < 0.995 
          < | \cos \theta_{e^-} |,
       \end{eqnarray}
         \label{Stag_cond}
       \esub
 \item When a photon is exchanged from both the $e^\pm$ legs in
       the $t$-channel, both of the outgoing $e^{\pm}$ tend to escape
       detection (no-tag event). The selection conditions are
       \begin{eqnarray}
        | \cos \theta_{e^{\pm}} | > 0.995.
         \label{Notag_cond}
       \end{eqnarray}
\end{enumerate}
\label{eeHcondition}
Although the above classification misses a part of the phase space
where $|\cos\theta_{e^{\pm}}|<0.995$ and $\pT (e^{\pm})<1\GEV$, we find
that less than 0.001\% (0.02\%) of the total $e^+ e^-H$ events escape from
this region of phase space at $\sqrt{s}=500\GEV (250\GEV)$.

For the $ZH$ production case (\ref{ZH_cond}) and for the double-tagged
$eeH$ events (\ref{Dtag_cond}), we assume that the whole 3-body phase
space is observable, and the inverse of the covariance matrix $V^{-1}$
is calculated by integrating over the phase space with the selection
cuts. In case of no-tagged events (\ref{Notag_cond}), most of the events
are due to $\gamma \gamma$ fusion, and it is sensitive to the 
$H \gamma \gamma$ coupling. We estimate the cross section by using the
equivalent real photon approximation in section \ref{EPF}. Since typical
transverse momentum of the Higgs boson is smaller than the experimental
resolution, the only observable kinematic variable is the Higgs boson
rapidity
\begin{eqnarray}
  y_{\rm H} = \frac{1}{2} \,\,
  \ln \frac{\EH + {\pH}_Z}{\EH -{\pH}_Z}
\end{eqnarray}
where ${\pH}_Z$ is the momentum component along the $e^-$ beam direction. 
The cross section for the no-tag $(ee)H$ events can then be expressed as 
\begin{eqnarray}
  \frac{d \sigma}{d \yH} = \Sigma_{\rm SM}(\yH) 
  + \sum_i \,\, c_i \,\, \Sigma_i(\yH)
\end{eqnarray}
and accordingly, the inverse covariance matrix elements are calculated
in terms of $\Sigma_{\rm SM}(\yH)$ and $\Sigma_i(\yH)$. In our study, 
we find in section \ref{EPF} that 
$\Sigma_i(\yH)={\rm constant} \,\, \times \,\, \Sigma_{\rm SM}(\yH)$,
and hence only the total production cross section constrains the 
$H\gamma\gamma$ coupling.

The single-tag $(e)eH$ events, where either $e^-$ or $e^+$ escapes
detection, are sensitive to both  $H\gamma\gamma$ and $HZ\gamma$
couplings. We can again use the equivalent real photon approximation for
the emission from untagged $e^{\pm}$, and the differential cross section
is expressed as
\begin{eqnarray}
 d \sigma = 
  d \hat{\sigma} (\hat{s}=sz) \;\,\, D_{\gamma/e}(z) \; \,\, dz\, ,
  \label{EPF_Xsec}
\end{eqnarray}
where $D_{\gamma/e}(z)$ is the number density of an equivalent real
photon of momentum fraction $z$, and the subprocess cross section for
$\gamma e\rightarrow H\,e$ can be expressed as
\begin{eqnarray}
 \frac{d \hat{\sigma}}{d\cos \hat{\theta}} = 
  \hat{\Sigma}_{\rm SM} (\cos \hat{\theta}) 
  + \sum_i\,\, c_i\,\, \hat{\Sigma}_i (\cos \hat{\theta})
  \label{EPF_DifXsec}
\end{eqnarray}
where $\hat{\theta}$ is the scattering angle in the observed $e\,\, H$
rest frame. By inserting eq.(\ref{EPF_DifXsec}) into
eq.(\ref{EPF_Xsec}), we find
\bsub
\begin{eqnarray}
 d \sigma &=& 
  \big[ D_{\gamma/e}(z) \,\,\hat{\Sigma}_{\rm SM}(\cos \hat{\theta}) +
  \sum_i\,\, c_i \,\, D_{\gamma/e}(z)\,\, \hat{\Sigma}_i(\cos \hat{\theta}) 
  \big]  \;\,\,  dz \; \,\, d\cos \hat{\theta}, \\
 &\equiv& \big[ \Sigma_{\rm SM}(z,\cos \hat{\theta}) +
  \sum_i \,\, c_i \,\, \Sigma_i(z,\cos \hat{\theta}) \big] 
  \; \,\,dz \;\,\,  d\cos \hat{\theta} 
\end{eqnarray}
\esub
The inverse covariance matrix is then obtained by replacing 
$\Sigma_{\{ {\rm SM},i \}} (\Phi_3)$ by 
$\Sigma_{\{ {\rm SM},i \}} (z,\cos \hat{\theta})$ in eq.(\ref{Vinv}),
and $d\Phi_3$ by $dz\, d\cos \hat{\theta}$.

When $f=\nu_e$, $\nu_{\mu}$ or $\nu_{\tau}$, we can measure neither
$p_1$ nor $p_2$, but only the sum $p_1+p_2$ from the four-momentum
conservation. In this case, only the Higgs boson energy and momentum
are observable when Higgs boson decays to $b \bar b$ pair, and the
observable cross section in the laboratory frame is
\begin{eqnarray}
  d \sigma = {\rm B} (H\to b \bar b) \,\,
   \big[ \Sigma_{\rm SM} (\EH,\,\cos \thetaH) 
    + \sum_i\,\, c_i \,\, \Sigma_i (\EH,\,\cos \thetaH) \big] 
  \,\, d\EH \,\,d\cos \thetaH \, .
  \label{Wfusion_Xsec}
\end{eqnarray}
The covariance matrix is obtained by using the above distributions. In
practice, the Higgs boson energy $\EH$ from $ZH$ production is peaked in
the region
\begin{eqnarray}
 \big\vert m_{\nu\,\bar\nu}-\mz \big\vert = 
  \left\vert \big(s+\mhsq-2\,\,\sqrt{s} \,\,\EH \big)^{1/2}-\mz \right\vert 
 <  \,\, 5 \,\,\Gamma_Z,
 \label{ZHprodcond}
\end{eqnarray} 
and only the $\cos\thetaH$ dependence of the distributions is effective.

In case of $f=\nu_e$, in addition, the $W$ boson fusion diagram in
fig.\ref{diagram}(b) contributes. Since this amplitude is rather large
at high energies, $\sqrt{s}\,\, \gsim \,\, 500 \GEV$, we make a careful
study of the $W$-fusion contribution by using the Higgs-boson rapidity
$\yH$ and the transverse momentum $\pTH$; 
\begin{eqnarray}
 d \sigma = {\rm B} (H\to b \bar b) \,\,
  \left[ \Sigma_{\rm SM}(\yH,\,\pTH) 
   + \sum_i \,\,c_i \,\,\Sigma_i(\yH,\,\pTH) \right]
  \;\,\, d\yH \;\,\, d\pTH\,\, .
  \label{Wfusion_dsigma}
\end{eqnarray}
The $W$-fusion events are selected by requiring 
\begin{eqnarray}
 \big\vert m_{\nu\,\bar\nu} -\mz \big\vert  = 
  \left\vert \big(s+\mhsq-2\,\,\sqrt{s} \,\, \EH \big)^{1/2}-\mz \right\vert 
  > \,\,5 \,\, \Gamma_Z,
  \label{Wfusion_cond1}
\end{eqnarray}
and in order to avoid contamination with the $\gamma \gamma$ fusion
events, we impose an additional constraint
\begin{eqnarray}
  \pTH > 10\GEV\, .
  \label{Wfusion_cond2}
\end{eqnarray}

When $f=q=u,d,s,c,b$, we cannot distinguish $f$ from $\bar{f}$
efficiently. In reference \cite{kniehl}, impacts of partial
identification of $b$ from $\bar{b}$ has been studied, and they are
found to be negligibly small. In this study, therefore, we do not
distinguish $q$ from $\bar{q}$ at all, and we do not distinguish quark
flavors. In this limit, the hadronic decays of the $Z$ boson gives two
jets with momenta $p_1$ and $p_2$ for $q$ and $\bar{q}$, respectively,
but we cannot distinguish the events from those where $q$-jet has the
momentum $p_2$ and $\bar{q}$-jet has $p_1$. The appropriate optimal
observables are then obtained as follows:
\begin{eqnarray}
 d\sigma = \frac{1}{2}\,\, 
  \left\{ \Sigma_{\rm SM}(\Phi_3) + \Sigma_{\rm SM}(\tilde{\Phi}_3)
   + \sum_i \,\,c_i\,\,\Bigg [\Sigma_i(\Phi_3) 
   + \Sigma_i(\tilde{\Phi}_3)\Bigg] \right\} \,\,d \Phi_3
  \label{dsigma_qq}
\end{eqnarray}
where $\tilde{\Phi}_3$ is obtained from the phase space point $\Phi_3$
by interchanging $p_1$ and $p_2$. The three body phase space volume
$d\Phi_3$ is divided by 2 in order to account for the double
counting. The inverse covariance matrix is calculated as in equation
(\ref{Vinv}), while $\Sigma$'s are replaced by
$\Sigma(\Phi)+\Sigma(\tilde{\Phi})$, and the phase space measure is
divided by 2.

\subsection{cross section for each process}
\label{OptPhaseSpace-3}
\begin{figure}[t!]
 \begin{center}
  \includegraphics[width=12cm,height=15cm,angle=-90]{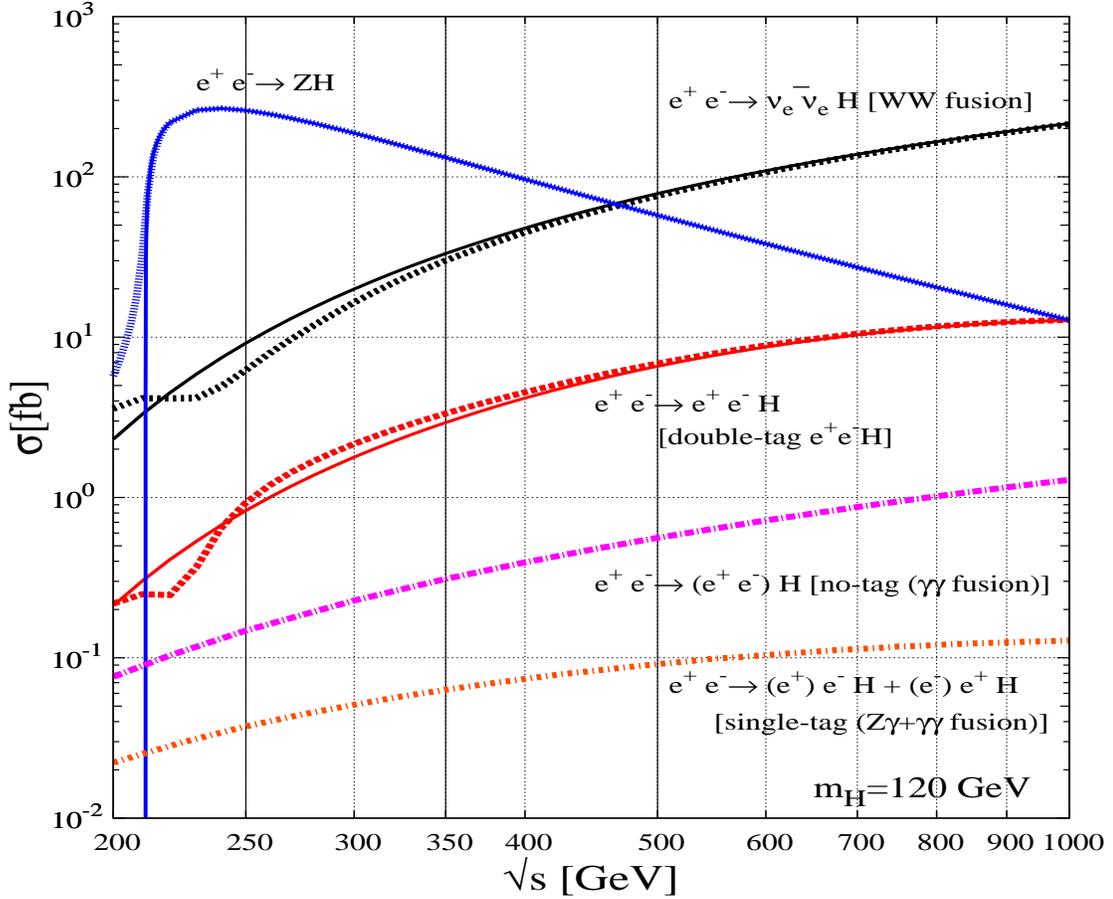}
  \caption{\small Total cross sections versus $e^+e^-$ collision energy
  $\sqrt{s}$ for the five processes that are sensitive to the $HVV$
  couplings at ILC. All the curves are for $\mh=120\GEV$. The $ZH$
  production cross section is the sum over all $Z\to f\bar f$ decay
  modes with $\vert m_{\nu\bar\nu}-m_Z\vert <5\,\Gamma_Z$, while the
  solid thin curves shows the $\Gamma_Z=0$ limit. The $\nu\bar\nu\,H$
  and the double tag $e^+e^-\, H$ events satisfy 
  $\vert m_{f\bar f}-\mz\vert >5\,\Gamma_Z$, and the tagged $e^{\pm}$
  has $\vert \cos\theta_{e^{\pm}}\vert<0.995$ and 
  ${p_{\rm T}}_{e^{\pm}}>1\GEV$, while  $\pTH>10\GEV$ is imposed on
  $\nu\bar{\nu}H$  process. The solid thin curves for 
 $e^+e^-\to \nu_e\bar\nu_e \,H$ and $e^+e^-\to e^+e^- \, H$ give the
  cross sections calculated from the $t$-channel $W$ and $Z$ boson
  exchange amplitudes only without imposing the invariant mass cut.}
 \label{rootsvariation}
 \end{center}
\end{figure}
Before we start examining the above processes one by one, we present here 
the total cross section of all the five processes for $\mh=120\GEV$ in 
fig.\ref{rootsvariation} as  a function of the c.m.\ energy
$\sqrt{s}$. All the curves show the SM prediction, which are obtained by
setting all $c_i$ to zero. As for $e^+e^- \rightarrow ZH$, we plot the
cross section for the sum of all the $Z$ boson decay modes when 
$\vert m_{f\bar f}-m_Z\vert<5\,\Gamma_Z$. The $\Gamma_Z=0$ limit result
is given by the thin curve. The cross section for 
$e^+e^- \rightarrow \nu_e \bar{\nu}_e H$ is obtained by requiring 
$\vert m_{\nu\bar \nu}-m_Z\vert>5\,\Gamma_Z$, eq.(\ref{Wfusion_cond1}),
and $\pTH >10$ GeV, eq.(\ref{Wfusion_cond2}), in order to suppress the
contribution from $Z \rightarrow \nu \bar{\nu}$ and 
$\gamma\gamma \rightarrow H$, respectively. The contribution from the
$WW$ fusion ($t$-channel $W$ exchange) process only without the 
$m_{\nu\bar{\nu}}$ cut is given by the thin curve, in order to show the
relevance of the interference effects and the $m_{\nu\bar{\nu}}$ cut at
low energies. The cross section for the process 
$e^+e^- \rightarrow e^+e^- H$ is separated into four cases; 
$Z\to e^+e^-$ $(\vert m_{ee}-\mz\vert<5\,\Gamma_Z)$, double-tag 
$(\vert m_{ee}-\mz\vert>5\,\Gamma_Z)$, single-tag and no-tag events,
where the final $e^+$ and $e^-$ are tagged when
$|\cos\theta_{e^{\pm}}|<0.995$ and ${p_{\rm T}}_{e^{\pm}}>1\GEV$. The
thin curve for the double tag $e^+e^-\,H$ cross section gives the
contribution from the $ZZ$ fusion only without the $m_{ee}$ cut. The
sign of the interference effect is opposite between the
$\nu_e\bar{\nu_e}H$ and $e^+e^-H$ process, because of the opposite
relative sign of the coupling factors between the $t$-channel and
$s$-channel amplitudes. In addition, when 
$\sqrt{s}\lsim \mh+\mz+5\,\Gamma_Z$, only the $m_{f\bar f}-\mz<0$ region
contribute with positive (negative) interference in $\nu_e\bar{\nu_e}H$
($e^+e^-H$) events, while at high $\sqrt{s}$ the negative (positive)
interference from the $m_{f\bar f}-\mz>0$ region dominates because the
magnitude of the $t$-channel amplitudes grow with $m_{f\bar f}$.

For the SM cross sections, we use the physical masses,
$\mw=80.423\GEV$, and $\mz=91.1876\GEV$ \cite{pdg2006}, and the
$\overline{\rm MS}$ couplings at the $\mz$ scale,
$\hat{\alpha}(\mz)=1/128.0$ and $\sin^2\hat{\theta}_{\rm W}(m_Z)=0.2312$ for
the gauge couplings, except when we use the equivalent real photon
distribution which is evaluated with $\alpha=1/137$. All the calculations
are done in the leading order of the perturbation theory, since none of
our results (errors and correlations) are sensitive to small differences
in the cross sections. We leave the important task of evaluating the
impacts of initial state radiation, both from  bremsstrahlung and
beamstrahlung, for  future studies.

We study the significance of each process in constraining the $HVV$
couplings quantitatively at four representative energies, $\sqrt{s}$ =
250, 350, 500$\GEV$ and $1\TEV$. The statistical errors are estimated for a
nominal integrated luminosity of $L_0=100\ifb$ at each energy. For
$\sqrt{s}$ = 250 and $350\GEV$, we consider only the $ZH$, 
$\nu_e \bar{\nu}_e H$ and double-tag $e^+e^-H$ processes, because the
single-tag and no-tag $e^+e^-H$ cross sections do not give sufficient
number of events in the SM. It should be noted that the $WW$-fusion
overtakes the $ZH$ production contribution at $\sqrt{s}=500\GEV$, and
even the cross section of the double-tag $eeH$ events from  $ZZ$ fusion
becomes comparable to that of the $ZH$ production at
$\sqrt{s}=1\TEV$. The single and no-tag $e^+e^-H$ events, which are
sensitive to the $HZ\gamma$ and $H\gamma\gamma$ couplings, respectively,
give sufficiently large cross sections only at $\sqrt{s}$ = $500\GEV$
and $1\TEV$. When we combine results from all the processes and from
all the energies, we examine the impacts of higher luminosity, by giving
$L=500\ifb$ at $\sqrt{s}$ = $500\GEV$ and at $\sqrt{s}=1\TEV$.
\begin{figure}[tbh!]
 \begin{center}
  \begin{minipage}[t]{0.33\textwidth}
   \includegraphics[width=12cm,height=5.5cm,angle=-90]
   {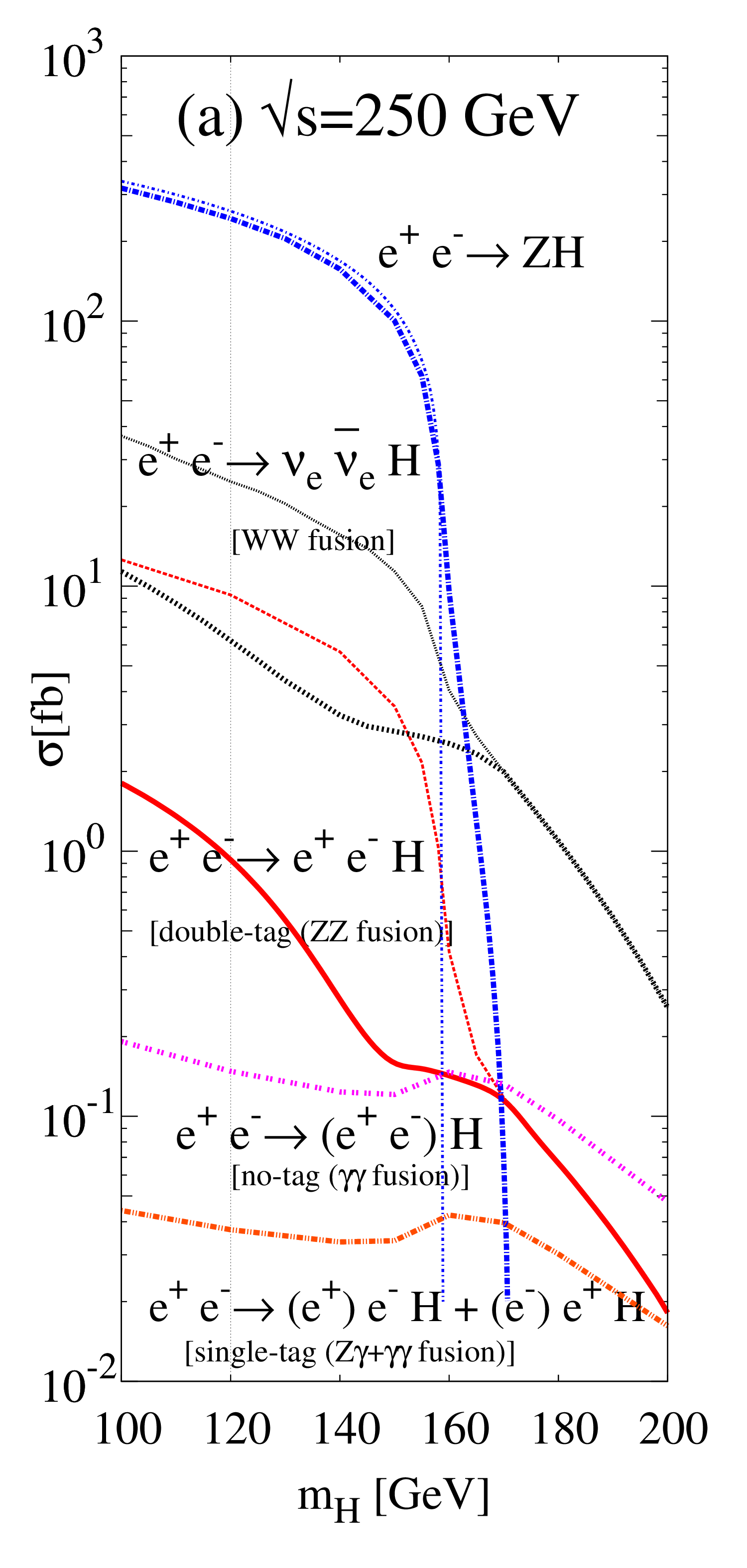}
  \end{minipage}
  \vspace{5mm}
  \begin{minipage}[t]{0.30\textwidth}
   \includegraphics[width=12cm,height=5cm,angle=-90]
   {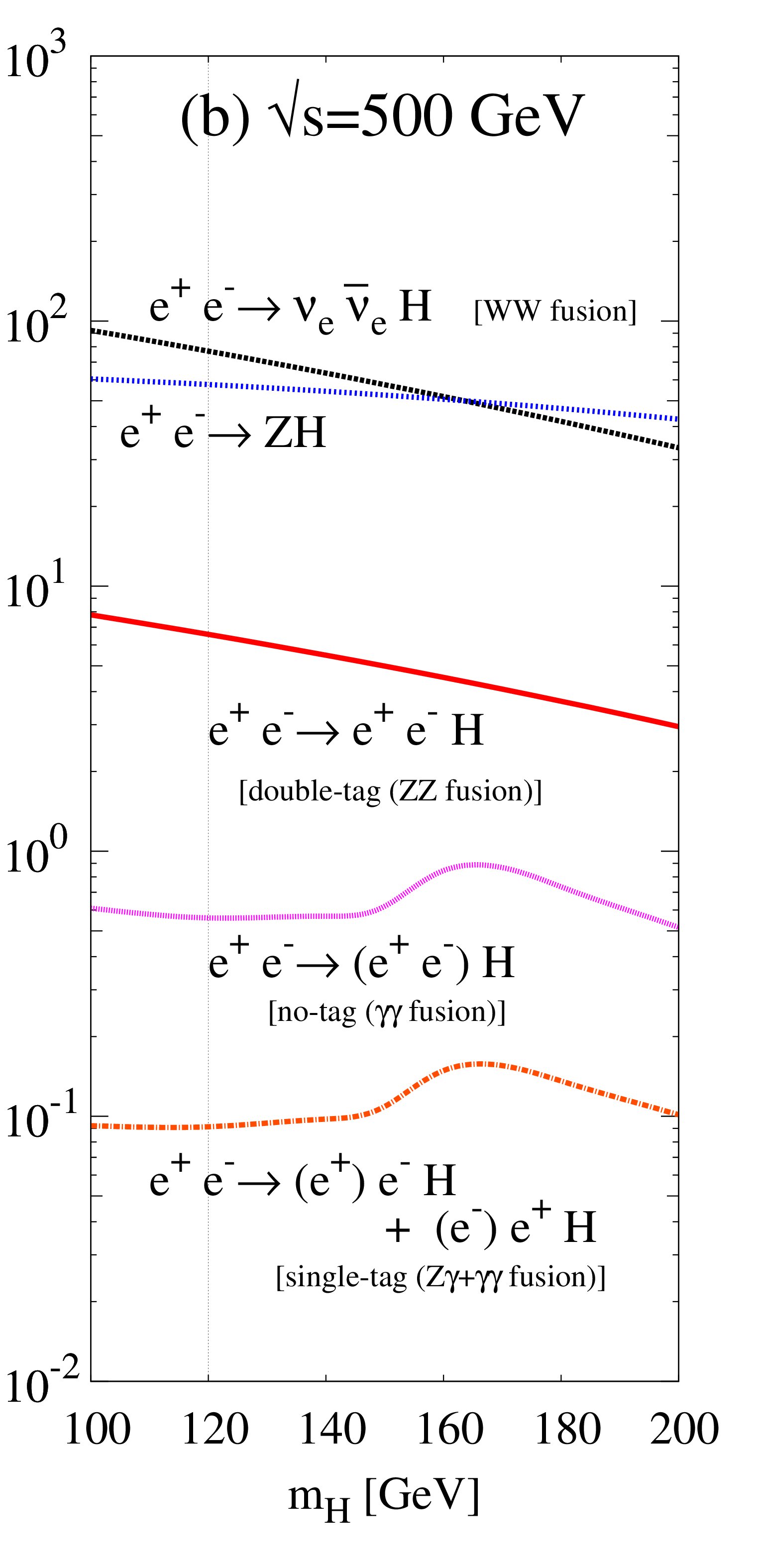}
   \end{minipage}
  \vspace{5mm}
  \begin{minipage}[t]{0.30\textwidth}
   \includegraphics[width=12cm,height=5cm,angle=-90]
   {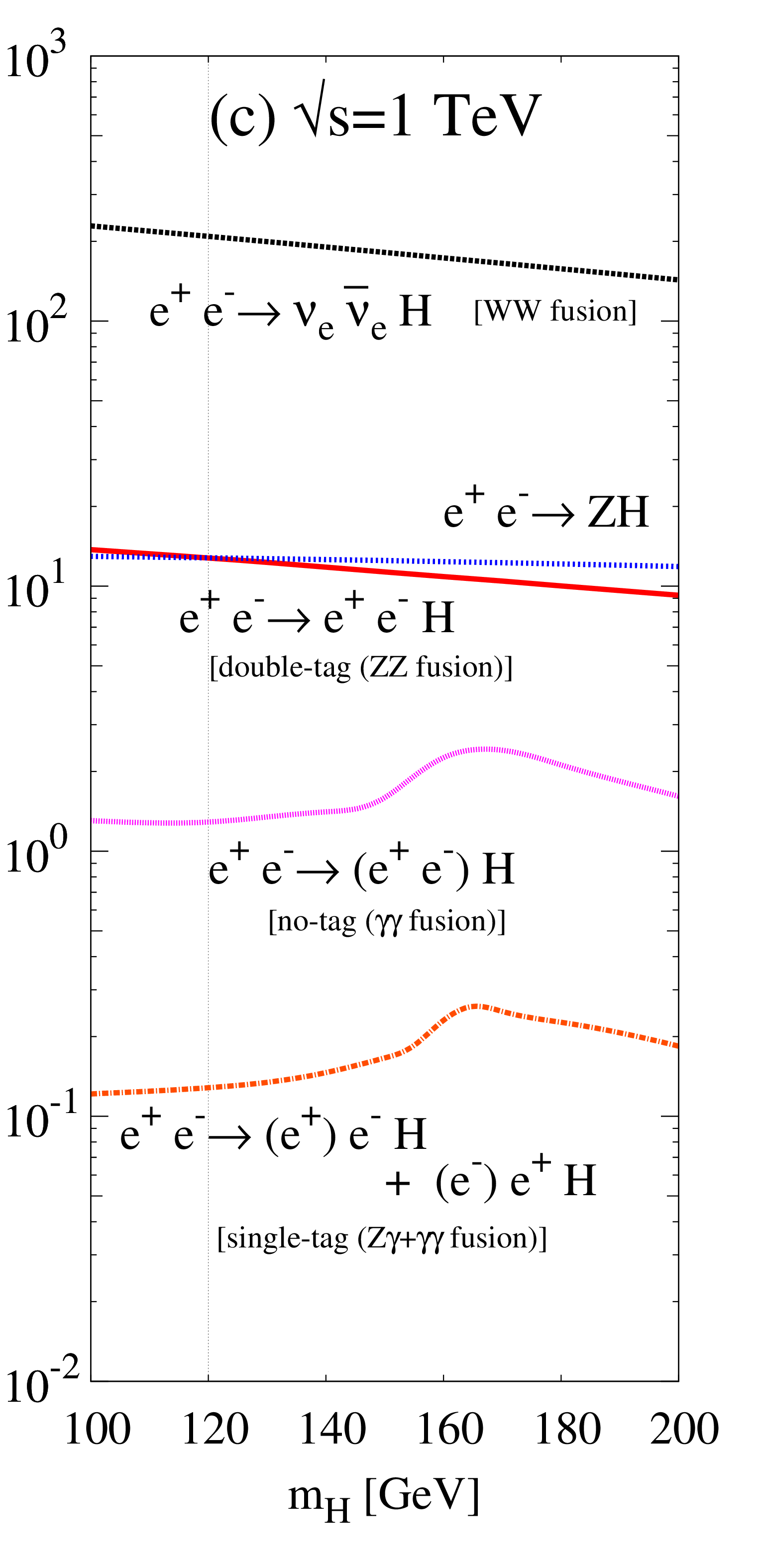}
  \end{minipage}
  \vskip -1cm
  \caption{\small Total cross sections versus $\mh$ for the five
  processes which are sensitive to the $HVV$ couplings at ILC, at
  (a) $\sqrt{s}=250\GEV$, (b) $500\GEV$  and (c) $1\TEV$. The tagged
  $e^{\pm}$ has $|\cos\theta_{e^{\pm}}|<0.995$ and 
  ${p_{\rm T}}_{e^{\pm}}>1\GEV$ in the laboratory frame. 
  $\vert m_{f\bar f} -m_Z\vert <5\,\,\Gamma_Z$ for $ZH$ production and
  $\vert m_{f\bar f} -m_Z\vert >5\,\,\Gamma_Z$ for $\nu_e\bar\nu_e\,H$
  and double tag $e^+e^-\,H$. The thin curves in (a) for 
  $\nu\bar\nu\, H$ and $e^+e^-\,H$ show the cross sections when the
  $Z\to f\bar{f}$ exclusion cut is removed, and that for $ZH$ shows the
  $\Gamma_Z=0$ limit. $\pTH>10\GEV$ is imposed on $\nu\bar{\nu}H$
  process.}
  \label{mhvariation}
 \end{center}
\end{figure}

The total cross sections are shown in fig.\ref{mhvariation} as functions
of the Higgs boson mass between $100\GEV$ and $200\GEV$ for (a)
$\sqrt{s}=250\GEV$, (b) $500\GEV$ and (c) $1\TEV$. The cross sections do
not depend strongly on $\mh$ for $\mh<200\GEV$, except for the dominant
$ZH$ production cross section at $\sqrt{s}=250\GEV$ which drops sharply
from 100~fb at $\mh\sim 150\GEV$ down to 1~fb at $\mh =165 \GEV$, and
vanishes at $m_H=\sqrt{s}-m_Z+5\,\Gamma_Z\sim 170\GEV$ because of the
final state cut $\big\vert m_{f\,\bar f} -m_Z\big\vert< 5\,\,\Gamma_Z$;
see eq.(\ref{ZH_cond}). The thin curve shows the zero-width limit. The
$ZH$ production process is dominant up to $\mh\sim 165 \GEV$, above
which the $WW$-fusion process becomes a main process, and its
cross section is above $\sim$1~fb up to $\mh\sim180 \GEV$.

In fig.\ref{mhvariation}a, we show two curves for the $\nu_e \bar{\nu_e} H$
and double-tag $eeH$ processes. The thick curves gives the cross
sections when the $Z\to f\bar{f}$ exclusion cut eq.(\ref{Wfusion_cond1})
is applied, and hence they may be regarded as those of $WW$ and $ZZ$
fusion events. On the other hand, the thin curves give the total cross
sections without the $Z\to f\bar{f}$ exclusion cut, and hence they receive 
contribution from the $ZH$ production amplitudes with 
$Z\to \nu_e \bar{\nu_e}$ or $e^+e^-$ transitions. Along the thick curves
for the $WW$ and $ZZ$ fusion events, the exclusion cut affects the
fusion cross sections below
$\mh=170\GEV\sim\sqrt{s}-\mz+5\,\,\Gamma_{Z}$, which remain almost
constant down to $\mh=150\GEV \sim \sqrt{s}-\mz-5\,\,\Gamma_{Z}$, where
the increase of the overall phase space is compensated by the increase
in the cut-out phase space region. 

At $\sqrt{s}= 500\GEV$ in fig.\ref{mhvariation}b, all the cross
sections of the $t$-channel processes increase, and that of the
$s$-channel $ZH$ production process decreases. The $ZH$ production and
the $WW$-fusion processes have almost the same cross section, over 30~fb
up to $\mh=200\GEV$, and the cross section of the double-tag $eeH$ event
is about one order of magnitude smaller than that of 
$\nu_e\bar\nu_e\, H$. The cross sections of the no-tag and the
single-tag $eeH$ events are still small, whose maximum is 1~fb and
0.2~fb, respectively, for $\mh \sim 2\,\,\mw$, where the one-loop
$H\gamma\gamma$ and $HZ\gamma$ vertices receive the $W$ pair threshold
enhancement.

At $\sqrt{s}= 1\TEV$ in fig.\ref{mhvariation}c, the cross section of the
$WW$-fusion process is above 100~fb, while those of the $ZH$ production
and the double-tag $eeH$ processes are $\sim$10~fb, almost independent
of $\mh$ up to $\mh=200\GEV$. As in fig.\ref{mhvariation}b for $\sqrt{s}=500$ GeV, the
cross section of the no-tag and single-tag $eeH$ processes take the
maximum values of  2~fb and 0.2~fb, respectively, around
$\mh\sim2\,\,m_W$. 

\section{$e^+e^- \rightarrow \nu_e \bar{\nu}_e H$ WW-fusion process}
\label{WWFusion} 
The $HWW$ coupling is best measured in WW-fusion production of the Higgs
boson, see fig.\ref{diagram}b, in the process
\begin{eqnarray}
  e^{-}\left(k_{1}, \frac{\sigma_{1}}{2}\right) + e^{+}\left(k_{2} ,
  \frac{\sigma_{2}}{2}\right) &\rightarrow & 
  \nu_e \left( \pone, \frac{\lambda_{1}}{2} \right) 
  + \bar{\nu}_e \left( \ptwo, \frac{\lambda_{2}}{2} \right) + H(\pH) ,
  \label{nunuh}
\end{eqnarray}
where $k_i$ and $p_i$ denote four-momenta of each particles,
$\sigma_i/2$ demote $e^{\pm}$ helicities, and $\lambda_i/2$ demote
$\nu_e$ or $\bar{\nu}_e$ helicities. The WW-fusion process contribute
only to one helicity amplitude, 
$\sigma_1=-\sigma_2=\lambda_1=-\lambda_2=-\,$. Contamination of the
$HZZ$ couplings via $s$-channel $ZH$ production followed by 
$Z \rightarrow \nu \bar{\nu}$ decays, see fig.\ref{diagram}a, can
be avoided by choosing suitable cuts on the observed Higgs boson energy
eq.(\ref{ZHprodcond}). Therefore, the process with ``a Higgs + missing
energy'' in $e^+e^-$ annihilation can probe $HWW$ couplings
independently of the other vector boson couplings. Once the branching
fraction of $H\rightarrow WW^*$ decay is known, the measurement of the
$HWW$ coupling determines the total decay width of the Higgs boson in a
model independent manner. It then allows us to translate all the Higgs
boson branching ratio measurements into the measurements of the partial
widths, from which we can determine the magnitude of various Higgs boson
couplings. It is also worth repeating here, that the sensitivity of the
$HWW$ coupling measurement is expected to be better at high energies,
because the cross section grows as $\log (s/m_H^2)$ with energy, in
contrast to the $ZH$ production cross section which decreases as $1/s$;
see fig.\ref{rootsvariation}. 

\subsection{Helicity Amplitudes and Backgrounds}
\noindent 
The helicity amplitudes for the process given in
eq.(\ref{nunuh}) can be written as:
\begin{eqnarray}
 M_{\sigma_{1} \sigma_{2}}^{\lambda_{1} \lambda_{2}}  &=&  
  \frac{g^2_W}{2} \,\,\delta_{\sigma_1 -} \,\,\delta_{\sigma_2 +}\,\,
  \delta_{\lambda_1 -}\,\, \delta_{\lambda_2 +}\,\,
  j_{e^{-} (\sigma_1,\lambda_1)}^{\mu}\,\, \Gamma_{\mu \nu}^{HWW} 
 \,\, j_{e^{+} (\sigma_2,\lambda_2)}^{\nu}\,\,  D_{W}(t_{1}) \,\, D_{W}(t_{2}),
\label{WW_ME}
\end{eqnarray}
\noindent 
where $D_{W}(t_i)=1/(t_i-m_{W}^{2})$ are the propagator factors with
$t_i=(k_i-p_i)^2$, (i=1,2) and
\bsub
\begin{eqnarray}
 j_{e^{-} (-,-)}^{\mu} &=& \overline{u} \big(p_1,\, -\frac{1}{2} \big)  
  \,\, \gamma^\mu \,\, P_L \,\, u \big(k_1,\,-\frac{1}{2} \big), \\
 j_{e^{+} (+,+)}^{\nu} &=& \overline{v} \big(p_2,\, +\frac{1}{2} \big) 
  \,\, \gamma^\nu \,\, P_L \,\, v \big(k_2,\,+\frac{1}{2} \big),
\end{eqnarray}
\label{WWcurrent}
\esub
\noindent
\hskip -1.4mm
are the leptonic charged currents. The explicit  form of the non-zero
components of the massless currents in the laboratory frame  are given
in eqs.(\ref{app2:current}) in Appendix \ref{app:current}. The $HWW$
coupling can be read off from eqs.(\ref{HVVFormFactor}) and (\ref{WWh})
and expressed as
\begin{eqnarray}
 \Gamma_{\mu\nu}^{\rm HWW} &=& \gz\,\, \mz \Bigg[
  \left\{ 1+ \aw -\frac{2}{\mzsq} \,\,
   \left( \frac{1}{2}\,\,\left(\mhsq-t_1-t_2\right)\,\, \bw -\mhsq\,\, 
    \dw \right)
  \right\}\,\, g^{\mu\nu} \nonumber \\
 & & \hskip 3cm
  + \frac{2}{\mzsq} \,\, \left(\bw-\dw\right)\,\, 
  q_{2\mu} \,\, q_{1\nu} \Bigg],
\end{eqnarray}
\noindent 
where $q_i^{\mu}=k_i^{\mu}-p_i^{\mu}$ (i=1,2). By contracting the
$t$-channel currents with the generalized $HWW$ vertex using
eqs.(\ref{FHGfuncDef}) in Appendix \ref{app:current}, we can write the
helicity amplitude (\ref{WW_ME}) as 
\begin{eqnarray}
 M_{-+}^{-+} &=& M_{\rm SM}\,\,  
  \left[ 1+\aw +\frac{1}{\mzsq }\left\{ \bw\,\, \left( t_1+t_2-\mhsq 
                    + \frac{s\,\, {\cal G}}{4} \right) 
  + \dw \,\, \left( 2\, \,\mhsq -\frac{s\,\,{\cal G}}{4} \right)\right\}
            \right] \nonumber\\
 \end{eqnarray}
\noindent 
where
\bsub
\begin{eqnarray}
 M_{\rm SM} &=& \frac{g^3_W\,\, \mw}{2} \,\, D_W(t_1) \,\, D_W(t_2) \,\,
  {\cal F}(\theta_1,\theta_2),\\
 {\cal F}(\theta_1,\theta_2) &=& 2\,\,s\,\, \sqrt{x_1 \,\,x_2}\,\, 
  \cos\frac{\theta_1}{2} \,\, \sin\frac{\theta_2}{2}, \\
 {\cal G}(\theta_1,\theta_2,\phi) &=& 
  \left[ 2 -x_1 \,\,(1 +\cos\theta_1)
   + x_1\,\, \sin\theta_1 \,\, \cot\frac{\theta_2}{2} \,\, e^{-i\phi}
  \right] \nonumber \\
 & \times & 
  \left[ 2 -x_2 \,\,(1 -\cos\theta_2)
   + x_2 \,\, \sin\theta_2 \,\, \tan\frac{\theta_1}{2} \,\, e^{i\phi} 
  \right].
\end{eqnarray}
\label{WW_HA}
\esub
\noindent 
\hskip -.15 cm 
Here $\phi=\phi_1-\phi_2$, and $x_i$, $\theta_i$, $\phi_i$ are the
energy fraction, polar and azimuthal angles, respectively, of $\nu_e$
$(i=1)$  and $\bar\nu_e$ $(i=2)$ in the laboratory frame; see
eq.(\ref{oldlabmom}) in Appendix \ref{app:phasespace}.

The SM contribution to the squared matrix element is
\begin{eqnarray}
 F_{\rm SM} = 
 \big \vert {M_{\rm SM}}_{-+}^{-+}\big\vert^2 
  &=& \big\vert {M_{\rm SM}}\big\vert^2 
  = \frac{g^6 \,\, \mwsq}{4} \,\, \bigg\vert D_W(t_1) \,\, D_W(t_2) \,\,
  {\cal F} (\theta_1, \theta_2)\bigg\vert^2 , 
\end{eqnarray}
and the distributions $F_{i} $  for each non-standard effective
couplings are 
\bsub
\begin{eqnarray}
 F_{\aw} &=& 2 \,\, {\rm Re} \big[ {M_{\aw}}_{-+}^{-+} \,\, 
  \left({M_{\rm SM}}_{-+}^{-+}\big)^\star \right]= 2 \,\,F_{\rm SM} 
  \label{WW_Faw} \\
 F_{\bw} &=& 2 \,\, {\rm Re} \big[ {M_{\bw}}_{-+}^{-+} \,\, 
  \left({M_{\rm SM}}_{-+}^{-+}\big)^\star \right]
  = \frac{2}{\mzsq} \left[ t_1+t_2-\mhsq +\frac{s}{4}\,\, {\rm Re} \,
                    {\cal G} \right]  \,\,\, F_{\rm SM} \label{WW_Fbw} \\
 F_{\dw} &=& 2 \,\, {\rm Re} \big[ {M_{\dw}}_{-+}^{-+} \,\, 
  \left({M_{\rm SM}}_{-+}^{-+}\big)^\star \right]
  = \frac{2}{\mzsq} \,\, \left[ 2\,\, \mhsq  -\frac{s}{4} \,\, {\rm Re}\,
                          {\cal G} \right] \,\,\, F_{\rm SM}. \label{WW_Fdw}
\end{eqnarray}
\label{WW_Fci}
\esub

As explained in section \ref{OptPhaseSpace}, since we cannot observe
$\nu_e$ and $\bar\nu_e$ momenta, we cannot make use of all the
distributions $F_i$ in eq.(\ref{WW_Fci}) to constrain the couplings
$c_{iWW}$. In fact, we can measure only the Higgs boson momenta, $\EH$
and $\cos\thetaH$, or $\pTH$ and $\yH$. In order to obtain the
observable weight functions, we perform the integration over the
$\nu_e$ and $\bar{\nu}_e$ momenta by using the 3-body phase space
parametrization in eq.(\ref{3bdyphsp-2}) as
\begin{eqnarray}
 d\Phi_3 &=& \frac{\sqrt{s}}{ 512 \,\, \pi^4} \, \,\,
  \bar{\beta}\,\,\, \Big( \frac{m^2_{\nu \bar{\nu}}}{s},\,\,\frac{\mhsq}{s}
                   \Big) \,\,\,
  d(\cos\theta^*) \, \,\,d\phi^* \, d\pTH \, \,\,d\yH\,\, ,
\end{eqnarray}
where $m_{\nu \bar{\nu}}^2=s+\mhsq-2\sqrt{s}\,\,\EH$, and 
$\EH=\sqrt{\pTH^2+\mhsq}$. The observable differential cross section
with initial $e^-$ and $e^+$ beam polarizations  $P$ and $\bar{P}$,
respectively, is
\begin{eqnarray}
 \frac{d^2 \sigma(P,\,\, \bar{P})}{d\yH \,\, \, d\pTH} &=& 
  {\rm B} (H \to b \bar b) \,\,
  \left[ \Sigma_{\rm SM}(P,\,\, \bar{P};\,\,\, \yH,\,\,\pTH) +
   \displaystyle \sum_i \,\,\, c_i\,\,\,  \Sigma_{c_i}(P,\bar{P};\,\,\, 
   \yH,\,\, \pTH) \right].
  \label{WW_DiffXsec}
\end{eqnarray}
Since only the left-handed $e^-$ and right-handed $e^+$ contribute to the
cross section, the weight functions are 
\begin{eqnarray}
&& \Sigma_i (P,\,\, \bar{P};\,\,\, \yH,\,\, \pTH)\nonumber\\ 
&&\hskip 0.5 cm = \nonumber \frac{1}{2\,\, s}
 \left( \frac{1-P}{2} \right)\,\, \left(\frac{1+\bar{P}}{2}\right)\,\, 
  \frac{1}{512\,\, \pi^4}\,\,\bar\beta
  \Big(\frac{m_{ff}^2}{s},\,\, \frac{\mhsq}{s}\Big)\,\,
  \int F_i (x_1,\,x_2,\,\theta_1,\,\theta_2,\, \phi) \,\,\,
  d\cos\theta^*\,\,  d\phi^* \nonumber \\ \label{WW_DIST}
\end{eqnarray}
where $i=$SM, $\aw$, $\bw$ and $\dw$. The covariance matrix for the
measurements of $c_i\,=\,{\aw,\,\,\bw,\,\, \dw}$ with an integrated
luminosity of $L_{(P,\bar{P})}$ is now expressed as, 
\begin{eqnarray}
 \left[V^{-1}_{(P,\bar{P})}\right]_{ij} &=& 
  {\rm B} (H \to b \bar b) \,\, L_{(P,\bar{P})} \,\, 
  \int \frac{\Sigma_i(P,\, \bar{P};\, \yH,\,\pTH) \,\,  
  \Sigma_j(P,\,\bar{P};\,\yH,\,\,\pTH) }
  { \Sigma_{SM}(P,\,\bar{P};\,\yH,\,\pTH) }\,\,  
  d\yH \,d\pTH. \label{WW_Vinv1}
\end{eqnarray}
In the following numerical studies, we set B$(H\to b \bar b)=0.9$ for
the branching fraction at $\mh=120\GEV$.

Before proceeding to the sensitivity analysis of the anomalous
couplings, we briefly remind ourselves of  the potential backgrounds to
this measurement. 
\begin{enumerate}
 \item No tag events in the process $e^+e^- \to e^+e^-H$ via the
       $t$-channel $\gamma \gamma$, $\gamma Z$ or $ZZ$ fusion
       contribute, but the Higgs-boson cannot have large
       ${p_T}_H$ in order for the $e^\pm$ to escape detection. We impose
       a selection cut $\pTH > 10\GEV$, eq.(\ref{Wfusion_cond2}),
       which is sufficient to suppress the no-tag $(ee)H$ contribution
       even at $\sqrt{s} \sim 1\TEV$. Contribution from $\gamma \gamma$
       fusion is negligibly small at $\pT\gsim$ a few GeV, as we will
       show in section \ref{EPF}, and that from $ZZ$ fusion is estimated
       to  about 5.4\%, 4.8\% and 3.9\% of the signal respectively, in
       the three smallest $\pTH$ bins, $10-20$, $20-30$ and $30- 40\GEV$
       at $\sqrt{s}=1\TEV$.
  \item The $s$-channel $ZH$ production process, where the $Z$ boson
        decays into a  $\nu \bar{\nu}$ pair, can also be a
        background to this measurement at low energies. Since the
       background events have missing mass peaked at $\mz$, we impose
       the $E_H$ cut, eq.(\ref{Wfusion_cond1}), which removes those
       events whose missing mass lies within 5 $\Gamma_Z$ of the $Z$
       boson mass.
 \end{enumerate}
Although small number of background events will survive the above cuts,
especially in the presence of initial state radiation and finite
detector resolution, we expect that their effect can be controlled and
that they will not affect the main conclusions of the following
analysis.

\subsection{Sensitivity analysis of the $HWW$ couplings}
\subsubsection{Sensitivity at ILC-I for  $\sqrt{s}=500$ GeV }
We first perform the binned analysis for single and double distribution
at $\sqrt{s}=500$ GeV. Since the weight functions of
eq.(\ref{WW_DiffXsec}) depend only on $\yH$ and $\pTH$, we calculate the
expected number of events in a  $(\yH,\,\,\pTH)$ bin  with the bin width
of $\Delta \yH=0.05$ and $\Delta \pTH\, =\,10 \GEV$. 

We show in figure \ref{fig:pT} the weight functions $\Sigma$'s
integrated over the rapidity, $\yH$.
\begin{figure}[htb!]
 \begin{center}
  \includegraphics[width=7.5cm,height=10cm,angle=-90]{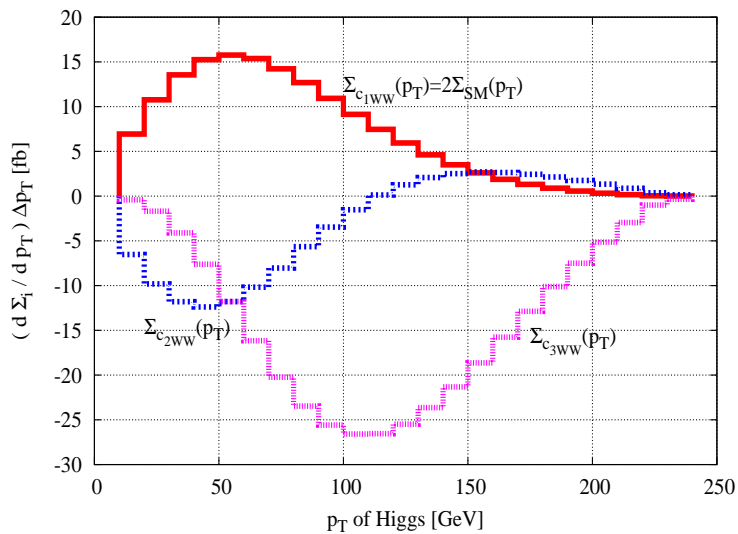}
  \caption{\small Histogram showing the $\pT$ distributions of the Higgs
  boson, where the differential cross section is integrated over $\yH$
  in each $\pTH$ bin of $10\GEV$ width at $\sqrt{s}=500\GEV$ for
  $\mh=120\GEV$. $\Sigma_{\rm SM}$ gives the SM distribution, and
  $\Sigma_{c_i}$ shows the coefficients of the non-standard $HWW$
  couplings $c_i=(\aw, \bw, \dw)$}
  \label{fig:pT}
 \end{center}
\end{figure}
This plot shows that the contribution from the operators corresponding
to the couplings, $\aw$, $\bw$ and $\dw$ can in principle be
differentiated  by using the $\pTH$ distribution, because they have
different shapes. $\Sigma_{\aw}$ and $\Sigma_{\dw}$ has a peak at
different $\pTH$, while $\Sigma_{\bw}$ changes sign. It is not clear,
however, how well they can be distinguished.

The covariance matrix method gives a quantitative answer  to this
question. Using the $\pTH$ distribution, we obtain the matrix elements
of the inverse covariance matrix as
\begin{eqnarray}
 (V^{-1})_{ij} = {\rm B}(H\to b \bar b) \,\, L_0 \,\, 
  \sum_{k}\,\,  \frac{\Sigma_i(\pTH^{(k)})\,\, \Sigma_j(\pTH^{(k)})}
  {\Sigma_{\rm SM}(\pTH^{k})} \,\,\, \Delta \pTH,
\end{eqnarray}
where $\pTH^{(k)}$ denotes the center of each $\pTH$ bin. Since the
covariance matrix is the same for the unpolarized case $P=\bar P=0$, and
for the polarized $e^-$ beam case with equi-partition of the total
luminosity into $P=\vert P\vert$ and $P=-\,\vert P\vert$, as in
eq.(\ref{polequipart}), we show our results without specifying the beam
polarization. It should be noted, that, because of the simple
polarization dependence of the $WW$ fusion cross section, as shown in
eq.(\ref{WW_DIST}), the covariance matrix for the general case of
asymmetric partition of the total luminosity is simply obtained by
adjusting an overall normalization factor which is proportional to the
total number of events. 

We find for $\sqrt{s}=500\GEV$ and $L_0=100\ifb$, 
\begin{eqnarray}
 \begin{array}{c}
  \Delta \aw = \,\,\pm\,\, .15 \\
  \Delta \bw = \,\,\pm\,\, .16 \\
  \Delta \dw = \,\,\pm \,\,.045 \\
 \end{array} \label{wlim1}
 & &
 \left(
  \begin{array}{ccc}
   1 \\
   .9986  &  1 \\
   .9989  &  .9982  & 1 \\
  \end{array}
 \right).
 \label{WW_500_Result_pTH-1}
\end{eqnarray}
We observe that the magnitudes of the correlation matrix elements are
quite large, implying that there is at least one combination of the
three couplings which cannot be measured accurately as compared to the
others. In order to identify which combination of the couplings are
measured accurately and inaccurately, we obtain the eigenvalues
$\lambda_i$ and the associated  eigenvector $\vec a_i$ of the covariance
matrix, in terms of which $V^{-1}$ and $V$ are expressed as  
\begin{eqnarray}
 V^{-1} &=& \sum_i\,\, \lambda_i^{-1} \,\, \vec{a_i}\,\, \vec{a_i}^{T}\,\, , 
  \hskip 1cm  {\rm or} \hskip 1cm
  V = \sum_i \,\, \lambda_i \,\,\vec{a_i} \,\, \vec{a_i}^{T}\,\, .
  \label{eigenV}
\end{eqnarray}
The resolving power of the measurements can be studied best by showing
the eigenvectors and their errors $\pm\,\sqrt{\lambda_i}$, 
\bsub
\begin{eqnarray}
 .22\,\,\aw +\,\,.061\,\,\bw -\,\,.97\,\,\dw &=& \,\,\pm \,\,.0020 \\
 .69\,\,\aw -\,\,.71\,\,\bw  +\,\,.12\,\,\dw &=& \,\,\pm \,\,.0058 \\
 .68\,\,\aw +\,\,.70\,\,\bw  +\,\,.20\,\,\dw &=&\,\, \pm \,\,.22 \,\, .
\end{eqnarray}
\label{WW_500_Result_pTH-2}
\esub
\noindent 
\hskip -.14 cm 
As anticipated, we find that the error of the combination
(\ref{WW_500_Result_pTH-2}c) is two orders of magnitude larger than
those of the other two. In fact, if we sum over the weight functions
given in eq.(\ref{WW_DIST}), for the combination of
eq.(\ref{WW_500_Result_pTH-2}c), we find that the sum almost
cancel out and hence this combination is poorly measured.

As above,  whenever the correlation matrix elements are large,  the
eigenvectors and their errors reveal much more information. Henceforth,
we present our results in terms of the eigenvectors and their errors of
the covariance matrix $V$, whenever they are more informative. 

In order to obtain the maximum  information from experiments, we further
study the two dimensional distributions of $\yH$ and $\pTH$ variables.
\begin{figure}[tbh!]
 \begin{center}
\hskip -4 cm
  \begin{minipage}[t]{0.30\textwidth}
  \par\hskip 1.25 cm  {(a) $\Sigma_{\aw}=2\,\Sigma_{\rm SM}$}
\vskip -1 cm
   \includegraphics[width=8cm,height=9cm,angle=-90]
  {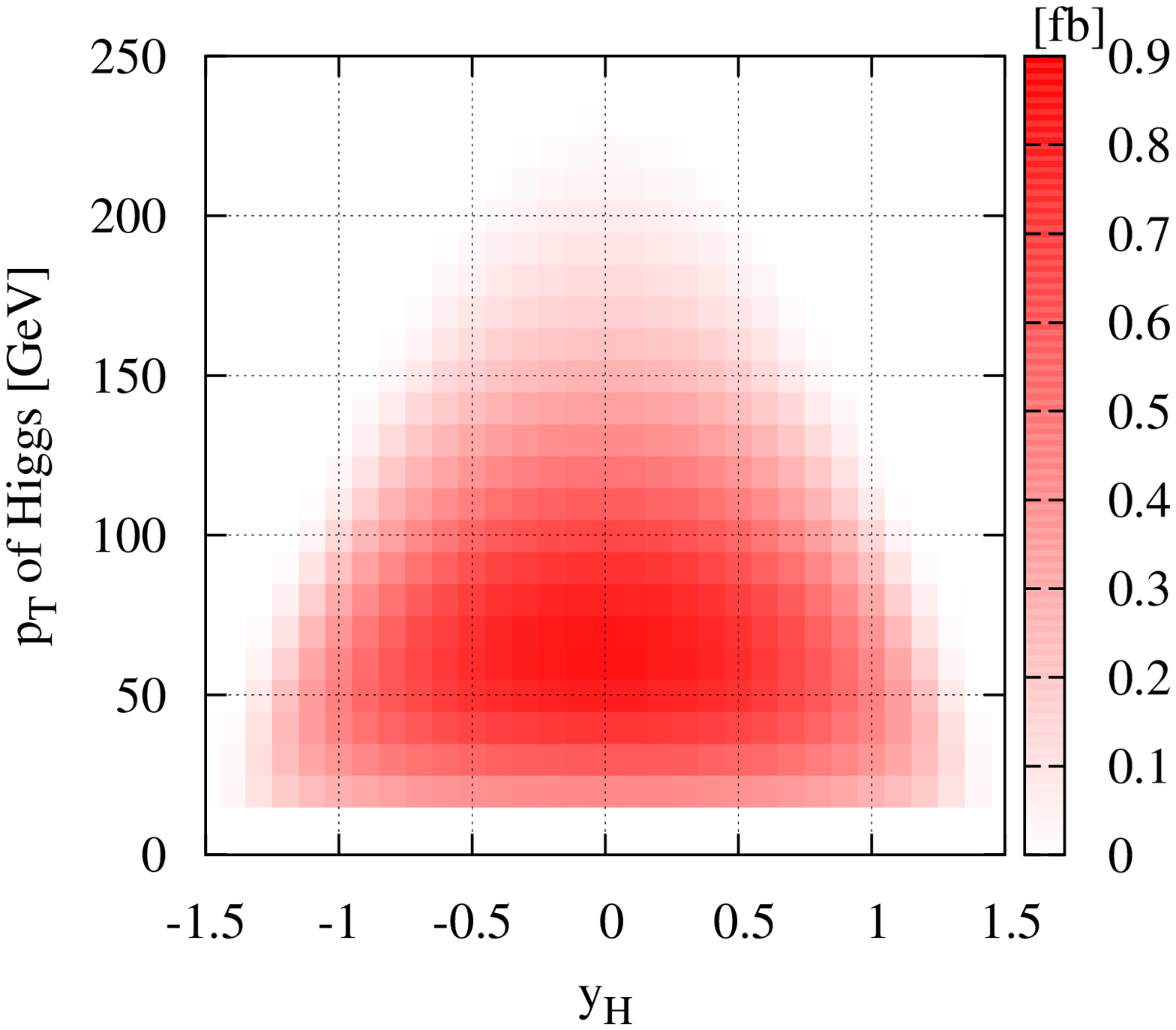}
  \end{minipage}
\hskip 0.25 cm
  \begin{minipage}[t]{0.30\textwidth}
   \par \hskip  2 cm {(b) $\Sigma_{\bw}$}
\vskip -1 cm
   \includegraphics[width=8cm,height=9cm,angle=-90]
  {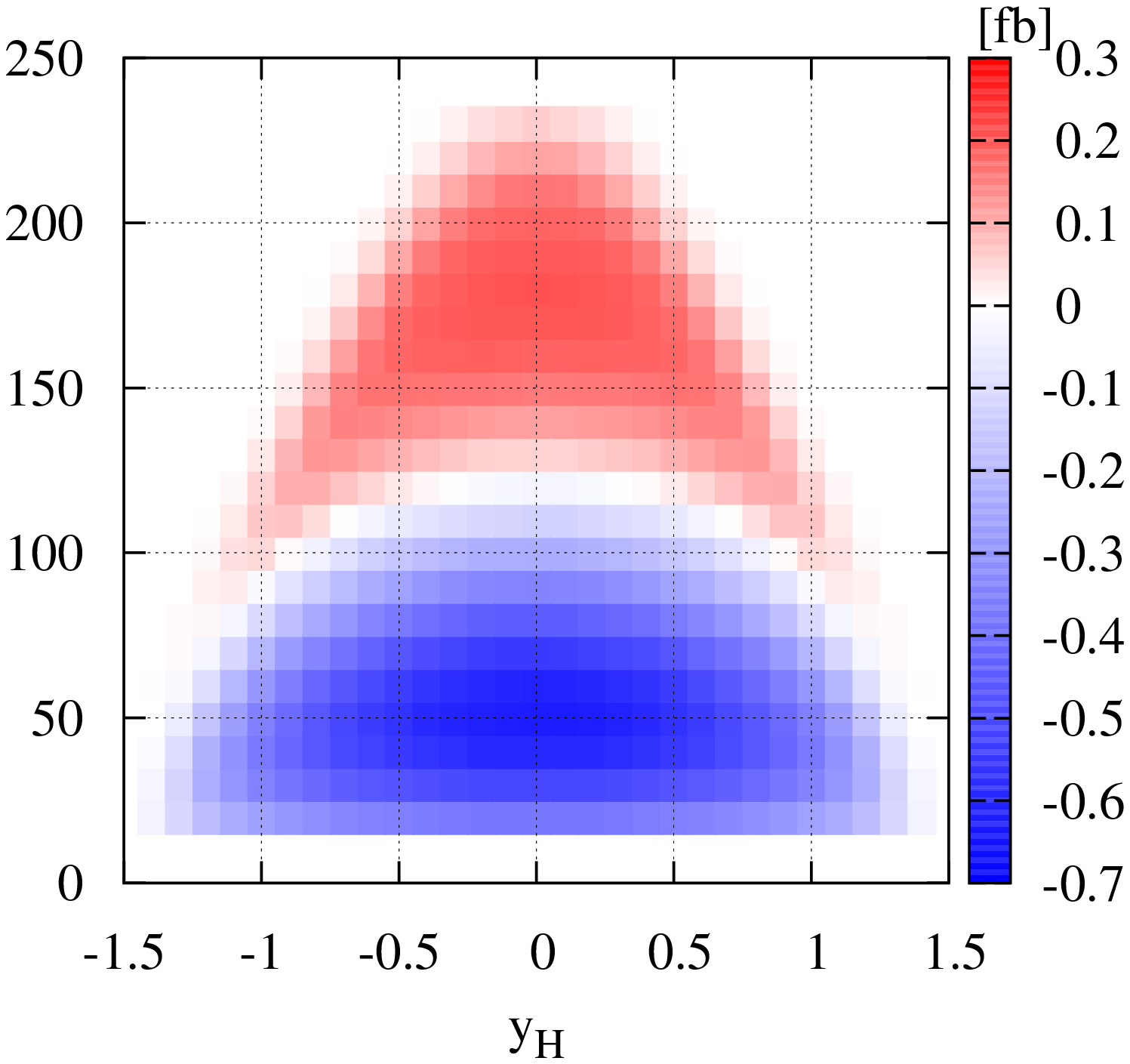}
  \end{minipage}
\hskip 0.25 cm
  \begin{minipage}[t]{0.30\textwidth}
   \par \hskip  2.5 cm{(c) $\Sigma_{\dw}$}
\vskip -1 cm
   \includegraphics[width=8 cm,height=9cm,angle=-90]
   {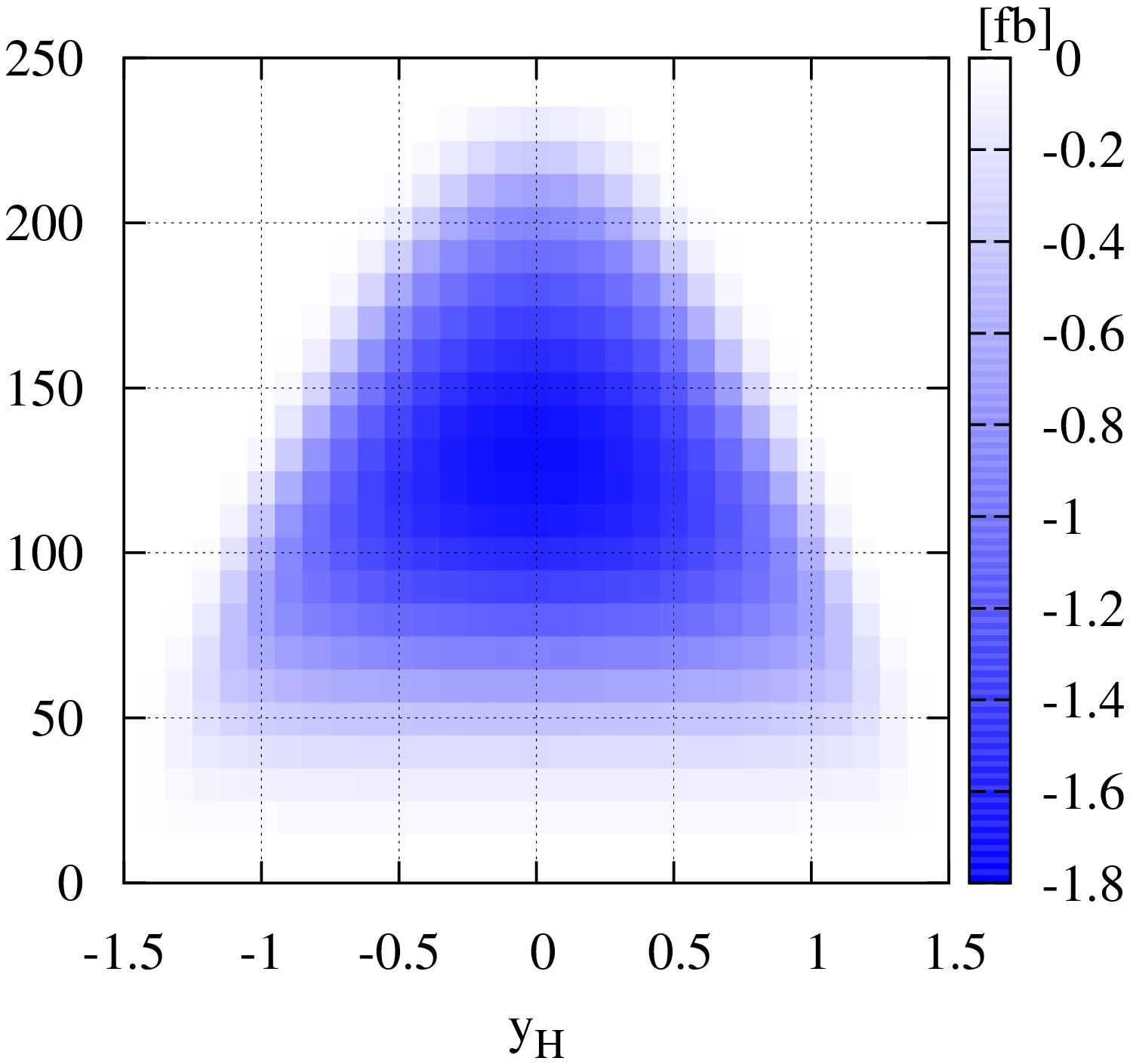}
  \end{minipage}
\vskip -0.75 cm
  \caption{\small Scatter plot  in the $\yH$-$\pTH$ plane for the
  coefficient of (a) $\aw$, (b) $\bw$ and (c) $\dw$. The  measure gives
  $( d^2 \sigma /d\yH d\pTH ) \Delta \yH \Delta \pTH$ in units of fb for
  the bin size of $\pTH=10\GEV$ and $\yH=0.05$, at $\sqrt{s}=500\GEV$
  for $\mh=120\GEV$.}
  \label{fcont}
 \end{center}
\end{figure}
\noindent 
We present the scatter plots of the weight functions of $c_{1WW}$,
$c_{2WW}$ and $c_{3WW}$ in figs.\ref{fcont}a, \ref{fcont}b and
\ref{fcont}c, respectively, on the $\yH$, $ \pTH$ plane. We find that
adding the rapidity distribution does not help much in resolving the
degeneracy, because the $\yH$ dependence of  the weight functions for
$\aw$, $\bw$ and $\dw$ are all similar. Only fig.\ref{fcont}b gives some
hint of an additional independent measurement as it has non-factorisable
dependencies in terms of $\yH$ and $\pTH$, unlike the other two cases.

By using the double differential distributions, we obtain the optimal
covariance matrix from 
\begin{eqnarray}
 \big(V^{-1}\big)_{ij} &=& {\rm B}(H\to b \bar b) \,\, L_0 \,\, 
  \sum_{l,\, m} \,\, 
  \left[ \frac{\Sigma_i \big(\yH^{(l)},\,\,\pTH^{(m)}\big) \,\,
   \Sigma_j \big(\yH^{(l)},\,\,\pTH^{(m)}\big) }
   { \Sigma_{\rm SM} \big(\yH^{(l)},\,\,\pTH^{(m)}\big)} \,\,
   \Delta \yH \,\,\Delta \pTH
  \right] ,
\end{eqnarray}
where we set the bin size $\Delta \yH=0.1$ and $\Delta \pTH=20\GEV$.
For the cross section of 77~fb (see figure \ref{rootsvariation}) at
$\sqrt{s}=500\GEV$, we expect 7700 events with $L_0=100\ifb$, when $e^-$
and $e^+$ beams are unpolarized, $P=\bar{P}=0$. The eigenvectors and
their errors are
\bsub
\begin{eqnarray}
 .24\,\,\aw\,\, +.045\,\,\bw -\,\,.97\,\,\dw &=& \,\,\pm \,\,.0020, 
  \label{w500a} \\
 .69\,\,\aw -\,\,.71\,\,\bw  +\,\,.13\,\,\dw &=& \,\,\pm \,\,.0059, \\
 .68\,\,\aw +\,\,.70\,\,\bw  +\,\,.20\,\,\dw &=& \,\,\pm \,\,.15,
\end{eqnarray}
\label{w500}
\esub
\noindent 
\hskip -0.14cm  
or in the standard representation
\begin{eqnarray}
 \begin{array}{c}
  \aw  =  \pm\,\,.099 \\
  \bw  =  \pm\,\,.10 \\
  \dw  =  \pm\,\,.029 \\
 \end{array}
 & &
 \left(
  \begin{array}{ccc}
   1 \\
   .997  &  1 \\
   .997  &  .995  & 1 \\
  \end{array}
 \right).
 \label{WWfusion_500-2}
\end{eqnarray}

When we compare the optimal result (\ref{WWfusion_500-2}) with the
previous one (\ref{WW_500_Result_pTH-1}) that used only the $\pTH$
distribution, we find more than 30\% improvements in the errors of all
the couplings and slight decrease in the correlations. While by
comparing the eigenvectors and their errors of the optimal results
(\ref{w500}) with (\ref{WW_500_Result_pTH-2}), we find that the only
effect of using the additional information is to reduce the error of the
least constrained combination by about 30\%. Neither the eigenvectors
nor the errors of the two accurately constrained combinations are
affected much by the optimization.

\subsubsection{Sensitivity at $\sqrt{s}=250$ GeV and 350 GeV with
   $t$-channel only}
The cross section for the $WW$ fusion is small at lower energies; 9.2~fb
at $\sqrt{s}=250\GEV$ and 33.2~fb at $350\GEV$ as can be seen from
the solid thin curve in fig.\ref{rootsvariation}, which further reduces
to 5.8~fb and 31.4~fb, respectively, after imposing the
$Z\to\nu\bar\nu$ exclusion cut $|m_{\nu\bar\nu}-\mz|>5\,\Gamma_Z$. Not
only the cross section is small but also contributions from the $ZH$
production amplitudes and the interference terms, as well as the effects
due to the $Z\to\nu\bar\nu$ exclusion cut, are significant at low
energies, as can be seen from the dashed thick curve in
fig.\ref{rootsvariation}. We first show the results of \lq
theoretical\rq~ studies based on the $WW$ fusion events only, which helps
us understanding the energy dependence of the $WW$ fusion amplitudes
when compared with the results at $\sqrt{s}$ = 500 GeV and 1 TeV. A more
realistic study will be given in the next subsection.

At $\sqrt{s}=250\GEV$ we will have only 580 events from $ WW$ fusion
with $L_0=100\ifb$ and with 100\% efficiency. Thus our integral
approximation for the covariance matrix, eq.(\ref{WW_Vinv1}), may not be
reliable. In order to examine the sensitivity of our results on the bin
size, we enlarge the bin size to $\Delta \yH=0.15$ and $\Delta
\pTH=30\GEV$ and count only the contributions from these bins with more
than 10 events for the nominal luminosity of $L_0=100\,\ifb$. We find
the eigenvectors and their $1\sigma$ errors at $\sqrt{s}=250\GEV$ with
$L_0=100\ifb$ and $P=\bar{P}=0$ to be 
\bsub
\begin{eqnarray}
 .61\,\,\aw -\,\,.045\,\,\bw -\,\,.79\,\,\dw &=& \,\,\pm\,\, .016, \\
 .58\,\,\aw -\,\,.65\,\,\bw  +\,\,.49\,\,\dw &=& \,\,\pm\,\, .029, \\
 .54\,\,\aw +\,\,.76\,\,\bw  +\,\,.38\,\,\dw &=& \,\,\pm\,\, 1.1.
\end{eqnarray}
\label{w250}
\esub
\noindent
\hskip -1.4mm
We observe that the nature of the eigenvectors are similar to those at
$\sqrt{s}=500\GEV$ in eq.(\ref{w500}), and also the hierarchy between
the largest error and the other two persists. The overall magnitude of
the errors are larger than those of eq.(\ref{w500}) at
$\sqrt{s}=500\GEV$ by a factor of 8, a half of which can be attributed
to the reduction of the cross section by a factor of
$13.5\sim(3.7)^2$. We confirm that both the eigenvectors and the
eigenvalues of eq.(\ref{w250}) obtained with the large bin size do not
differ much from those in the integral (infinitesimal bin size) limit.
The errors of the two most accurately measured combinations are
especially insensitive to the details of our binning procedure. 

We also examine the case at $\sqrt{s}=350\GEV$ with $L_0=100\ifb$, since
it is natural to study the top quark property in detail near the
$t\bar{t}$ threshold. We will have about 3,140 events and the
eigenvectors and their 1$\sigma$ errors for $P=\bar{P}=0$ are :
\bsub
\begin{eqnarray}
 .44\,\,\aw -\,\,.068\,\,\bw -\,\,.89\,\,\dw &=&\,\, \pm\,\, .0050, \\
 .63\,\,\aw -\,\,.68\,\,\bw  +\,\,.37\,\,\dw &=&\,\, \pm \,\,.011, \\
 .63\,\,\aw +\,\,.73\,\,\bw  +\,\,.26\,\,\dw &=& \,\,\pm \,\,.44.
\end{eqnarray}
\label{w350}
\esub
\noindent
\hskip -1.4mm
Again, we observe the similar hierarchy pattern.

\subsubsection{Sensitivity at $\sqrt{s}=250$ GeV and 350 GeV with
   $ZH$ interference}
Since the cross sections of $t$-channel $WW$ fusion process are small
at $\sqrt{s}$ = 250 and 350 GeV, contribution from the $ZH$ production
amplitudes is rather significant despite the $Z\to\nu\bar\nu$ exclusion cut 
$\vert m_{\nu\bar\nu}-\mz\vert>5\, \Gamma_Z$. At $\sqrt{s}=250\GEV$, the
cross section grows from 5.8~fb to 6.2~fb mainly because of the 0.7~fb
contribution from the $s$-channel amplitude squared, which overcomes the
-0.3~fb contribution from the destructive interference between the
$t$-channel and $s$-channel amplitudes.
On the other hand, the cross section at $\sqrt{s}=350\GEV$ decreases
slightly from 31.4~fb to 30.1~fb because of the destructive
interference. In addition, $ZH$ production followed by
$Z\to\nu_\mu\bar\nu_\mu$ and $Z\to\nu_\tau\bar\nu_\tau$ contributes with
1.4~fb at $\sqrt{s}=250\GEV$, and 1.0~fb at $\sqrt{s}=350\GEV$, after
the $Z\to \nu\bar\nu$ exclusion cut. These events are experimentally
indistinguishable from the $WW$ fusion $\nu_e\bar\nu_e\, H$
process. Therefore the total cross sections become 7.6~fb and 31.1~fb at
$\sqrt{s}=250\GEV$ and $350\GEV$, respectively.

\begin{figure}[htb!]
 \begin{center}
  \includegraphics[width=9cm,height=12cm,angle=-90]{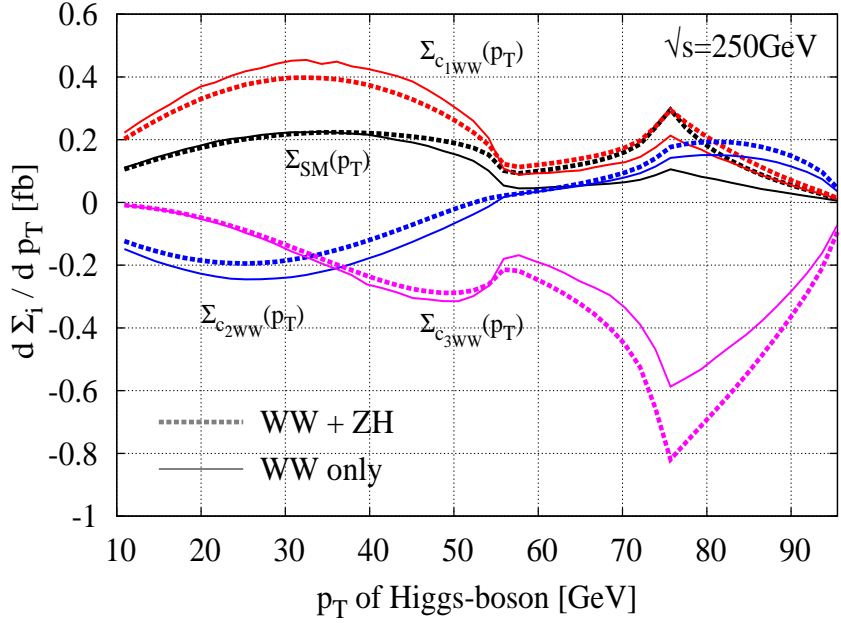}
  \caption{\small $\pT$ distribution of the weight function for the
  $\nu\bar\nu H$ process, where the differential distributions are
  integrated over $\yH$ at $\sqrt{s}=250\GEV$ for $\mh=120\GEV$. 
  $\Sigma_{\rm SM}$ gives the SM distribution, and
  $\Sigma_{c_i}$ shows the coefficients of the non-standard $HWW$
  couplings $c_i\equiv\aw, \bw, \dw$. The thick dashed lines show the
  weight functions that include both the $t$- and $s$-channel
  contributions, while the thin lines are those with the $t$-channel
  contribution only. The thick dashed curve for the SM distribution,
  $\Sigma_{SM}$, is the sum over all the neutrino species 
  $(\nu_e,\, \nu_\mu,\, \nu_\tau)$.}
  \label{fig:pT250}
 \end{center}
\end{figure}
We show in fig.\ref{fig:pT250} the $\pTH$ distribution of the SM,
$\Sigma_{SM}$, and the weight functions $\Sigma_i$ at
$\sqrt{s}=250\GEV$. The thin lines show the $WW$ fusion contribution
only, which are the $\yH$ integral of the 2-dimensional weight functions
used to obtain the results of eq.(\ref{w250}) and eq.(\ref{w350}). The
thick dashed lines are obtained after the interference between the
$t$-channel $WW$ fusion and the $s$-channel $ZH$ production amplitudes
are taken into account. Since we can safely neglect the non-standard
contributions in the sub-dominant $ZH$ production amplitudes, we can
constrain the $HWW$ couplings by using the two dimensional weight
functions $\Sigma_i(\yH,\pTH)$ of eq.(\ref{WW_DIST}), including the
interference contributions, where only the SM amplitudes in
eq.(\ref{WW_Fci}) are replaced by the sum of the $WW$ fusion and the
$ZH$ production amplitudes. It should also be noted that the thick
dashed SM curve in fig.\ref{fig:pT250} includes the contributions
from all the neutrino flavors, which cannot be distinguished from the
signal. The limited phase space at low energies gives rise to the
complex $\pT$ dependence at 56~GeV and 75~GeV, which are the Higgs boson
momenta when $m_{\nu\bar\nu}=m_Z+5\,\Gamma_Z$ and
$m_{\nu\bar\nu}=m_Z-5\,\Gamma_Z$, respectively. We find that there are
large destructive interference effects for $\aw$ and $\bw$ at low
$\pTH$. On the contrary the $\bw$ and $\dw$ curves show strong
constructive interferences at high $\pTH$.

By using the two-dimensional weight functions, we find the eigenvectors
and their $1\sigma$ errors at $\sqrt{s}=250\GEV$ with $L_0=100\ifb$ and
$P=\bar{P}=0$ to be
\bsub
\begin{eqnarray}
 .51\,\,\aw +\,\,.085\,\,\bw -\,\,.86\,\,\dw &=& \,\,\pm\,\, .017, \\
 .71\,\,\aw -\,\,.61\,\,\bw  +\,\,.36\,\,\dw &=& \,\,\pm\,\, .030, \\
 .49\,\,\aw +\,\,.79\,\,\bw  +\,\,.37\,\,\dw &=& \,\,\pm\,\, .51 \, .
\end{eqnarray}
\label{w250stcomb}
\esub
\hskip -.14 cm 
Comparing the results of eq.(\ref{w250stcomb}) with those of
eq.(\ref{w250}) from the $t$-channel contribution only, we find that the
errors of the first two eigenvectors are almost the same despite the growth
of the SM cross section. It is because the general growth of the
magnitude of the SM distribution and the weight functions at
$\pTH>55\GEV$ is compensated by the decrease of the weight functions at
low $\pTH$. The third error decreases because it becomes more difficult
to obtain a combination of the three weight functions which cancel out.

Similarly, we present the results for $\sqrt{s}$ = 350 GeV with
$L_0=100\ifb$, where we have included the $s$-channel $ZH$ contribution.
\bsub
\begin{eqnarray}
 .39\,\,\aw +\,\,.032\,\,\bw -\,\,.92\,\,\dw &=&\,\, \pm\,\, .0048, \\
 .66\,\,\aw -\,\,.71\,\,\bw  +\,\,.25\,\,\dw &=&\,\, \pm \,\,.010, \\
 .64\,\,\aw +\,\,.71\,\,\bw  +\,\,.29\,\,\dw &=& \,\,\pm \,\,.14\, .
\end{eqnarray}
\label{w350stcomb}
\esub
\hskip -.14 cm 
Comparing eq.(\ref{w350stcomb}) with eq.(\ref{w350}), the errors of the
first two eigenvectors are reduced slightly, while the largest error is
reduced by a factor 3. Again it is a consequence of the complicated
interference patterns for the three weight functions.

\subsubsection{Sensitivity at $\sqrt{s}=1$ TeV }
Fig.\ref{rootsvariation} shows that the $WW$ fusion cross section grows
with the c.m.\ energy, and reaches 210~fb at $\sqrt{s}=1\TEV$. Precision
studies of Higgs boson properties, including the  $HWW$ couplings, will
be one of the major motivations of the energy upgrade to ILC-II. We find
the eigenvectors and their errors for $L_0=100\ifb$ and $P=\bar{P}=0$,
\bsub
\begin{eqnarray}
 .14\,\,\aw +\,\,.038\,\,\bw -\,\,.990\,\,\dw  &=& \,\,\pm \,\,.00079 \\
 .71\,\,\aw -\,\,.70\,\,\bw  +\,\,.072\,\,\dw &=& \,\,\pm \,\,.0034 \\
 .69\,\,\aw +\,\,.71\,\,\bw  +\,\,.12\,\,\dw  &=&\,\, \pm \,\,.070
\end{eqnarray}
\label{w1tev}
\esub
\noindent 
\hskip -1.4mm
The errors are now a factor of 2 to 3 smaller than those at $500\GEV$ in
eq.(\ref{WW_500_Result_pTH-2}). The reduction factor is bigger than the
naive expectation from the cross section ratio 
$(77{\rm fb}/210{\rm fb})^{1/2} \sim 0.6$. This is because of the $s$
dependence of the weight functions of $\bw$ and $\dw$, in
eqs.(\ref{WW_Fbw}) and (\ref{WW_Fdw}), which grows linearly with $s$. On
the other hand, the reduction is not so strong as we would expect from
the linear growth of the weight functions. We find that the power of the
weight functions that grow with $s$ for $\bw$ and $\dw$ in
eqs.(\ref{WW_Fbw}) and (\ref{WW_Fdw}), respectively, is greatly reduced
because of the non-observability of the $\nu_e$ and $\bar\nu_e$ momenta,
which results in the integration of the factor 
${\cal G}(\theta_1,\,\theta_2,\,\phi)$ in eq.(\ref{WW_DIST}) for the
observable weight functions. 

It is instructive to study, just as a reference, how well we could have
measured  the $HWW$ couplings if $\nu_e$ and $\bar{\nu}_e$ momenta were
measurable. We would then be able to use the dependence of the weight
functions on the full 3-body phase space, $x_1$ or $x_2$,
$\cos\theta_1$, $\cos\theta_2$, $\phi_1-\phi_2$ in the $e^+e^-$ c.m.\
frame; see Appendix \ref{app:phasespace} for our phase space
parameterizations. By using the integral approximation of
eq.(\ref{WW_Vinv1}), we find at $\sqrt{s}=1\TEV$,
\bsub
\begin{eqnarray}
 .084\,\,\aw +\,\,.14\,\,\bw  -\,\,.986\,\,\dw   &=& \,\,\pm \,\, .00083 \\
 .099\,\,\aw -\,\,.986\,\,\bw  -\,\,.013\,\,\dw  &=& \,\,\pm \,\,.0017 \\
 .992\,\,\aw  +.\,\,087\,\,\bw +\,\,.097\,\,\dw  &=& \,\,\pm \,\,.0064.
\end{eqnarray}
\label{w1tev-2}
\esub
\hskip -1.4mm 
It is remarkable that,  no hierarchy among the three eigenvalues
survives and that all the three couplings are measured accurately and
rather independently,  as is clear from the single coupling dominance
 of the three weight functions in
eq.(\ref{WW_Fci}). It is perhaps most remarkable here that the error of the
most accurately measured combination of the three couplings in a
realistic environment, 0.00075 in eq.(\ref{w1tev}a), is not much
different from the corresponding one, 0.00079 in
eq.(\ref{w1tev-2}a).

\subsubsection{Beam Polarization Effects}
All the above results are obtained for unpolarized $e^+$ and $e^-$
beams, $P=\bar{P}=0$. Since only left-handed $e^-$ and right-handed
$e^+\,(\sigma_1=-\sigma_2=-)$ contribute to the $WW$ fusion process,
the polarization dependence of our results for 
$e^+e^-\to \nu_e\bar\nu_e\,H$, can be obtained in a straight forward
manner. If, for instance, an integrated luminosity of $L_{(P,\bar{P})}$
is devoted to experiments with $e^-$ beam with polarization $P$ and
$e^+$ beam with polarization $\bar{P}$, then the total covariance matrix
is simply
\begin{eqnarray}
 \bigg[V^{-1}_{(P,\bar{P})}\bigg]_{ij} = \frac{L_{(P,\,\bar{P})}}{L_0} \,\,  
  (1- P)\,\,(1+\bar P)\,\, \bigg[V^{-1}_{(0,\,0)}\bigg]_{ij} \,\, ,
  \label{WW_Lum_and_Vinv}
\end{eqnarray}
\noindent 
where $L_0=100\ifb$ is the nominal integrated luminosity, adopted
throughout this report. It is clear from this expression that if we
distribute the luminosity equally to all four combination of $e^{\pm}$
polarizations ; 
\begin{eqnarray}
 L_{(\vert P\vert,\vert\bar{ P}\vert)} = 
  L_{(\vert P\vert,-\vert\bar{ P}\vert)} = 
  L_{(-\vert P\vert,\vert\bar{ P}\vert)} = 
  L_{(-\vert P\vert,-\vert\bar{ P}\vert)} = \frac{L_{\rm total}}{4}\,\, ,
  \label{WW_Lum}
\end{eqnarray}
then the total covariance matrix is identical to that of the unpolarized
case. The same applies for our choice of  eq.(\ref{polequipart}). On the
other hand, if we devote the entire luminosity with fixed polarization,
such as 
\begin{eqnarray}
 L_{(-\vert P\vert,\vert\bar{ P}\vert)} = L_{\rm total},
\end{eqnarray}
then the inverse of the covariance matrix will be larger by a factor of
$(1+\vert P\vert)(1+\vert\bar{ P}\vert)$, simply because of the large
$WW$ fusion cross section. 

\section{ZH Production Process}
\label{ZHProd}
In $ZH$ production via $s$-channel $\gamma$ and $Z$ exchange, we
can study $HZ\gamma$ and $HZZ$ vertices. Unlike the $WW$ fusion process
all the final states are observable when the $Z$ boson decays into a
pair of charged leptons or quark jets. The technique of optimal
observables provides us with high discriminating power among the couplings.

\subsection{Helicity Amplitudes, Backgrounds and Event Selection}
The momentum and helicity assignment in $ZH$ production followed by 
$Z \rightarrow f\bar{f}$ decay is
\begin{eqnarray}
  e^{-}\left(k_{1}, \frac{\sigma_1}{2}\right) 
   + e^{+}\left(k_{2}, \frac{\sigma_2}{2}\right) &\rightarrow& 
   Z(p_Z,\lambda) + H(\pH), \nonumber \\
 & &\hskip 0.75 cm  \drsh \,\,\,\, \,f \left( p_1,\frac{\lambda_1}{2}\right)
  + \bar{f} \left( p_2,\frac{\lambda_2}{2}\right)
  \label{ZH_process}
\end{eqnarray}
We neglect the mass of $e^{\pm}$ and the outgoing fermions, and the
fermionic chirality conservation tells $\sigma_1=-\sigma_2=\alpha$, and
$\lambda_1=-\lambda_2=\beta$. Then the helicity amplitudes
\begin{eqnarray}
&& M^{e^+e^- \rightarrow ZH \rightarrow f\bar{f}H}_{\alpha,\beta} 
  = \nonumber \\
 & &\hskip 0.5 cm  j^{\mu}_{1 \alpha} \,\,
  \Bigg[ g^{\gamma ee}_{\alpha} \,\, D_{\gamma}(s) \,\, 
  \Gamma^{HZ\gamma}_{\mu\nu} 
  + g^{Zee}_{\alpha} \,\, D_{Z}(s)\,\, \Gamma^{HZZ}_{\mu\nu} \Bigg]\,\, 
  \frac{\sum_{\lambda}\,\, \epsilon^{\nu *}(p_Z,\lambda)\,\, 
  \epsilon_{\rho}(p_Z,\lambda)}
  {p_Z^2 - \mzsq + i\,\,\mz \,\,\Gamma_Z} \,\, g^{Zff}_{+\,\beta}\,\,  
  j_{2 \beta}^\rho
  \label{ZH_ME-1}
\end{eqnarray}
where 
\bsub
\begin{eqnarray}
 j^\mu_{1\alpha}&=&
  \bar{v}(k_2,\,\,-\,\alpha /\,2)\,\,\gamma^\mu \,\, P_{\alpha}\,\, 
  u(k_1,\,\,\alpha /\,2) \label{anncurrent}\\
 j^\nu_{2\beta}&=&
  \bar{u}(p_1,\,\,\beta/\,2)\,\,\gamma^\nu \,\,P_{\beta}\,\,
  v(p_2,\,\,-\,\beta/\, 2) \label{zcurrent}
\end{eqnarray}
\esub
\noindent
are the $e^+e^-$ annihilation current (\ref{anncurrent}) and the $Z
\rightarrow f\,\,\bar{f}$ current (\ref{zcurrent}), respectively. Here
$\alpha$ and $\beta$ denotes the sign of the $e^-$ and $f$ helicity,
respectively, and $P_{\alpha}$ is the chirality projection operator
$P_{\alpha}=(1+\alpha\,\, \gamma_5)/2$, sometimes referred to as
$P_-=P_L$ and $P_+=P_R$. The propagators and the couplings are denoted
as
\bsub
\begin{eqnarray}
 D_{\gamma}(s) &=& 1/\,s, \;\;\;\; 
  D_Z (s) = 1/\,(s-\mzsq+i\,\,\mz\,\,\Gamma_Z), \\
 g_{\alpha}^{\gamma ff} &=& e \,\, Q_f, \;\;\;\; 
  g_{\alpha}^{Zff} = \gz \,\, \left( T_3^{f\alpha} -Q_f \,\,\swsq\right),
  \label{ZH_Zffcoupling}
\end{eqnarray}
\label{ZH_PropCoup}
\esub
\hskip -.14 cm 
where $Q_f$ and $T_3^{f\alpha}$ are the electric charge and the weak
isospin of the fermion $f$ with chirality $\alpha$.

In the vicinity of the $Z$ boson resonance, $m_{f\,\bar f}\sim m_Z$, 
we can factorize the amplitudes of eq.(\ref{ZH_ME-1}) into the 
$e^+e^- \rightarrow ZH$ production part and the 
$Z \rightarrow f\bar{f}$ decay part, summed over the decaying $Z$ boson
helicities ;
\begin{eqnarray}
 M^{e^+e^- \rightarrow ZH \rightarrow f\bar{f}H}_{\alpha \beta} =
  \sum_{\lambda} \,\, T^{ZH}_{\alpha \lambda}\,\, D_Z(p_Z^2) \,\,
  T^{Z \rightarrow f\bar{f}}_{\lambda \beta}
\end{eqnarray}
The production amplitudes $T^{ZH}_{\alpha \lambda}$ are obtained in the
$e^+e^-$ collision c.m.\ frame. Here $\lambda \equiv (+,\,\,0,\,\,-)$ is
the helicity of the produced $Z$ boson.  By using the generic $HZZ$ and
$HZ\gamma$ vertices of eqs.(\ref{ZZh2}) and (\ref{ZAh1}), respectively,
and by using the $e^+e^-$ annihilation currents of
eq.(\ref{schancurrent}) in Appendix \ref{app:current}, we find for
$p_Z^2=m_Z^2$
\begin{eqnarray}
 T^{ZH}_{\alpha \lambda}
  &=& j^{\mu}_{1 \alpha} \,\, 
  \Bigg[ g^{\gamma ee}_{\alpha} \,\, D_{\gamma}(s) \,\,
  \Gamma^{HZ\gamma}_{\mu\nu} 
   + g^{Zee}_{\alpha} \,\, D_{Z}(s) \,\, \Gamma^{HZZ}_{\mu\nu} \biggr]\,\, 
  \epsilon^{\nu *}(p_Z,\lambda) \\
  &=& \hat{M}_{\alpha}^{\lambda} \,\,\, 
   d^1_{\alpha,\lambda}(\thetaZ),\label{prodamp}
\end{eqnarray}
where
\bsub
\begin{eqnarray}
 \hat{M}_{\alpha}^{\lambda=0} &=& \displaystyle \sum_{V=\gamma,Z}\,\,
  g_{\alpha}^{Vee} \,\,\gz\,\, \sqrt{2\,\,s} \,\,E_Z \,\,D_V(s) \,\,
  \left( h_1^{ZV} + h_2^{ZV} \,\, \frac{\sqrt{s} \,\,E_Z\,\, \beta_Z^2}{\mzsq}
  \right), \\
 \hat{M}_{\alpha}^{\lambda=\pm} &=& \displaystyle \sum_{V=\gamma,Z}\,\,
  g_{\alpha}^{Vee} \,\,\gz \,\,\sqrt{2\,\,s}\,\,\mz \,\,D_V(s) \,\, h_1^{ZV},
\end{eqnarray}
\label{ZH_ME-2}
\esub
\noindent
\hskip -1.4mm
with $\beta_Z=\sqrt{1- m_Z^2 / \, E_Z^2}$ and
\bsub
\begin{eqnarray}
 h_1^{ZZ} &=& 1 + \az + \cz\,\, \frac{s+\mzsq}{\mzsq} 
  + \big(\bz-\cz\big)\,\,\frac{s+\mzsq-\mhsq}{\mzsq}, \label{ZH_h1Z} \\
 h_2^{ZZ} &=& -\,\,2\,\,\big(\bz-\cz\big), \\
 h_1^{Z\gamma} &=& 2\,\, \cg\,\, \frac{s}{\mzsq} +
  \big(\bg-\cg\big)\,\,\frac{s+\mzsq-\mhsq}{\mzsq}, \\
 h_2^{Z\gamma} &=& -\,\,2\,\,\big(\bg-\cg\big). \label{ZH_h2a}
\end{eqnarray}
\label{ZH_FF}
\esub
\noindent 
\hskip -1.4mm
In eq.(\ref{prodamp}), the $d$ functions 
\begin{eqnarray}
 d^1_{\alpha,\lambda=0}(\theta) = - \frac{\alpha \sin\theta}{\sqrt{2}}&,& 
  \hskip 0.3cm
  d^1_{\alpha,\lambda=\pm}(\theta)=\frac{1+\lambda\alpha\cos\theta}{2}
\end{eqnarray} 
\noindent 
dictate the overlap of the initial ($e^+e^-$) and the final ($ZH$)
state angular momentum states in the $J=1$ channel, in terms of the
opening angle $\theta$ between the electron and the $Z$-boson momenta
in the $e^+e^-$ c.m.\ frame.

The $Z$ boson decay amplitude is expressed in the $Z \to\bar{f}$ rest
frame simply as,
\begin{eqnarray}
 T^{Z \rightarrow f\bar{f}}_{\lambda\beta} = 
  g^{Zff}_{\beta}\,\, \epsilon^{\rho}(p_Z,\lambda) \,\,j_{2 \beta \rho} = 
  g^{Zff}_{\beta}\,\, \mz \,\,d^1_{\lambda,\beta}(\theta^*) \,\, e^{i\,\lambda \,\phi^*},
\end{eqnarray}
\noindent 
for  $\lambda_1=-\lambda_2=\beta$ massless quarks and leptons, and 
again in the zero-width limit, $p_Z^2=\mzsq$. Here $\theta^*$ and
$\phi^*$ are the polar and  azimuthal angles of the fermion $f$ in the
decaying $Z$ boson rest frame, where the polar axis is chosen along the
$Z$ boson momentum direction in the $e^+e^-$ rest frame, and $\phi^*$ is
measured from the $e^+e^- \rightarrow ZH$ scattering plane; see Appendix
{\ref{app:phasespace}}.

From the above helicity amplitudes, we can calculate the weight
functions $\Sigma_{c_i}$. The  differential cross section, with the
$e^-$ polarization $\alpha$,  summed over the final state helicities
$\beta $ is
\bsub
\begin{eqnarray}
 d\sigma_{\alpha}^f &=& \frac{1}{2\,\,s} \,\,\sum_{\beta} \,\,\biggr| 
  \sum_{\lambda}\,\,T_{\alpha\lambda}^{ZH} \,\,D_Z(p_Z^2) \,\,
  T^{Z \rightarrow f\bar{f}}_{\lambda\beta} \biggr|^2 \,\,
  d \Phi_3 \\
 &=& \frac{1}{2\,\,s}\,\, \sum_{\beta}\,\, \biggr| 
  \sum_{\lambda} \,\,T_{\alpha\lambda}^{ZH}\,\,
  T_{\lambda\beta}^{Z \rightarrow f\bar{f}} \biggr|^2 \,\,
 \bigg\vert  D_Z(p_Z^2) \bigg\vert^2  \,\, d \Phi_3. 
\end{eqnarray}
\esub
We can simplify the phase space $d\Phi_3$ in the small width limit as 
\bsub
\begin{eqnarray}
 \big\vert D_Z(p_Z^2) \big\vert^2  \,\,d\Phi_3 
  &=& \left\vert p_Z^2-\mzsq+i\,\,\mz\,\,\Gamma_Z \right\vert^{-2}
  \,\,\,\, \frac{dp_Z^2}{2\,\,\pi} \,\,
  \,\, d\Phi_2(Z \rightarrow f\bar{f}) \,\, \,\,d\Phi_2(ZH) \\
 &\rightarrow& \frac{\pi\,\, \delta(p_Z^2-\mzsq)}{\mz\,\,\Gamma_Z} \,\,\,\,
  \frac{dp_Z^2}{2\,\,\pi} \,\,\,\, d\Phi_2(Z \rightarrow f\bar{f}) 
  \,\,  \,\,d\Phi_2(ZH) \\
 &=& \frac{1}{2\,\mz\,\Gamma_Z}\,d\hat{\Phi}_3.
\end{eqnarray}
\esub
\noindent
Here $d\hat{\Phi}_3$ is the reduced 3-body phase space element,
\begin{eqnarray}
 d\hat{\Phi}_3 &\equiv& d\Phi_2(Z\rightarrow f\bar{f})
  \,\,\,\, d\Phi_2(ZH)
  = \big(8\,\,\pi\big)^{-3}\,\,\,\,\bar{\beta}
  \left(\frac{\mzsq}{s},\,\,\frac{\mhsq}{s}\right)\,\,\,\,
  d\cos\thetaZ \,\, d\cos\theta^* \,\, d\phi^*\,\, ,\label{red3bdy}
\end{eqnarray}
\noindent 
where $\bar{\beta}(a,b)$ is given by eq.(\ref{betafunction}) in
Appendix~\ref{app:phasespace}.

The differential cross section is now expressed as
\begin{eqnarray}
 d \sigma_{\alpha}^f 
  = \frac{1}{2\,\,s} \,\,\sum_{\lambda,\lambda'} 
\,\,  T_{\alpha \lambda}^{ZH}\,\, (T_{\alpha \lambda'}^{ZH})^*\,\, 
  \rho_{\lambda\lambda'}^f \,\,d\hat{\Phi}_3
\end{eqnarray}
In this equation, we introduced the $Z \rightarrow f \,\,\bar{f}$ decay
density matrix by summing over the helicities and colors of $f$ and
$\bar{f}$,
\begin{eqnarray}
 \rho_{\lambda\lambda'}^f =  
  \frac{1}{2\,\,\mz\,\,\Gamma_Z} \,\,\sum_{\rm color}\,\, \sum_{\beta}\,\,
  T_{\lambda \beta}^{Z\rightarrow f\bar{f}}\,\,\,\, 
  \left(T_{\lambda' \beta}^{Z\rightarrow f\bar{f}}\right)^*\,\, ,
\end{eqnarray}
\noindent 
which is normalized as 
\begin{eqnarray}
 \int \rho_{\lambda\lambda'}^f \,\,\, d\Phi_2 (Z \rightarrow f\bar{f}) 
  = \frac{\Gamma(Z \rightarrow f\bar{f})}{8\,\,\pi \,\,\Gamma_{\rm Z}} 
  \,\,\,\,\delta_{\lambda\lambda'}
  = \frac{{\rm B}_f}{8\,\,\pi} \,\,\,\,\delta_{\lambda\lambda'}\,\, .
\end{eqnarray}
By using eqs.(\ref{schan1})-(\ref{schan3}) in Appendix~\ref{app:current}
for the $Z\to f\bar f$ currents and the $Z$ boson polarization vectors, 
we obtain a compact expression for the $Z$ boson decay density matrix
\begin{eqnarray}
 \rho_{\lambda\lambda'}^f &=& 12\,\,\pi\,\, {\rm B}_f\,\,
  \left[
  \left(
   \begin{array}{ccc}
    \frac{1+\cos^2\theta^*}{2} & 
     -\frac{\sin\theta^*\,\,\cos\theta^*}{\sqrt{2}} e^{i\,\phi^*} &
     \frac{\sin^2\theta^*}{2} e^{2\,i\,\phi^*} \\
    -\frac{\sin\theta^*\,\,\cos\theta^*}{\sqrt{2}} e^{-\,i\,\phi^*} &
     \sin^2\theta^* &
     \frac{\sin\theta^*\,\,\cos\theta^*}{\sqrt{2}} e^{i\,\phi^*} \\
    \frac{\sin^2\theta^*}{2} e^{-\,2\,i\,\phi^*} &
     \frac{\sin\theta^*\,\,\cos\theta^*}{\sqrt{2}} e^{-\,i\,\phi^*} &
     \frac{1+\cos^2\theta^*}{2} \\
   \end{array}
   \right)\right.
 \nonumber \\
& & \hskip 3cm  
+ {\rm A}_f 
  \left. \left(
   \begin{array}{ccc}
    -\cos\theta^* & 
     \frac{\sin\theta^*}{\sqrt{2}} e^{i\,\phi^*} &
     0 \\
    \frac{\sin\theta^*}{\sqrt{2}} e^{-\,i\,\phi^*} &
     0 &
     \frac{\sin\theta^*}{\sqrt{2}} e^{i\,\phi^*} \\
    0 &
     \frac{\sin\theta^*}{\sqrt{2}} e^{-\,i\,\phi^*} &
     \cos\theta^* \\
   \end{array}
  \right) \right]\,\, ,
\label{ZH_densityM}
\end{eqnarray}
\noindent 
for $\lambda,\,\,\lambda^\prime=-,\,\,0,\,\,+$. Here ${\rm B}_f$ is the
$Z\rightarrow f\bar{f}$ decay branching fraction,
\begin{eqnarray}
 {\rm B}_f = \frac{\Gamma(Z\rightarrow f\bar{f})}{\Gamma_{\rm Z}} = 
  \frac{\mz \,\,N_f}{16\,\,\pi\,\,\Gamma_{\rm Z}} \,\,
  \left[\big(g_L^{Zff}\big)^2+\big(g_R^{Zff}\big)^2 \right],
  \label{Bf}
\end{eqnarray}
with $N_f=1$ for leptons and $N_f=3$ for quarks, and
\begin{eqnarray}
 {\rm A}_f = 
  \frac{\big(g_L^{Zff}\big)^2-\big(g_R^{Zff}\big)^2}
  {\big(g_L^{Zff}\big)^2+\big(g_R^{Zff}\big)^2}
  \label{Af}
\end{eqnarray}
is the left-right asymmetry parameter of the $Zff$ couplings. In the
following analysis, we use the tree-level expression for the asymmetry
parameter eq.(\ref{Af}) for the charged lepton decays, whereas we
replace the branching fraction eq.(\ref{Bf}) by the observed
values \cite{pdg2006} in simulating  the number of events.

Let us now obtain the weight functions for polarized $e^\pm$
beams. The differential cross section for polarized $e^+\,e^-$ beams is
expressed as
\begin{eqnarray}
 d\sigma^f(P,\bar{P}) &=& \bigg[ 
  \Sigma_{\rm SM}^f\big(P,\,\bar{P};\,\,\hat{\Phi}_3\big) +
  \sum_{c_i} \,\, c_i\,\, \Sigma_{c_i}^f\big(P,\,\bar{P};\,\,\hat{\Phi}_3\big)
  \bigg] \, \,\, d\hat{\Phi}_3
\end{eqnarray} 
\noindent 
where the weight functions are obtained by summing over the $e^-$
helicity $\alpha$,
\bsub
\begin{eqnarray}
 \Sigma_{\rm SM}^f\big(P,\,\bar{P};\,\,\hat{\Phi}_3\big) &=&
  \frac{1}{2\,s}\, \sum_{\alpha}\, 
  \left( \frac{1+\alpha\, P}{2} \right) \,
  \left( \frac{1-\alpha\, \bar{P}}{2} \right) \nonumber\\
 & & \,\, \sum_{\lambda,\lambda'} \,(\hat{M}_{\rm SM})_{\alpha}^{\lambda} \,
  (\hat{M}_{\rm SM})_{\alpha}^{\lambda'*}\,
  d_{\alpha,\lambda}^1(\thetaZ)\, d_{\alpha,\lambda'}^1(\thetaZ)\,
  \rho_{\lambda\lambda'}^f\, , \label{ZH_DistSM} \end{eqnarray}
\begin{eqnarray}
 \Sigma_{c_i}^f(P,\bar{P};\hat{\Phi}_3) &=&
  \frac{1}{2\,\,s}\,\, \sum_{\alpha}\,\, 
  \left( \frac{1+\alpha\,\, P}{2} \right) \,\,
  \left( \frac{1-\alpha\,\, \bar{P}}{2} \right) \nonumber\\
 & & \hskip -1.0cm \sum_{\lambda,\lambda'} \,\,
  \left[ \bigg\{ \big(\hat{M}_{c_i}\big)_{\alpha}^{\lambda}\,\,
   \big(\hat{M}_{\rm SM}^*\big)_{\alpha}^{\lambda'} 
   + \big(\hat{M}_{\rm SM}\big)_{\alpha}^{\lambda} \,\,
   \big(\hat{M}_{c_i}^*\big)_{\alpha}^{\lambda'} \bigg\}\,\,
   d_{\alpha,\lambda}^1(\thetaZ) \,\,d_{\alpha,\lambda'}^1(\thetaZ)
   \,\,  \rho_{\lambda\lambda'}^f \right]\,. \label{ZH_DistCi}
\end{eqnarray}
\esub
Here the $ZH$ production matrix elements, 
$\big(\hat{M}_{\rm SM}\big)_{\alpha}^{\lambda}$ and 
$\big(\hat{M}_{c_i}\big)_{\alpha}^{\lambda}\equiv\big(\hat{M}_{c_{iZV}}\big)_{\alpha}^{\lambda} $,  are obtained from eqs.(\ref{ZH_ME-2})-(\ref{ZH_FF}):
\bsub
\begin{eqnarray}
 \big(\hat{M}_{\rm SM}\big)_{\alpha}^{\lambda} = 
  \big(\hat{M}_{\az}\big)_{\alpha}^{\lambda}
  &=& \frac{\mz \,\,\sqrt{2\,\,s}}{s-\mzsq} \,\,\gz\,\, 
  g_{\alpha}^{Zee}\,\, \gamma_{\lambda}, \\
 \big(\hat{M}_{c_{kZV}}\big)_{\alpha}^{\lambda} 
  &=& \big(\hat{M}_{\rm SM}\big)_{\alpha}^{\lambda}\,\,
  R_V^{\alpha}(s) \,\,f_{kV}^{\lambda}(s) \hskip 0.5cm (k=2,3),
\end{eqnarray}
\label{ZH_ProME}
\esub
\noindent
\hskip -1.4mm
with
\bsub
\begin{eqnarray}
 \gamma_{\pm} &=& 1, \hskip 0.5cm \gamma_0 = \frac{E_Z}{\mz}, \\
 R_Z^{\alpha} &=& 1, \hskip 0.5cm
  R_{\gamma}^{+} = -\,\cot \theta_{\rm W}\,\, \left( \frac{s-\mzsq}{s} \right),
  \hskip 0.5cm R_{\gamma}^{-} = \tan 2\,\theta_{\rm W} 
  \left( \frac{s-\mzsq}{s} \right), \label{RZdef}\\
 f_{2Z}^{\lambda}(s) &=& \frac{s-\mhsq}{\mzsq} +1 
  - \delta_{\lambda 0} \,\,\frac{2\,\,\sqrt{s}}{\mz} \,\,
  \frac{\gamma_0^2 -1}{\gamma_0}, 
  \hskip 0.1cm
  f_{2\gamma}^{\lambda}(s) = \frac{s-\mhsq}{\mzsq} +1
  - \delta_{\lambda 0}\,\, \frac{2\,\,\sqrt{s}}{\mz}\,\,
  \frac{\gamma_0^2 -1}{\gamma_0}, \\
 f_{3Z}^{\lambda}(s) &=& \frac{\mhsq}{\mzsq} 
  + \delta_{\lambda 0} \,\,\frac{2\,\,\sqrt{s}}{\mz}\,\,
  \frac{\gamma_0^2-1}{\gamma_0},
  \hskip 1.5cm
  f_{3\gamma}^{\lambda}(s) = \frac{s+\mhsq}{\mzsq} -1
  + \delta_{\lambda 0}\,\, \frac{2\sqrt{s}}{\mz}\,\,
 \frac{\gamma_0^2-1}{\gamma_0},
\end{eqnarray}
\label{ZH_ProMEsub}
\esub
\noindent
\hskip -1.8mm 
where $\delta_{00}=\delta_{\gamma\gamma}=\delta_{ZZ}=1$
and $\delta_{\pm 0}=\delta_{Z\gamma}=\delta_{\gamma Z}=0$.
The weight functions $\Sigma_{c_i}^f$ are now expressed as
\bsub
\begin{eqnarray}
 \Sigma_{\rm SM}^f(P,\,\bar{P};\,\,\hat{\Phi}_3) &=& \sum_{\alpha} \,
  \left( \frac{1+\alpha \,P}{2} \right) \,
  \left( \frac{1-\alpha\, \bar{P}}{2} \right) \,\, 
  \frac{\gzsq\,\big(g_{\alpha}^{Zee}\big)^2\,\mzsq}{(s-\mzsq)^2}\nonumber\\
 & & \sum_{\lambda,\lambda'}\,\, \gamma_{\lambda} \gamma_{\lambda'} \,\,
  d_{\alpha\lambda}^1(\thetaZ) \,\, d_{\alpha\lambda'}^{1}(\thetaZ) \,\,
  \rho_{\lambda\lambda'}^f \label{ZH_Dist2SM} \\
 \Sigma_{c_{kZV}}^f(P,\bar{P};\hat{\Phi}_3) &=& \sum_{\alpha} \,
  \left( \frac{1+\alpha \,P}{2} \right) \,
  \left( \frac{1-\alpha\, \bar{P}}{2} \right) \,\, 
  \frac{\gzsq\,\big(g_{\alpha}^{Zee}\big)^2\,\mzsq}{(s-\mzsq)^2}\nonumber\\
 & & \hskip -0.5cm
  \left[ \sum_{\lambda,\lambda'}\,\, \gamma_{\lambda} \,\,
   \gamma_{\lambda'} \,\, R_V^{\alpha} \,\,
   \bigg\{ f_{kV}^{\lambda}(s) + f_{kV}^{\lambda'}(s) \bigg\}\,\,\,\,
   d_{\alpha\lambda}^1(\thetaZ) \,\,d_{\alpha\lambda'}^{1}(\thetaZ) \,\,
   {\rm Re} \big(\rho_{\lambda\lambda'}^f\big) \right]\, ,\label{ZH_Dist2Ci}
\end{eqnarray}
\label{ZH_Dist2}
\esub
\hskip -1.7mm 
for $k=2,3$ and $V=Z,\gamma$. The weight function for $\az$ is simply
twice the SM one, eq.(\ref{ZH_Dist2}a). Since all our matrix elements
are real, only the real parts of the density matrix elements 
$\rho^f_{\lambda\lambda'}$ in eq.(\ref{ZH_densityM}) contribute to the
weight functions.  

The weight functions depend both on the beam polarization and on the
final fermion species. For brevity, we give expressions for covariance
matrices for electron polarization $P$, with unpolarized positron 
$(\bar P=0)$. For $f=e$, $\mu$ or $\tau$, we can use the full phase space
information of $\hat\theta,\,\,\theta^*$ and $\phi^*$, and accordingly
the covariance matrix is obtained as
\begin{eqnarray}
 \left[\big(V_P^{l}\big)^{-1}\right]_{ij} = L_0\,\,
  \int \frac{\Sigma_{c_i}^l(P,\,0;\,\,\hat{\Phi}_3) 
  \,\,\,\,\,\,\Sigma_{c_j}^l(P,\,0;\,\,\hat{\Phi}_3)}
  {\Sigma_{\rm SM}^l(P,\,0;\,\,\hat{\Phi}_3)}\,\,\,\, d\hat{\Phi}_3,
\end{eqnarray}
where $L_0=100\ifb$. 

On the other hand, when the outgoing fermion is invisible, i.e.\ , for
$f=\nu_e,\nu_{\mu}$ or $\nu_{\tau}$, we can observe only the
$\cos\thetaH=-\,\,\cos\theta_Z$ distribution, and the covariance matrix
is,
\begin{eqnarray}
 \left[\big(V_P^{\nu}\big)^{-1}\right]_{ij} = 
  {\rm B} (H \to b \bar b) \,\, L_0 \,
  \int \frac{\int \Sigma_{c_i}^{\nu}(P,\,0;\,\hat{\Phi}_3) \,
  d\cos\theta^*\,\, d\phi^* \,
  \int \Sigma_{c_j}^{\nu}(P,\,0;\,\hat{\Phi}_3) \,
  d\cos\theta^* \,\,d\phi^*}
  {\int \Sigma_{\rm SM}^{\nu}(P,\,0;\,\hat{\Phi}_3) \, 
  d\cos\theta^* \,\,d\phi^*} \,
  d\cos\thetaZ. \nonumber \\
\end{eqnarray} 
\noindent 
Here again, we multiply the branching fraction B$(H \to b \bar b)=0.9$
because the $H \to \tau\tau$ decay mode may not be useful in the
analysis.

When the outgoing fermions are quarks, $f=u,d,c,s,b$, we use only the
half phase space information for $\theta^*$ and $\phi^*$ because the
$q$-jet and $\bar{q}$-jet cannot be distinguished. We find
\begin{eqnarray}
 \left[\big(V_P^{q}\big)^{-1}\right]_{ij} = L_0 \int \, \frac{
  \bigg(\Sigma_{c_i}^{q}\big(P,\,0;\,\,\hat{\Phi}_3\big) 
  + \Sigma_{c_i}^{q}\big(P,\,0;\,\,\tilde{\hat{\Phi}}_3\big)\bigg)
  \bigg(\Sigma_{c_j}^{q}\big(P,\,0;\,\,\hat{\Phi}_3\big) 
  + \Sigma_{c_j}^{q}\big(P,\,0;\,\,\tilde{\hat{\Phi}}_3\big)\bigg)}
  {\bigg(\Sigma_{\rm SM}^{q}\big(P,\,0;\,\,\hat{\Phi}_3\big) 
  + \Sigma_{\rm SM}^{q}\big(P,\,0;\,\,\tilde{\hat{\Phi}}_3\big) \bigg)} \,\,
  \frac{1}{2} d\hat{\Phi}_3 \nonumber\\
\end{eqnarray}
where $\tilde{\hat{\Phi}}_3$ is the phase space obtained by
interchanging $q$-jet and $\bar{q}$-jet; see
eq.(\ref{dsigma_qq}). Finally, the electron beam polarization is taken
into account as
\begin{eqnarray}
 V^{-1} =\sum_{P=\pm0.8} \,\,\,\, \frac{L_P}{L_0} \,\,\,\, 
  \sum_{f=l,\nu,q}\,\,\,\, \bigg(V_P^f\bigg)^{-1},
\end{eqnarray}
where $L_P$ is the integrated luminosity for each electron beam
polarization.

We expect no significant background to this channel. But the $WW$ fusion
process, $e^+e^- \rightarrow \nu_e \,\bar{\nu}_e\,\, H$, can mimic the $HZ$
production when the $Z$ decays into neutrinos. This background is
avoided by demanding the missing mass to lie within five times 
the $Z$ boson width, see eq.(\ref{ZHprodcond}). After the  cut, we
confirm numerically that our results are not affected by the
interference effects, in the $\nu_e\bar\nu_eH$ channel.

\subsection{Sensitivity analysis of $HZZ$ and $HZ\gamma$ coupling}
We first note that the five CP conserving dimensionless 
effective couplings   $\az$, $\bz$, $\cz$, $\bg$ and $\cg$ coming from the
dimension-six operators, contribute to the process only through the  four
form factors of eq.(\ref{ZH_FF}). In particular, only two combinations
of the three couplings $\az,\,\,\bz$  and $ \cz$ appear in the from
factors of the  $HZZ$ and $HZ\gamma$ vertices. Consequently, we cannot
determine the couplings independently at a fixed collision energy.

In the following analysis, we show the eigenvectors for the four
combinations of the effective couplings,
\begin{eqnarray} 
 \bzp = \bz+\frac{\mzsq}{s+\mzsq}\,\,\az, \hskip 0.2cm
  \czp = \cz+\frac{\mzsq}{s+\mzsq}\,\,\az, \hskip 0.2cm 
  \bg, \hskip 0.2cm {\rm and }\,\,\,\, \cg,
\end{eqnarray}
which can be measured simultaneously at each energy. Note that the
unmeasurable combination 
\begin{eqnarray}
 \az - (\bz+\cz)\,\,\, \frac{\mzsq}{s+\mzsq}
\end{eqnarray}
changes with the collision energy $\sqrt{s}$, and hence moderate
constraint on the coupling $\az$, which measures the strength of the
standard $HZZ$ coupling, can be obtained after combining experiments at
different energies as discussed in ref.\cite{kniehl}.

The covariance matrices are expressed  in terms of the eigenvectors
and the errors. With un-polarized electron beam, $|P|=0$, we find for
$\sqrt{s}=250\GEV$ and $L=100\ifb$,
\bsub
\begin{eqnarray}
 .88\,\,\bzp   +.41\,\,\czp  +.097\,\,\bg +.15\,\,\cg & = & \pm .00048, \\
 .45\,\,\bzp   -.86\,\,\czp  -.026\,\,\bg -.24\,\,\cg & =&  \pm .0042, \\
 .068\,\,\bzp  +.28\,\,\czp  -.54\,\,\bg  -.79\,\,\cg & =&  \pm .0054, \\
 .046\,\,\bzp  +.10\,\,\czp  +.83\,\,\bg  -.54\,\,\cg & =&  \pm .072.
\end{eqnarray}
\label{ZH_ResultUnPol250}
\esub
\hskip -.16 cm 
The smallness of the magnitudes of the coefficients of the $HZ\gamma$
couplings in the first two eigenvectors and that of the coefficients of
the $HZZ$ couplings in the latter eigenvectors reveals that the $HZZ$
couplings and $HZ\gamma$ couplings are measured rather
independently. Inspection of the first two eigen-vectors in
eq.(\ref{ZH_ResultUnPol250}), which are most tightly constrained, tells
that the couplings $\bzp$ and $\czp$ can be measured much more
accurately than the other two couplings, $\bg$ and $\cg$, and also their
sum in eq.(\ref{ZH_ResultUnPol250}a) can be measured almost one order of
magnitude better than their difference in
eq.(\ref{ZH_ResultUnPol250}b). The errors of the $HZ\gamma$ couplings,
$\bg$ and $\cg$, which can be deciphered from the third and the fourth
eigenvectors,  are  significantly larger; those of their sum in
eq.(\ref{ZH_ResultUnPol250}c) and their difference in
eq.(\ref{ZH_ResultUnPol250}d) are, respectively, more than one and two
orders of magnitude larger than the smallest error in
eq.(\ref{ZH_ResultUnPol250}a).  This can be attributed to the
cancellation of the contributions from the  $HZ\gamma$ couplings
between the left handed and right handed  electron contributions, where
the interference term between the $\gamma $ and $Z$ exchange amplitudes
change sign,
\begin{eqnarray}
 g_L^{\gamma ee}\,\,g_L^{Zee}+g_R^{\gamma ee}\,\, g_R^{Zee} 
  = -\,e\,\, \left( g_L^{Zee}+\,g_R^{Zee} \right) 
  = e\,\,g_Z\,\, \left(1/2 -2\,\,\sin^2\theta_W \right)
  \ll e\,\,g_Z\,\, .\label{signchange}
\end{eqnarray}

Expressing eq.(\ref{ZH_ResultUnPol250}) in terms of the covariance
matrix, we reproduce identical results given in eq.(5.4) of
ref.\cite{kniehl}, even though we do not take into account $\tau
$-polarization and partial identification of both $b$ and $\bar b$
jets. Small improvements of the covariance matrix due to these
additional measurements can be hidden in the numerical uncertainty of
the correlation matrix elements in the standard expression in terms of
the errors and their correlations. 

The constraint on the couplings become significantly stronger when we
introduce the $e^-$ beam polarization. With $\big\vert P\big\vert$=80\%,
$L_{(P=0.8)}=L_{(P=-0.8)}=50\ifb$ at $\sqrt{s}=250\GEV$, we find
\bsub
\begin{eqnarray}
 .056\,\,\bzp   +.027\,\,\czp  +.53\,\,\bg   +.85\,\,\cg  & = & \pm .00024, \\
 .90\,\,\bzp    +.43\,\,\czp   -.032\,\,\bg  -.053\,\,\cg & = & \pm .00049, \\
 -.094\,\,\bzp  +.20\,\,\czp   -.83\,\,\bg   +.51\,\,\cg  & = & \pm .0034, \\
 .42\,\,\bzp    -.88\,\,\czp   -.19\,\,\bg   +.12\,\,\cg  & = & \pm .0043.
\end{eqnarray}
 \label{ZH_ResultPol250}
\esub
\hskip -.14 cm 
It is salient that the errors of the couplings  $\bzp$ and $\czp$, that
of their sum in eq.(\ref{ZH_ResultPol250}b) and that of their difference
(\ref{ZH_ResultPol250}d), do not improve at all by introducing the beam
polarization, from the corresponding ones in
eqs.(\ref{ZH_ResultUnPol250}a) and (\ref{ZH_ResultUnPol250}b).  On the
other hand, the errors of the two eigenvectors with dominant $\bg$ and
$\cg$ components are reduced by a factor of 20. This observation can be
explained as follows:
\begin{enumerate}
 \item The $\bz$ and $\cz$ coupling contribution depends on the beam
       polarization exactly the  same way as the SM. In particular, the
       weight functions for $\bz$ and $\cz$ in eq.(\ref{ZH_Dist2Ci})
       have the same polarization dependence as the SM contribution in
       eq.(\ref{ZH_Dist2SM}), essentially because  $R_Z^\alpha =1$ in
       eq.(\ref{ZH_Dist2Ci}). 
 \item  When a photon propagates in the $s$-channel, the $\bg$ and $\cg$
       contributions to the amplitude do not change by the beam
       polarization since the electromagnetic interactions are chirality
       blind. But the interference of the $\gamma $ exchange and the  SM
       $Z$ boson exchange amplitude changes sign; see
       eq.(\ref{signchange}). 
\end{enumerate}

We  present our results for $\sqrt{s}=350\GEV$ for $L=100\ifb$ and
$\big\vert P\big\vert =80\%$.
\renewcommand{\thefootnote}{\fnsymbol{footnote}}$^{\footnotemark[1]}$
\footnotetext[1]{From now on, we do not report our results for
$\big\vert P\big\vert =0$ in order to save space. Individual results
with $\big\vert P\big\vert =0$ will be available on request from the
authors.}
\bsub
\begin{eqnarray}
 .029\,\,\bzp +.030\,\,\czp +.34\,\,\bg  +.94\,\,\cg  &=& \pm .00015, \\
 .71\,\,\bzp  +.70\,\,\czp  -.013\,\,\bg -.040\,\,\cg &=& \pm .00040, \\
 .67\,\,\bzp  -.68\,\,\czp  -.28\,\,\bg  +.100\,\,\cg &=& \pm .0014, \\
 -.21\,\,\bzp +.21\,\,\czp  -.90\,\,\bg  +.32\,\,\cg  &=& \pm .0012.
\end{eqnarray}
 \label{ZH_ResultPol350}
\esub
\hskip -.14 cm 
We first note that the eigenvectors and their ordering remain similar to
those at $\sqrt{s}=250\GEV$. We observe that $\czp$ is more accurately
measured at high energies where the amplitude for the longitudinal $Z$
boson $(\lambda=0)$ grows, mainly because of the $\gamma_0=E_Z/\mz$
factor in eq.(\ref{ZH_ProME}b). The weight functions for the couplings
$\czp$ and $\cg$ grows as $\gamma_0^2$ from the the dominant $\lambda=0$
amplitudes; see eq.(\ref{ZH_ProMEsub}d). In contrast, the weight
functions for $\bzp$ and $\bg$ do not increase at high energies, since
the $\lambda=0$ amplitude tends to be canceled by the $\lambda=\pm$
(transverse) modes; see eq.(\ref{ZH_ProMEsub}c). This growth of the
weight functions at high energies explains the general decrease of all
the errors at $\sqrt{s}=350\GEV$ in eq.(\ref{ZH_ResultPol350}) as
compared to those at $\sqrt{s}=250\GEV$ in eq.(\ref{ZH_ResultPol250}),
despite the decrease of the total cross section from 260~fb to 130~fb.

At $\sqrt{s}=500\GEV$ for $L_0=100\ifb$ with $P=\pm 0.8$, we find
\bsub
\begin{eqnarray}
 .015\,\,\bzp +.035\,\,\czp +.18\,\,\bg   +.98\,\,\cg  &=& \pm .000099, \\
 .38\,\,\bzp  +.92\,\,\czp  -.0085\,\,\bg -.037\,\,\cg &=& \pm .00028,  \\
 .24\,\,\bzp  -.097\,\,\czp +.95\,\,\bg   -.17\,\,\cg  &=& \pm .00086, \\
 .89\,\,\bzp  -.37\,\,\czp  -.25\,\,\bg   +.046\,\,\cg &=& \pm .0010.
\end{eqnarray}
 \label{ZH_ResultPol500}
\esub
\hskip -.14 cm 
Despite smaller cross-section at $\sqrt{s}$ = 500 GeV, all the errors in
eq.(\ref{ZH_ResultPol500}) are smaller than those in
eq.(\ref{ZH_ResultPol350}) for $\sqrt{s}$ = 350 GeV.

The energy dependence of the measurement at higher energies
$\left(\sqrt{s}\gg m_Z\right)$ can be read off from Table
\ref{Tab:ZH_MEdepend}, where we  show the high energy limit of the
matrix elements  $\big(\hat M_{c_i}\big)^\lambda_\alpha$ of
eq.(\ref{ZH_ProME}).
\begin{table}[htb!]
 \begin{center}
  \begin{ruledtabular}
   \begin{tabular}{||c|cc|cc||}
    \multirow{2}{*}{$c_i$}&\multicolumn{2}{c|}
    {$\,\,\,\,Z_T(\lambda=\pm)\,\,\,\,\,\,$}&
    \multicolumn{2}{c||}{$\,\,\,\,Z_L(\lambda=0)\,\,\,\,\,\,$} \\ 
    & Energy Dependence & 
    coefficient &
    Energy Dependence 
    & coefficient \\\hline
    SM & $1/\gamma_0$  & $ \gz \,\, g_{\alpha}^{Zee}$ & 1 & 
    $\gz \,\,g_{\alpha}^{Zee}$  \\\hline
    $\bz$ & $\gamma_0$  & $4\,\,\gz\,\, g_{\alpha}^{Zee}$ & $1/\,\gamma_0$ & 
    $\gz\,\, g_{\alpha}^{Zee}\,\, \left(5-\,\mhsq/\,\mzsq\right)$ \\\hline
    $\cz$ & $1/\,\gamma_0$  & $ \gz\,\, g_{\alpha}^{Zee}\,\, \mhsq /\,\mzsq$ & 
    $\gamma_0$ &   $4\,\, \gz\,\, g_{\alpha}^{Zee}$ \\\hline
    $\bg$ & $\gamma_0$ & $4\,\,e$ &$1/\,\gamma_0$ & 
    $ e\,\, \left(5-\,\,\mhsq /\,\mzsq\right)$ \\  \hline
    $\cg$ & $\gamma_0$ &   $4\,\,e$ &$\gamma_0$ &
    $8\,\,e$
   \end{tabular} 
  \end{ruledtabular}
 \end{center}
 \caption{\small The high energy limit of the  matrix elements
 $\big(\hat M_{c_i}\big)^\lambda_\alpha$ in  eq.(\ref{ZH_ProME}). Here
 $\gamma_0 =E_Z/\, \mz$.}
\label{Tab:ZH_MEdepend}
\end{table}
Since the weight functions are the interference between the SM
amplitudes and the $c_i$ amplitudes, we immediately find that $\cz$ and
$\cg$ are measured most accurately at high energies from the $Z_L$
contribution, because their weight functions grow linearly with
$\gamma_0=E_Z/m_Z$. The couplings $\bz$ and $\bg$ are best measured from
the $Z_T$ contribution, whose weight functions do not vanish at high
energies.

Finally we present the results for  $\sqrt{s}=1\TEV$ for $L_0=100\ifb$
with $P=\pm \,\,0.8$; 
\bsub
\begin{eqnarray}
  .004 \,\bzp +.039 \,\czp +.047 \,\bg +.998 \,\cg &=& \pm .000046
   \label{ZH_ResultPol1000-1} \\
  .092 \,\bzp +.995 \,\czp -.003 \,\bg -.039 \,\cg &=& \pm .00013
   \label{ZH_ResultPol1000-2} \\
  -.16 \,\bzp +.014 \,\czp -.986 \,\bg +.047 \,\cg &=& \pm .00086
   \label{ZH_ResultPol1000-3} \\
  .98 \,\bzp  -.092 \,\czp -.16 \,\bg  +.007 \,\cg &=& \pm .0012.
   \label{ZH_ResultPol1000-4}
\end{eqnarray}
 \label{ZH_ResultPol1000}
\esub
\noindent
\hskip -1.6mm
All the four couplings are now measured rather independently, as we can
observe from the single coupling dominance in all the four
eigenvectors. The error of $\cg$ and $\czp$ becomes half of those at
$500\GEV$ in eq.(\ref{ZH_ResultPol500}), because their weight functions
are proportional to $\gamma_0$. On the other hand, the errors of $\bzp$
and $\bg$ do not improve  because the weight functions remain constant
at high energies.

\section{$e^+e^- \rightarrow e^+e^- H$ double-tag $ZZ$-fusion processes}
\label{ZZ} 
In this section we study the $HZZ$, $HZ\gamma $ and $H\gamma \gamma$
couplings via $t$-channel Vector Boson Fusion (VBF) process, 
$e^+e^- \rightarrow e^+e^-H$; see fig.\ref{diagram}b. The merit of
the $t$-channel VBF processes is that the cross-section grows with
$\sqrt{s}$, and also, we expect sensitivity to the coupling $\az$, which
cannot be measured independently from the other couplings in the
$s$-channel process, $e^+e^- \rightarrow ZH$.

\subsection{Helicity Amplitudes and Background}
The momentum and helicity assignments are the same as those in
eq.(\ref{nunuh}) for the $W$-boson fusion process :
\begin{eqnarray}
  e^{-}\left(k_{1}, \frac{\sigma_1}{2} \right) 
   + e^{+}\left(k_{2}, \frac{\sigma_2}{2} \right) \rightarrow 
   e^- \left( p_1,\frac{\lambda_1}{2} \right)
   + e^+ \left( p_2,\frac{\lambda_2}{2} \right)
   + H(\pH). 
\label{eeH_process}
\end{eqnarray}
Here again we neglect the mass of $e^{\pm}$, and only two diagrams of
fig.\ref{diagram}a ($ZH$ production) and fig.\ref{diagram}b (VBF)
contribute. The electron chirality conservation tells
$\sigma_i=\lambda_i$ for the VBF amplitudes, and $\sigma_1=-\sigma_2$
and $\lambda_1=-\lambda_2$ for $ZH$ production amplitudes. When 
$\sigma_1=\lambda_1=-\sigma_2=-\lambda_2$, the two amplitudes
interfere. The $ZH$ production contribution is suppressed by requiring
the invariant mass of the $e^+ e^-$ pair to be away from $\mz$,
eq.(\ref{Wfusion_cond1}).

The helicity amplitudes for the VBF process are given as
\begin{eqnarray}
 & & \hskip -1cm M_{\sigma_1 \sigma_2}^{\lambda_1 \lambda_2} 
  (e^+ e^- \rightarrow e^- e^+ H) \nonumber \\
 & & = j_{e^- (\sigma_1,\lambda_1)}^{\mu} \,\,
  \biggl[ g^{Zee}_{\sigma_1} \,\, D_Z(t_1) \,\, \Gamma_{\mu\nu}^{ZZH} 
  \,\,D_Z(t_2)\,\, g^{Zee}_{-\sigma_2}
  + g^{Zee}_{\sigma_1} \,\, D_Z(t_1)\,\, \Gamma_{\mu\nu}^{Z\gamma H}\,\,
  D_{\gamma}(t_2) \,\, g^{\gamma ee}_{-\sigma_2} \nonumber \\
 & & \,\,\,\, 
  + \,\, g^{\gamma ee}_{\sigma_1} \,\, D_{\gamma}(t_1) \,\, 
  \Gamma_{\mu\nu}^{\gamma ZH} \,\, D_Z(t_2) \,\, g^{Zee}_{-\sigma_2} 
  + g^{\gamma ee}_{\sigma_1} \,\, D_{\gamma}(t_1)\,\, 
  \Gamma_{\mu\nu}^{\gamma \gamma H} \,\, D_{\gamma}(t_2) \,\, 
  g^{\gamma ee}_{-\sigma_2} \biggr] \,\, j_{e^+ (\sigma_2,\lambda_2)}^{\nu},
\end{eqnarray}
\noindent 
for $\lambda_1=\sigma_1$, and $\lambda_2=\sigma_2$. The propagator
factors $D_V(t_i)$ and the $Vee$ couplings $g_{\alpha}^{Vee}$ are given
in eq.(\ref{ZH_PropCoup}). The $t$-channel currents
\bsub
\begin{eqnarray}
j_{e^- (\alpha\alpha)}^{\mu} 
&=&\bar{u}(p_1,\,\,\alpha /\,2)\,\,\gamma^\mu \,\, P_{\alpha}\,\, 
u(k_1,\,\,\alpha /\,2)\,\, ,\label{fusioncurrent1}\\
j_{e^+ (\beta\beta)}^{\nu}&=&\bar{v}(k_2,\,\,\beta/\,2)\,\,
\gamma^\nu \,\,P_{-\,\beta}\,\,v(p_2,\,\,\beta/\, 2)\,\, \label{fusioncurrent2}
\end{eqnarray}
\label{fusioncurrent}
\esub
\hskip -1.4mm 
are evaluated in the laboratory frame,
\bsub
\begin{eqnarray}
 j_{e^- (\alpha\alpha)}^{\mu} &=& \sqrt{s \,\,x_1} \,\,
  \left (\cos \frac{\theta_1}{2},\,\, 
   \sin \frac{\theta_1}{2}\,\, e^{i\alpha \,\phi_1},\,\, 
   -i\,\,\alpha\,\,\sin \frac{\theta_1}{2}\,\, e^{i\alpha \phi_1},\,\, 
   \cos \frac{\theta_1}{2}\right) , \label{tchancurrent1} \\
 j_{e^+ (\beta\beta)}^{\nu} &=& \sqrt{s \,\, x_2} \,\, 
  \left( \sin \frac{\theta_2}{2},\,\,
   \cos \frac{\theta_2}{2}\,\, e^{-\beta i\phi_2},\,\, 
   i\,\,\beta \,\,\cos \frac{\theta_2}{2}\,\, e^{-i\beta \phi_2},\,\, 
   -\,\sin \frac{\theta_2}{2} \right),
  \label{tchancurrent2}
\end{eqnarray}
  \label{tchancurrent}
\esub
\noindent
\hskip -1.4mm
in the $m_e=0$ limit, where the four-momenta of the final electron and
positron are parametrized as in eq.(\ref{oldlabmom}). The helicity
amplitudes are then expressed as
\begin{eqnarray}
 M_{\alpha \beta}^{\alpha \beta} &=& \gz\,\, \mz\,\,
  \sum_{V_1=Z,\gamma}\,\, \sum_{V_2=Z,\gamma}\,\,  
  g_{\alpha}^{V_1 ee} \,\, D_{V_1}(t_1) \,\, D_{V_2}(t_2) \,\,
  g_{-\beta}^{V_2 ee} \,\, {\cal F}\big(\theta_1,\theta_2\big) 
  \nonumber \\
 & & \hskip 1cm \otimes
  \left[ \delta_{\alpha\beta}\,\,
   \left( h_1^{V_1 V_2} \,\,{\cal H} +\frac{h_2^{V_1 V_2}}{\mzsq}\,\,
    \frac{s}{8}\,\, {\cal G'} \right)
   + \delta_{\alpha,-\beta}\,\,
   \left( h_1^{V_1 V_2} +\frac{h_2^{V_1 V_2}}{\mzsq}
    \,\,\frac{s}{8} \,\,{\cal G} \right) \right],
  \label{tchanhelamp}
\end{eqnarray}
\noindent 
where the functions ${\cal F}$, ${\cal H}$, ${\cal G}$ and ${\cal G}'$
are given in eq.(\ref{FHGfuncDef}) and (\ref{FHGfunc}) of Appendix
\ref{app:current}. The above helicity amplitudes can be expressed as
\begin{eqnarray} 
 M_{\alpha \beta}^{\alpha \beta} &=& (M_{\rm SM})_{\alpha\beta}^{\alpha\beta}
  + \sum_i \,\, c_i\,\, (M_{c_i})_{\alpha\beta}^{\alpha\beta},
\label{eeH_ME}
\end{eqnarray}
where
\bsub
\begin{eqnarray}
 (M_{\az})_{\alpha\beta}^{\alpha\beta} &=& 
  (M_{\rm SM})_{\alpha\beta}^{\alpha\beta} \nonumber \\
 &=& \gz \, \mz \, g_{\alpha}^{Zee} \,
  g_{-\beta}^{Zee} \, D_Z(t_1) \, D_Z(t_2) 
  \,\,{\cal F}(\theta_1,\theta_2)\,\,  
  \Bigg( \delta_{\alpha\beta}\,\, {\cal H}(\theta_1,\theta_2,\phi)
  + \delta_{\alpha,-\,\beta} \Bigg), \\
 (M_{c_{2 V_1 V_2}})_{\alpha\beta}^{\alpha\beta} &=& 
  (M_{\rm SM})_{\alpha\beta}^{\alpha\beta} \,\,
  \frac{D_{V_1}(t_1)} {D_Z(t_1)}\,\, \frac{D_{V_2}(t_2)}{D_Z(t_2)}  
  \left[ f_{2V_1V_2} + \frac{s}{4\mzsq} \,\, \frac{\delta_{\alpha\beta}
   \,\,{\cal G'} +\delta_{\alpha,-\,\beta}\,\,{\cal G}}
   {\delta_{\alpha\beta}\,\, {\cal H}\big(\theta_1,\theta_2,\phi\big)
   + \delta_{\alpha,-\,\beta}}
  \right], \\
 (M_{c_{3 V_1 V_2}})_{\alpha\beta}^{\alpha\beta} &=& 
  (M_{\rm SM})_{\alpha\beta}^{\alpha\beta} 
  \,\, \frac{D_{V_1}(t_1)}{D_Z(t_1)}\,\,  \frac{D_{V_2}(t_2)}{D_Z(t_2)}
  \left[ f_{3V_1V_2} - \frac{s}{4 \mzsq} \,\, 
   \frac{ \delta_{\alpha\beta} \,\, {\cal G'} 
   +\delta_{\alpha,\,-\beta}\,\, {\cal G}}
   {\delta_{\alpha\beta} \,\, {\cal H}\big(\theta_1,\theta_2,\phi\big)
   + \delta_{\alpha,-\,\beta}}
  \right],
\end{eqnarray}
\label{eeH_MEci}
\esub
\noindent
\hskip -1.4mm
with
\bsub
\begin{eqnarray}
 f_{1ZZ} &=& 1,  \hskip 1cm
  f_{2ZZ} = f_{2Z\gamma} = f_{2\gamma Z} = f_{2\gamma\gamma} = 
  -\frac{\mhsq -t_1 -t_2}{\mzsq}, \label{eeH_FF-1} \\
 f_{3ZZ} &=& \frac{\mhsq}{\mzsq}, \hskip 1cm
  f_{3Z\gamma} = \frac{\mhsq-t_1+t_2}{\mzsq}, \hskip 1cm
  f_{3\gamma Z} = \frac{\mhsq+t_1-t_2}{\mzsq}.
  \label{eeH_FF-2}
\end{eqnarray}
\label{eeH_FF}
\esub
\noindent
\hskip -1.4mm
It should be noted that in the notation of eq.(\ref{eeH_ME}), the
coefficients of the $HZ\gamma$ couplings should be interpreted as 
\begin{eqnarray}
 c_{kZ\gamma} (M_{c_{kZ\gamma}})_{\alpha\beta}^{\alpha\beta} +
 c_{k\gamma Z} (M_{c_{k\gamma Z}})_{\alpha\beta}^{\alpha\beta}
 =  c_{kZ\gamma} 
 \left[ (M_{c_{kZ\gamma}})_{\alpha\beta}^{\alpha\beta} + 
  (M_{c_{k\gamma Z}})_{\alpha\beta}^{\alpha\beta} \right]
\end{eqnarray}
for $k=2,3$, by using the expressions in eq.(\ref{eeH_MEci}b) and
(\ref{eeH_MEci}c).

The SM distribution $\Sigma_{\rm SM}$ and the weight functions
$\Sigma_{c_i}$ depend on the 3-body phase space in the laboratory frame
which is parametrized as in eq.(\ref{3bdyphsp-4}), 
\begin{eqnarray}
 d\Phi_3 = \frac{s}{1024 \,\, \pi^4} \,\,\,\,
  \frac{x_1\,\, (1-x_1-\mhsq/s)}{[1-x_1\,\, (1-\cos\theta_{12})/\,2]^2} 
  \,\,\,\,dx_1 \,\,\,\, d\cos\theta_1 \,\,\,\, d\cos\theta_2 \,\,\,\,  d\phi,
\end{eqnarray}
where $\phi=\phi_2-\phi_1$. The differential cross section with $e^-$
and $e^+$ beam polarizations, $P$ and $\bar{P}$, respectively, is
expressed as in eq.(\ref{dsigma}) where the weight functions are
\bsub
\begin{eqnarray}
 \Sigma_{\rm SM}(P,\,\bar{P};\,\,\Phi_3) &=& \frac{1}{2 \,s} \,\, 
  \sum_{\alpha,\beta} \,\, \left( \frac{1+\alpha \,\,P}{2}\right) \,\, 
  \left( \frac{1+\beta \,\,\bar{P}}{2} \right) \,\,
  | (M_{\rm SM})_{\alpha\beta}^{\alpha\beta} |^2 , \nonumber\\
 \Sigma_{c_i}\big(P,\,\bar{P};\,\,\Phi_3\big) &=& \frac{1}{2\,\,s} \,\, 
  \sum_{\alpha\beta} \,\, \left(\frac{1+\alpha \,\,P}{2} \right) \,\,
  \left( \frac{1+\beta \,\,\bar{P}}{2} \right) \,\,
  2\,\,{\rm Re} \,\left[
  \big (M_{c_i}\big)_{\alpha,\beta}^{\alpha\beta}\,\,\, 
  \big (M_{\rm SM}^*\big)_{\alpha\beta}^{\alpha\beta} \right] \, ,
\end{eqnarray}
\label{eeH_WF}
\esub
\noindent
\hskip -1.7mm
The six non-standard couplings, $c_i\equiv\az$, $\bz$, $\cz$, $\bg$,
$\cg$ and $\bgg$, contribute to the $t$-channel $Z$ and $\gamma$
exchange processes, and the inverse of the covariant matrix is then
evaluated as in eq.(\ref{Vinv-b}). As shown in fig.\ref{rootsvariation}
and noted in section \ref{OptPhaseSpace-3}, the interference
contribution from the $s$-channel $ZH$ production amplitudes is not
negligible at $\sqrt{s}=250\GEV$ and at $350\GEV$. We therefore replace
the SM amplitude in eq.(\ref{eeH_WF}) by the sum of the $t$-channel and
$s$-channel amplitudes at these energies. That is, we neglect
contributions from the anomalies in the suppressed $ZH$ production
amplitudes.

\subsection{Constraint on the $HZZ$, $HZ\gamma$ and $H\gamma\gamma$ couplings}

\subsubsection{Sensitivity at $\sqrt{s}=250\GEV$ and $350\GEV$ with
   the integration approximation}

We first present the \lq theoretical\rq~ result for the contribution
from the $t$-channel vector boson fusion, using the standard integration
procedure for $\sqrt{s}$ = 250 and 350 GeV. Later on we take into
consideration the interference effects from $s$-channel $ZH$ production
and adopt a more realistic approach to evaluate the inverse of the
respective covariance matrices.  

Our results for the purely $t$-channel process at $\sqrt{s}=250\GEV$ for
$L_0=100\ifb$ with $|P|=80\%$ are :
\bsub
\begin{eqnarray}
 .001 \,\az -.000 \,\bz -.001 \,\cz -.001 \,\bg +.99997 \,\cg -.008 \,\bgg 
  &=& \,\pm \, .0014, \hskip 1cm \label{ZZeigen250-1} \\
 .028 \,\az -.067 \,\bz +.005 \,\cz -.065 \,\bg -.008 \,\cg -.995 \,\bgg  
  &=& \,\pm \, .0040, \label{ZZeigen250-2} \\
 .009 \,\az -.097 \,\bz +.010 \,\cz +.993 \,\bg +.000 \,\cg -.058 \,\bgg
  &=& \,\pm \,.0099, \label{ZZeigen250-3} \\
 .28 \,\az +.54 \,\bz -.79 \,\cz +.056\,\bg -.002 \,\cg -.036 \,\bgg  
  &=& \,\pm \, .047, \label{ZZeigen250-4} \\
 .73 \,\az -.65 \,\bz -.19 \,\cz -.064 \,\bg -.000 \,\cg +.068 \,\bgg  
  &=& \,\pm \,.067, \label{ZZeigen250-5} \\
 .62 \,\az +.52 \,\bz +.58 \,\cz +.039 \,\bg +.000 \,\cg -.017 \,\bgg  
  &=& \,\pm \,.17. \label{ZZeigen250-6}
\end{eqnarray}
\label{ZZeigen250}
\esub
\noindent
\hskip -1.4mm
The cross section for the $t$-channel double-tag $eeH$ process is only
0.83~fb as can be seen from the solid thin curve in
fig.\ref{rootsvariation}. Imposing the exclusion cut
$|m_{e^+e^-}-\mz|>5\,\Gamma_Z$, it reduces to 0.55~fb, or 55 events for
$100\ifb$. Nevertheless, we present the above results in the integral
approximation (the small bin size limit) over the 3-body phase space, as
a reference to study the energy dependence. Binning effects for small
statistics is reported in the next sub-subsection.

It is remarkable that the two $HZ\gamma$ couplings and the
$H\gamma\gamma$ coupling are rather uniquely constrained in the first
three eigenvectors of eq.(\ref{ZZeigen250}) with small errors. This is
a consequence of the combined effect of the strong $e^-$ beam
polarization dependence of the $Z-\gamma$ interference term (even though
it is effective only for the $\gamma$-exchange between $e^-$ and
$H$) and also from the kinematic difference between the $\gamma$ and
$Z$ propagator factors when $\vert t_1\vert$ or $\vert t_2\vert$ or both
are significantly smaller than $\mzsq$. In fact, we confirm the latter
effects from the unpolarized beam ($P=0$) case where the three couplings
are still measured rather uniquely~\cite{YMthesis}; $\bgg$ is now
constrained most accurately with the error $\pm.0041$, $\cg$ has an
error of $\pm.011$, while $\bg$ is poorly constrained. The three $HZZ$
couplings are constrained rather weakly in the last three lines of
eq.(\ref{ZZeigen250}).

Similarly at $\sqrt{s}=350\GEV$ for $L=100\ifb$ with $|P|=80\%$, we find
\bsub
\begin{eqnarray}
 .002 \,\az -.000 \,\bz -.003 \,\cz +\,.001 \,\bg +.99998 \,\cg 
  -.006 \,\bgg &=& \,\pm \, .00085, \hskip 1cm \label{ZZeigen350-1} \\
 .046 \,\az -.17 \,\bz -.021 \,\cz -.21 \,\bg -.006 \,\cg 
  -.96 \,\bgg &=& \,\pm \, .0035, \label{ZZeigen350-2} \\
 .025 \,\az -.099 \,\bz +.006 \,\cz +.98 \,\bg -.003 \,\cg  
  -.19 \,\bgg &=& \,\pm \, .0053, \label{ZZeigen350-3} \\
 .29 \,\az +.41 \,\bz -.86 \,\cz +.030 \,\bg -.004 \,\cg 
  -.047 \,\bgg &=& \,\pm \, .016, \label{ZZeigen350-4} \\
 .47 \,\az -.82 \,\bz -.24 \,\cz -.057 \,\bg -.000 \,\cg 
  +.19 \,\bgg &=& \,\pm \, .025, \label{ZZeigen350-5} \\
 .83 \,\az +.34 \,\bz +.44 \,\cz +.004 \,\bg -.000 \,\cg    
  -.032 \,\bgg &=& \,\pm \, .054. \label{ZZeigen350-6}
\end{eqnarray}
\label{ZZeigen350}
\esub
\noindent
\hskip -1.4mm
Here the cross section is 2.9~fb which is almost a factor 3.5  times
larger than those at $250\GEV$; see fig.\ref{rootsvariation}. The effect
of the exclusion cut is negligible and the cross section is reduced to
2.8~fb which would mean 280 events with 100$\ifb$. Therefore, we naively
expect a factor of 2 improvements in the statistical error when compared
with the results eq.(\ref{ZZeigen250}). The results in
eq.(\ref{ZZeigen350}) show that the error of the $HZZ$ couplings
decreases to about 1/3, those of the $HZ\gamma$ couplings to $\sim 1/2$,
while the $H\gamma\gamma$ couping does not show a significant
improvement. The $H\gamma\gamma$ coupling measurement does not improve
because the amplitudes with $|t_i|<\mzsq$ do not increase much with
$\sqrt{s}$. We notice that the two couplings $\bgg$ and $\bg$ now have
some correlation, as can be seen from the combinations
$(0.21\bg+0.96\bgg)$ in eq.(\ref{ZZeigen350-2}) and $(0.98\bg-0.19\bgg)$ 
in eq.(\ref{ZZeigen350-3}). It is because of their similar behavior when
both $|t_1|$ and $|t_2|$ are larger than $\mz$. Such region of the phase
space is tiny at $\sqrt{s}=250\GEV$, but starts appearing at $350\GEV$,
and will be dominating at higher energies. 

\subsubsection{Sensitivity at $\sqrt{s}=250\GEV$ and $350\GEV$ with
   the table method including $s$-channel contributions}
As the $t$-channel cross sections are very small at $\sqrt{s}$ = 250 GeV
and 350 GeV, the interference effects due to the contribution from $ZH$
amplitudes along with the exclusion $|m_{e^+e^-}-\mz| >5\,\Gamma_Z$ can
be significant as can be seen from the thick dashed line, in
fig.\ref{rootsvariation}. To simulate the realistic experimental
situation we adopt the table method to calculate $V^{-1}$, which is more
realistic when the expected number of events is not large. Here we
examine the two dimensional weight functions in terms of the momentum
transfers $|t_1|$ and $|t_2|$.

At $\sqrt{s}$ = 250 GeV with $L_0=100\ifb$ and $\vert P\vert=80\%$, the
cross section of the double tag $e^+e^-\,H$ process becomes
0.93~fb, after including the contribution from the $s$-channel
amplitudes; see fig.\ref{rootsvariation}. Because of the smallness of
the signal events (93 events for $100\ifb$ with 100\% efficiency), we
integrate the weight functions over the azimuthal angle and
$m_{e^+e^-}$, and divide $|t_i|$'s into 3 regions each as follows:
$|t_i|<(45\GEV)^2$, $(45\GEV)^2<|t_i|<(90\GEV)^2$ and
$|t_i|>(90\GEV)^2$. With this binning, all the 18 bins (9 bins each for
$P=0.8$ and $P=-0.8$) have approximately 5 events. We obtain the
$V^{-1}$ matrix by summing over the contributions from 18 bins, and find 
\bsub
\begin{eqnarray}
 .006 \,\az -.0 \,\bz -.010 \,\cz +.31 \,\bg +.95 \,\cg +.051 \,\bgg 
  &=& \,\pm \, .0043, \hskip 1cm \label{ZZeigencomb250-1} \\
 .050 \,\az +.13 \,\bz +.10 \,\cz +.93 \,\bg -.31 \,\cg +.065 \,\bgg  
  &=& \,\pm \, .013, \label{ZZeigencomb250-2} \\
 .24 \,\az +.63 \,\bz +.60 \,\cz -.11 \,\bg +.064 \,\cg -.41 \,\bgg
  &=& \,\pm \,.016, \label{ZZeigencomb250-3} \\
 .26 \,\az +.29 \,\bz +.18 \,\cz -.14\,\bg -.004 \,\cg +.89 \,\bgg  
  &=& \,\pm \, .043, \label{ZZeigencomb250-4} \\
 .80 \,\az +.13 \,\bz -.56 \,\cz +.010 \,\bg -.005 \,\cg -.16 \,\bgg  
  &=& \,\pm \,.088, \label{ZZeigencomb250-5} \\
 .48 \,\az -.70 \,\bz +.53 \,\cz +.012 \,\bg -.001 \,\cg -.015 \,\bgg  
  &=& \,\pm \,.37. \label{ZZeigencomb250-6} 
\end{eqnarray}
\label{ZZeigencomb250}
\esub
\noindent
\hskip -1.4 mm 
In spite of the increase of the cross section, from 0.55~fb to 0.93~fb,
we observe that the errors of all the eigenvectors in
eq.(\ref{ZZeigencomb250}) are larger than these in eq.(\ref{ZZeigen250})
which are obtained by using the integral approximation. Especially, the
error of the eigenvector with dominant $\bgg$ component in
eq.(\ref{ZZeigencomb250-4}) becomes an order of magnitude larger than
the corresponding one in eq.(\ref{ZZeigen250-2}). The main reason for
the enhancement of the errors is the loss of information due to the
large bin size. For instance, the error of $\bgg$ is reduced
significantly if we divide the $|t_i|<(45\GEV)^2$ bin into 2 bins, while
that of $\cg$ is reduced if we divide the $|t_i|>(90\GEV)^2$ bin into 2
bins. In the very high luminosity limit, we can decrease the bin size,
and the integral limit of eq.(\ref{ZZeigen250}), corrected for the
interference effects, is obtained. In the following analysis,
however, we will use the result eq.(\ref{ZZeigencomb250}) as the
contribution from the double-tag $eeH$ events at $\sqrt{s}=250\GEV$ for
$100\ifb$ with $|P|=0.8$.
 
We repeat the same exercise at $\sqrt{s}$ = 350 GeV for $L_0=100\ifb$
with $|P|=0.8$. As can be seen from fig.\ref{rootsvariation} the cross
section increases to 3.3~fb from 2.9~fb due to constructive
interference. Here we considered 6 bins each for $\vert t_1\vert $ and
$\vert t_2\vert$, to make the number of event almost the same in all the
72 bins. We find :
\bsub
\begin{eqnarray}
 .004 \,\az +.005 \,\bz -.002 \,\cz +\,.11 \,\bg +.99 \,\cg 
  +.014 \,\bgg &=& \,\pm \, .0012, \hskip 1cm \label{ZZeigen350c-1} \\
 .007 \,\az +.006 \,\bz -.072 \,\cz +.97 \,\bg -.11 \,\cg 
  +.22 \,\bgg &=& \,\pm \, .0074, \label{ZZeigen350c-2} \\
 .21 \,\az +.31 \,\bz +.44 \,\cz +.21 \,\bg -.013 \,\cg  
  -.79 \,\bgg &=& \,\pm \, .011, \label{ZZeigen350c-3} \\
 .67 \,\az +.20 \,\bz -.70 \,\cz -.026 \,\bg -.000 \,\cg 
  -.14 \,\bgg &=& \,\pm \, .023, \label{ZZeigen350c-4} \\
 .71 \,\az -.31 \,\bz +.52 \,\cz -.046 \,\bg -.000 \,\cg 
  +.35 \,\bgg &=& \,\pm \, .079, \label{ZZeigen350c-5} \\
 .023 \,\az +.88 \,\bz +.19 \,\cz -.092 \,\bg -.000 \,\cg    
  +.43 \,\bgg &=& \,\pm \, .095. \label{ZZeigen350c-6}
\end{eqnarray}
\label{ZZeigencomb350}
\esub
\noindent
\hskip -.14 cm 
All the errors in eq.(\ref{ZZeigencomb350}) are approximately a factor
of 2 larger than the corresponding ones in eq.(\ref{ZZeigen350}), mainly
because of the loss of information due to the binning and also to the
reduction of the dimensionality of the weight functions from 4 ($x_i$,
$\cos\theta_1$, $\cos\theta_2$, $\phi$) to 2 ($|t_1|$, $|t_2|$).

\subsubsection{Sensitivity at $\sqrt{s}=500\GEV$ and $1\TEV$}

At $\sqrt{s}=500\GEV$ with $L=100\ifb$ with $|P|=80\%$, we find :
\bsub
\begin{eqnarray}
 .002 \,\az +.000 \,\bz -.007 \,\cz -.004 \,\bg +.99995 \,\cg -.005
  \,\bgg &=& \,\pm \, .00057, \hskip 1cm \label{ZZeigen500-1} \\
 .034 \,\az -.20 \,\bz  -.064 \,\cz -.61 \,\bg -.007 \,\cg
  -.76 \,\bgg &=& \,\pm \, .0029, \label{ZZeigen500-2} \\
 .044 \,\az -.28 \,\bz -.030 \,\cz +.78 \,\bg +.000 \,\cg 
  -.55 \,\bgg &=& \,\pm \, .0035, \label{ZZeigen500-3} \\
 .19 \,\az +.31 \,\bz -.93 \,\cz +.045 \,\bg -.007 \,\cg  
  -.029 \,\bgg &=& \,\pm \, .0067, \label{ZZeigen500-4} \\
 .34 \,\az -.84 \,\bz -.23 \,\cz -.098 \,\bg -.000 \,\cg 
  +.33 \,\bgg &=& \,\pm \, .013, \label{ZZeigen500-5} \\
 .92 \,\az +.27 \,\bz +.28 \,\cz +.012 \,\bg -.000 \,\cg 
  -.063 \,\bgg &=& \,\pm \, .031. \label{ZZeigen500-6}
\end{eqnarray}
\label{ZZeigen500}
\esub
\noindent
\hskip -1.4mm
The cross section now is 6.6~fb, or 660 events with $100\ifb$; see
fig.\ref{rootsvariation}. We therefore expect that the statistical
errors should be about 2/3 of those at $\sqrt{s}=350\GEV$. In fact, all
the errors of eq.(\ref{ZZeigen500}) are smaller than the corresponding
errors in eq.(\ref{ZZeigen350}) by a factor of 1.2 to 2.4. At this
energy, as mentioned above, kinematic difference in the $Z$ and $\gamma$
propagators diminishes, and the $H\gamma\gamma$ coupling $\bgg$ tends to
mix with $\bg$ and the $HZZ$ couplings.

And at $\sqrt{s}=1\TEV$ for $L=100\ifb$ with $|P|=80\%$, we find :
\bsub
\begin{eqnarray}
 .002 \,\az +.002 \,\bz -.020 \,\cz -.019 \,\bg +\,.9996 \,\cg  
  -.006 \,\bgg &=& \,\pm \, .0004 ,\hskip 1cm \label{ZZeigen1000-1} \\
 .008 \,\az +.033 \,\bz -.15 \,\cz -.95 \,\bg -.022 \,\cg 
  -.26 \,\bgg &=& \,\pm \,.0018 , \label{ZZeigen1000-2} \\
 .089 \,\az +.12 \,\bz -.97 \,\cz +.19 \,\bg -.017 \,\cg 
  -.091 \,\bgg &=& \,\pm \, .0022 , \label{ZZeigen1000-3} \\
 .035 \,\az -.67 \,\bz +.018 \,\cz +.17 \,\bg +.000 \,\cg
  -.72 \,\bgg &=& \,\pm \, .0025 , \label{ZZeigen1000-4} \\
 .15 \,\az -.72 \,\bz -.17 \,\cz -.17 \,\bg -.002 \,\cg  
  +.64 \,\bgg &=& \,\pm \, .0058 , \label{ZZeigen1000-5} \\
 .98 \,\az +.12 \,\bz +.11 \,\cz +.011 \,\bg +.000 \,\cg  
  -.063 \,\bgg &=& \,\pm \,.019 . \label{ZZeigen1000-6}
\end{eqnarray}
\label{ZZeigen1000}
\esub
\noindent
\hskip -1.4mm
The cross section is around 13~fb, or 1300 events with $100\ifb$. The
error of $\cg$ in eq.(\ref{ZZeigen1000-1}) decreases by a factor 1.4
from the corresponding one in eq.(\ref{ZZeigen500-1}) for $500\GEV$. The
other errors are reduced more significantly. At high energies, the
distinction among the weight functions becomes clear and $\cg$,
$\bg$, $\cz$ and $\az$ are measured rather independently, while only
$\bz$ and $\bgg$ are correlated. Since the $\gamma$ and $Z$ propagator
factors are similar at high energies, the distinction come from the extra
$(t_1,t_2)$ dependence in eq.(\ref{eeH_FF}) and from the azimuthal
angle dependence in the factor ${\cal G}$ and ${\cal G'}$; see
eq.(\ref{FHGfunc}) in appendix B. Most significant improvement is found
for the error of the eigenvector with dominant $\cz$ component, which
decreases by a factor of 3 from 0.0067 at $\sqrt{s}=500\GEV$ in
eq.(\ref{ZZeigen500-4}) to 0.0022 at $\sqrt{s}=1\TEV$ in
eq.(\ref{ZZeigen1000-3}).

Overall, we find that the $HZ\gamma$ couplings are most accurately
measured at high energies in the double-tag $eeH$ process. However, the
accuracies of the $\bg$ and $\cg$ measurements cannot compete with those
from the $ZH$ process, as can be seen e.g.\ by comparing
eqs.(\ref{ZZeigen1000-1}) and (\ref{ZZeigen1000-2}) with
eqs.(\ref{ZH_ResultPol1000-1}) and (\ref{ZH_ResultPol1000-3}),
respectively, at $\sqrt{s}=1\TEV$. The double-tag events are found to be
most important in distinguishing $\az$, the SM like $HZZ$ coupling, from
$\bz$ and $\cz$.

\section{no-tag $ee \rightarrow (ee)H$ process}
\label{EPF}
Although the double-tag $t$-channel $eeH$ process shows 
certain sensitivity to the $H\gamma\gamma$ coupling $\bgg$, the
sensitivity at $\sqrt{s}=500\GEV$ and $1\TEV$ with $L_0$ = 100$\ifb$ is
not sufficient for resolving the $H\gamma\gamma$ coupling of the SM from
the $W$-boson and top quark loops. We can expect better sensitivity to
the $H\gamma\gamma$ coupling when the $e^{\pm}$ are not tagged, because
of the fusion of almost real photons. The produced Higgs-boson has
little $p_T$, and can be distinguished from the $\nu_e \bar{\nu}_e H$
events.

\subsection{Cross section}
The momentum and helicity assignment for the initial and final particles
in no-tag $(ee)H$ process is the same as the double-tag $eeH$ process,
eq.(\ref{eeH_process}). The momentum and helicity of the intermediate
almost real photon is assigned as
\bsub
\begin{eqnarray}
 e^- \left( k_1,\frac{\sigma_1}{2} \right) &\rightarrow& 
  e^- \left( p_1,\frac{\lambda_1}{2} \right) + 
  \gamma \left( k'_1,\lambda'_1 \right),\\
 e^+ \left( k_2,\frac{\sigma_2}{2} \right)& \rightarrow &
  e^+ \left( p_2,\frac{\lambda_2}{2} \right) + 
  \gamma \left( k'_2,\lambda'_2 \right),
\end{eqnarray}
\esub
\noindent
where $\lambda'_1$ and $\lambda'_2$ is the helicity of the equivalent
real photon emitted from $e^-$ and $e^+$, respectively. The final
electron and positron escape detection by going into the beam pipe,
and an almost real photon of the virtuality $\vert q^2\vert \ll \mhsq$
is emitted in the same direction; see e.g.\ ref.\cite{WWa} for a
review. The $Z$ boson contribution is suppressed by a  small factor of
$|q^2|/\mzsq \ll 1$. The minimum and maximum magnitude of the square of
the electron or positron momentum transfer is calculated as
\begin{eqnarray}
 q_{\rm min}^2 = \frac{m_e^2 \,\,z^2}{1-z},\;\;\;\;\;\; 
  q_{\rm max}^2 =(1-z) \,\,\frac{s}{2}\,\,(1-\cos \theta_{\rm min}),
\label{qminqmax}
\end{eqnarray}
where $z=2\, E_\gamma /\sqrt{s}$ is the energy fraction of the photon,
and $\theta_{\rm min}$ is the polar angle below which the final
$e^{\pm}$ escapes detection into the beam pipe. In our analysis, we set
$\cos\theta_{\rm min}=0.995$; see eq.(\ref{Notag_cond}). The equivalent
real photon distribution can be split into two parts,
\begin{eqnarray}
 D_{\gamma/e^{\pm}}(z,Q^2) &=& 
  D_{\gamma/e^{\pm}}^{\sigma_i\lambda'_j=+}(z,Q^2) + 
  D_{\gamma/e^{\pm}}^{\sigma_i\lambda'_j=-}(z,Q^2),
  \label{NTeeH_EPA}
\end{eqnarray}
where the helicity preserving component
$(\sigma\lambda'=+)$ and the helicity flip component
$(\sigma\lambda'=-)$ are \cite{hagiphasespace}
\bsub
\begin{eqnarray}
 D_{\gamma/e^{\pm}}^{+}(z,Q^2) &=&
  \frac{\alpha}{2\,\,\pi}\,\, \biggl[ \frac{1}{z} \,\,
  \left( \ln \frac{Q^2}{q_{\rm min}^2}-1\right)+z \biggr], \\
 D_{\gamma/e^{\pm}}^{-}(z,Q^2) &=&
  \frac{\alpha}{2\,\,\pi} \,\,\biggl[ \frac{(1-z)^2}{z} 
 \,\, \left( \ln \frac{Q^2}{q_{\rm min}^2}-1\right) \biggr],
\end{eqnarray}
\label{NTeeH_EPA2}
\esub
\hskip -0.7mm respectively, with
\begin{eqnarray}
 Q^2 = {\rm min} \{\mhsq, \,\,q_{\rm max}^2 \}.
  \label{qmaxdef}
\end{eqnarray}
It is worth noting that the last term $z$ in the parenthesis of the
helicity preserving $(\sigma\lambda'=+)$ distribution
(\ref{NTeeH_EPA2}a) comes from the helicity flip $(\sigma\lambda=-)$
amplitude where the final $e^{\pm}$ helicity is opposite from the
initial one. Since contribution from the the momentum transfer much
smaller than $1\GEV^2$ dominates the photon distribution, we use
$\alpha=1/137$ in eq.(\ref{NTeeH_EPA2}). The real photon approximation
for the matrix elements cease to be valid when $Q^2=|t_i|$ becomes
${\cal O}(\mhsq)$. In our study, the maximum momentum transfer 
$q^2_{\rm max}$ stays below $\sim0.25\,\, \mhsq$ even at $\sqrt{s}=1$
TeV, and hence our predictions do not change significantly when we
replace the scale from $\mhsq$ by $0.1\,\,\mhsq$, in eq.(\ref{qmaxdef}).

The cross section for the no-tag $(ee)H$ events is now expressed as
\begin{eqnarray}
 \sigma^{(ee)H} &=& 
  \sum_{\alpha,\beta} \left( \frac{1+\alpha \, P}{2} \right)
  \left( \frac{1+\beta \, \bar{P}}{2} \right) \,
  \sum_{\lambda'_1,\lambda'_2} \,\,\int_0^1 dx_1\, \int_0^1 dx_2 \, \,
  D^{\alpha\lambda'_1}_{\gamma/e^-}(x_1,Q^2) \,\, 
  D^{\beta\lambda'_2}_{\gamma/e^+}(x_2,Q^2) \, \,
  \hat{\sigma}^{\gamma\gamma \rightarrow H}_{\lambda'_1\lambda'_2}. \nonumber\\
 \label{NTeeH_Xsec}
\end{eqnarray}
where
\begin{eqnarray}
\hat{\sigma}^{\gamma\gamma \rightarrow H}_{\lambda'_1\lambda'_2} (\hat{s}) 
 = \delta_{\lambda'_1\lambda'_2} \,\, \bigg(16\,\, \pi^2\bigg)\,\,
 \frac{\Gamma_{\gamma\gamma}}{\mh} \,\, \delta(\hat{s}-\mhsq).
\label{Xsec_aaH}
\end{eqnarray}
Here, $x_i$ and $\lambda'_i$ are the energy fractions and the helicities
of the equivalent real photon in $e^-(i=1)$ and $e^+(i=2)$,
respectively, $\alpha/2$ and $\beta/2$ are the $e^-$ and $e^+$
helicities, and $\hat{s}=x_1 x_2 s$ is the square of the $\gamma\gamma$
collision energy. Higgs-boson is produced only when
$\lambda'_1=\lambda'_2$. Therefore, the total cross section and the
rapidity $\yH=1/2\log(x_1/x_2)$ distribution of the Higgs-boson, which
has strong dependence on the helicity profile of the equivalent real
photon distribution, depends on the $e^{\pm}$ beam polarizations $P$
and $\bar{P}$. We find, however, that the polarization effects survive
only when both $e^-$ and $e^+$ have non-zero polarization.

The un-polarized cross section of the no-tag $(ee)H$ events in
fig.\ref{rootsvariation} for $\mh=120\GEV$ and those of
fig.\ref{mhvariation} for $\mh=100\GEV$ to $200\GEV$ are calculated by
using the above formula with
\begin{eqnarray}
 \Gamma_{\gamma \gamma} &=& 
  \frac{\alpha^2 \,\,m_H^3}{256\,\, \pi^3\,\, v^2} \,\,\, 
  \big\vert I\big\vert^2,
\label{Ggg1}
\end{eqnarray}
where the loop function {\it I} sums over the contributions from the
$W$-boson and the top-quark \cite{HiggsHunter}. The enhancement due to
the $W$-boson pair production threshold can be observed in
fig.\ref{mhvariation}. For $\mh=120\GEV$, the decay width of
eq.(\ref{Ggg1}) gives $\Gamma_{\gamma \gamma}^{\rm SM}=0.0073\MEV$, and
the cross section is 0.15 fb, 0.32 fb, 0.56 fb and 1.3 fb at
$\sqrt{s}=250$, 350, 500 and $1000\GEV$, respectively, as shown in
fig.\ref{rootsvariation}. When the integrated luminosity is $100\ifb$,
the expected number of events are 15, 32, 56 and 128,
respectively. Therefore, in this study, we consider the no-tag events
only at $\sqrt{s}=500\GEV$ and $1\TEV$.

\subsection{Error Estimation for $\bgg$}
The interaction Lagrangian for the effective $H\gamma\gamma$ vertex 
\begin{eqnarray}
 L_{\rm int}^{H\gamma\gamma} 
  = \frac{1}{v}\,\,\, \bgg \,\,\, H \,\,A_{\mu\nu}\,\,\,  A^{\mu\nu},
  \label{HAAIntLag}
\end{eqnarray}
contains both the SM loop contribution and new physics effects. We
parametrize them as
\begin{eqnarray}
 \bgg = \bggSM + \Delta\bgg.
  \label{bggCoup}
\end{eqnarray}
The SM contribution is approximated by
the loop function {\it I}, and
its numerical value for $\mh=120\GEV$ is
\begin{eqnarray}
 \bggSM = \frac{\alpha}{8\,\,\pi}\,\, I = - \,0.00187,
\label{bggValue}
\end{eqnarray}
with negligibly small imaginary part from the $b$ and $\tau$
contributions. Our approximation of the point-like coupling in the
effective Lagrangian eq.(\ref{HAAIntLag}) is valid as long as 
$\vert q^2 \vert \ll \mwsq,m_t^2$. The cross section of the no-tag
$(ee)H$ process for $\mh=120\GEV$ is expressed in terms of the
generalized $H\gamma\gamma$ coupling given in eqs.(\ref{HAAIntLag}) and
(\ref{bggCoup}) as
\bsub
\begin{eqnarray}
 \sigma^{ee \rightarrow (ee)H}(\sqrt{s}=500\GEV) 
  &=& 1.61 \times 10^5\,\,\,
  \bigg ( \bggSM + \Delta\bgg \bigg)^2\,\,\, {\rm fb}, \\
 \sigma^{ee \rightarrow (ee)H}(\sqrt{s}=1\TEV) 
  &=& 3.70 \times 10^5 \,\,\,
  \bigg( \bggSM + \Delta\bgg \bigg)^2 \,\,\, {\rm fb},
\end{eqnarray}
\label{DTeeH_Xsec_bgg}
\esub
\hskip -1.4mm
\noindent As mentioned earlier, this cross section
does not depend on the $e^-$ polarization $P$, as long as the $e^+$
polarization is zero, $\bar{P}=0$.

While evaluating the error of the cross section, we  take into
account of the contribution from $WW$-fusion $\nu_e \bar{\nu}_e H$
events, which cannot be distinguished from the no-tag $(ee)H$ events
when the Higgs boson has small transverse momentum. In the following
analysis, we make a very naive estimate for the background and the
errors by assuming that the Higgs boson $\pT$ can be resolved at $3\GEV$
accuracy and that the SM background contribution can be estimated from
the high $\pT$ distribution.  In fig.\ref{Number of event at low Higgs
pT}, we show the expected $\pT$ distribution of the Higgs boson from the
no-tag $(ee)H$ process as the red-solid rectangle, which is overlaid
above the $WW$-fusion contribution shown by the gray-shaded histograms; 
\begin{figure}[htb!]
  \begin{center}
    \begin{minipage}[t]{0.45\textwidth}
      \includegraphics[width=9cm,height=7cm,angle=-90]
     {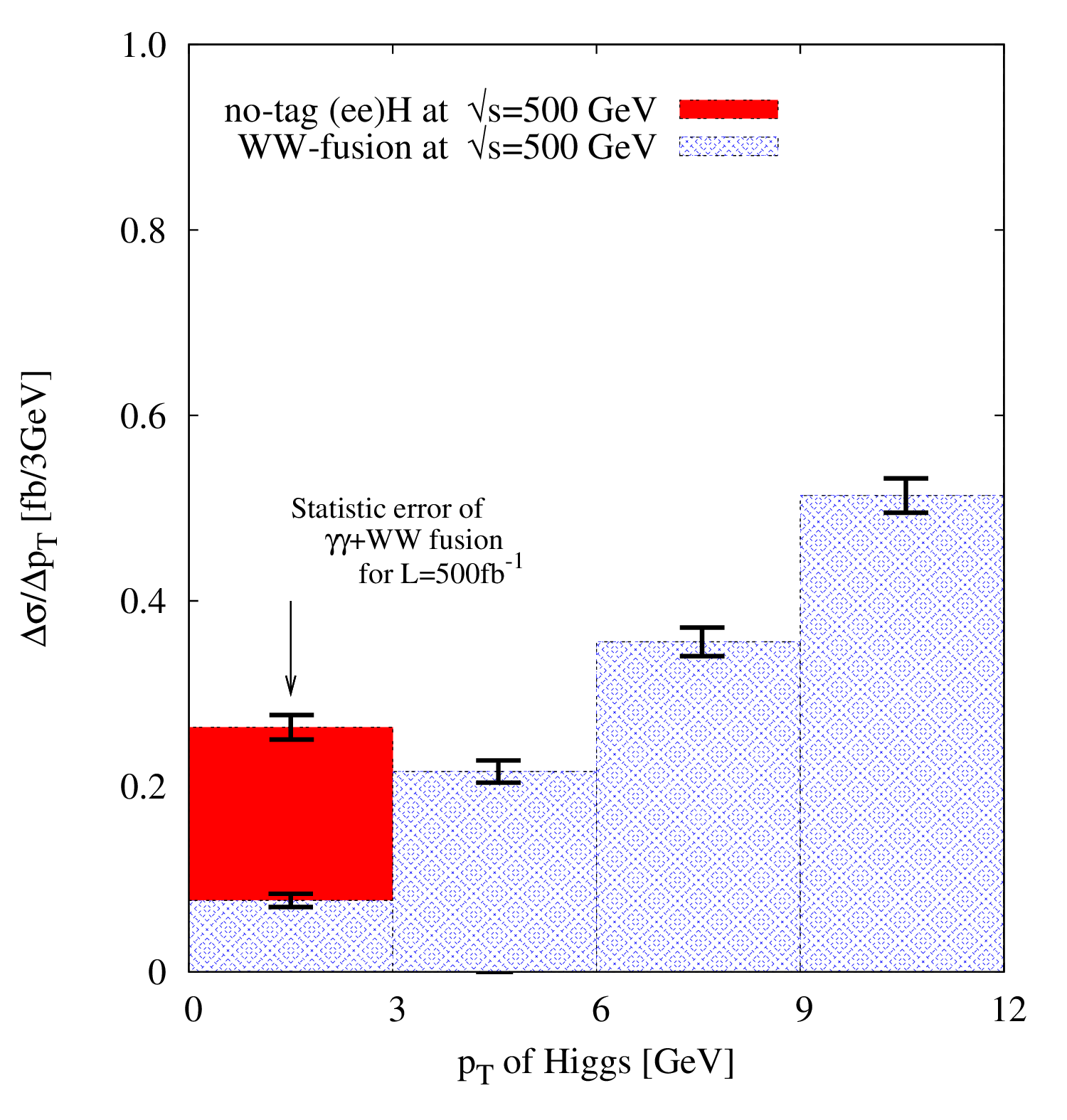} 
    \end{minipage}
    \vspace{5mm}
    \begin{minipage}[t]{0.45\textwidth}
      \includegraphics[width=9cm,height=7cm,angle=-90]
     {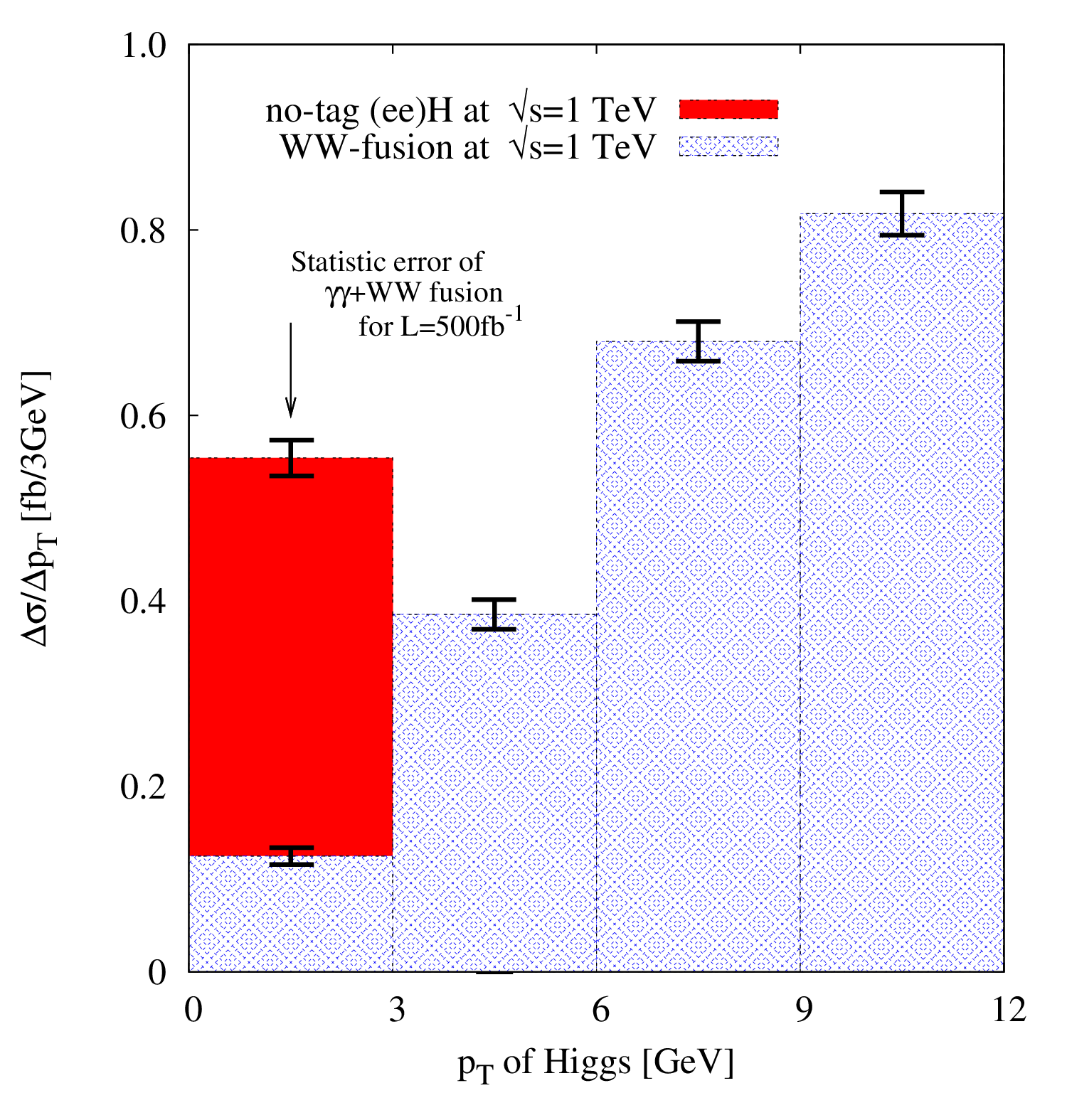} 
    \end{minipage}
    \caption{The histograms for the number of events at low Higgs $p_T$}
    \label{Number of event at low Higgs pT}
  \end{center}
\end{figure}
\noindent 
fig.\ref{Number of event at low Higgs pT}a is for $\sqrt{s}=500\GEV$
and fig.\ref{Number of event at low Higgs pT}b is for
$\sqrt{s}=1\TEV$. Also shown in the figures are the statistical errors
corresponding to the  background $WW$-fusion events, and that of the sum
of the no-tag $(ee)H$ and $WW$-fusion events for the lowest $\pT$
($\pT<3\GEV$) bin with $L_0=500\ifb$. We can then estimate the error of
the signal $(ee)H$ cross section as follows :
\begin{eqnarray}
 \sigma_{\rm sig} &=& \sigma_{\rm tot}-\sigma_{\rm BG} = 
  \bar{\sigma}_{\rm tot} -\bar{\sigma}_{\rm BG} \pm 
  \sqrt{ \frac{\bar{\sigma}^2_{\rm tot}}{N_{\rm tot}} 
  + \frac{\bar{\sigma}^2_{\rm BG}}{N_{\rm BG}}} \nonumber\\
 &=& \bar{\sigma}_{\rm sig} \pm 
  \sqrt{ \frac{\bar{\sigma}_{\rm tot}+\bar{\sigma}_{\rm BG}}{L} },
\end{eqnarray}
where $\bar{\sigma}$ represents the mean value and $L$ is the integrated
luminosity. The errors of the total and the background cross sections
in the lowest $\pT$ bin  are estimated to be 0.023 fb and 0.012 fb,
respectively, at $\sqrt{s}=500\GEV$, and 0.033 fb and 0.016 fb,
respectively, at $\sqrt{s}=1\TEV$. By taking the squared sum of the
errors, we obtain the following estimates :
\bsub
\begin{eqnarray}
 \sigma_{\rm sig} &=& 0.56 \pm 0.045 \,\,\,\, {\rm fb} \,\,\,\, \;\;\; 
  {\rm at} \;\;\, \sqrt{s}=500\GEV ; L=500\ifb, \\
 \sigma_{\rm sig} &=& 1.29 \pm 0.064\,\,\,\, {\rm fb} \,\,\,\, \;\;\;\; 
  {\rm at} \;\;\, \sqrt{s}=1\TEV ; L=500\ifb.
\end{eqnarray}
\label{xsecsignal}
\esub
We then find from eq.(\ref{DTeeH_Xsec_bgg}) 
\bsub
\begin{eqnarray}
 \bggSM + \Delta\bgg &=& -\,0.00187 \;
  \bigg\{
  \begin{array}{c}
   + 0.000074 \\
   - 0.000077 \\
  \end{array}
  \;\;\;\; {\rm at} \;\; \sqrt{s}=500\GEV ; L=500\ifb,
  \label{NTeeH_Res500} \\
 \bggSM + \Delta\bgg &=& -\,0.00187 \;
  \bigg\{
  \begin{array}{c}
   + 0.000046 \\
   - 0.000047 \\
  \end{array}
  \;\;\;\; {\rm at} \;\; \sqrt{s}=1\TEV ; L=500\ifb.
  \label{NTeeH_Res1000}
\end{eqnarray}
\label{NTeeH_Result}
\esub
\hskip -1.4mm 
The above results show that the no-tag $(ee)H$ process at $e^+e^-$
collider can measure the $H\gamma\gamma$ coupling with the 4\% and 2\%
accuracy with the integrated luminosity of $500\ifb$ at
$\sqrt{s}=500\GEV$ and $1\TEV$, respectively. In contrast, the
sensitivity of the double tag $ee\,H$ events on the $\bgg$ coupling in
eqs.(\ref{ZZeigen500}) and (\ref{ZZeigen1000}), gives the error of
0.0016 and 0.0014 at $\sqrt{s}$ = 500 GeV and 1 TeV, respectively, with
500 $\ifb$, which is barely enough to resolve the SM coupling at 1
$\sigma$ level. From this exercise, we can conclude that the no-tag
$(ee)H$ process  is about 20 to 30 times more sensitive to the
$H\gamma\gamma$   coupling than double-tag $eeH$ process,   at
$\sqrt{s}=500\GEV$  and $1\TEV$, if our naive estimation of the
background is valid.

This naive estimation  of the errors is based on the assumption that
the background contribution from the $WW$-fusion process can be
estimated from the high $\pT$ distribution. This assumption can be
tested experimentally  by using the different polarization dependence of
the signal and the background. The background $WW$-fusion process
scales as $1-P$, i.e.\ , it increases for $P<0$ and decreases for $P>0$,
while the signal remains  independent on $P$ as long as positron is
unpolarized. Once the positron polarization is available, both the total
cross section and the rapidity distribution of the no-tag $(ee)H$ events
depend  on the sign of the product $P\cdot\bar P$, which should be a
distinctive signature of the $\gamma\gamma$ fusion.

Although the sensitivity to the $H\gamma\gamma$ coupling is expected to
be measured far more accurately with the photon-linear collider 
\cite{gammacollider,HKZaacollider}, our results show the sensitivity
limit when it is not realized.

\section{Single-tag $e^+e^- \rightarrow (e^{\pm})e^{\mp}H$ process}
\label{SingleTag}
This process essentially measures $e^{\pm}\gamma \rightarrow e^{\pm}H$, 
where the initial photon comes from the other $\e^{\mp}$ beam as an
equivalent real photon. Since we have studied both double-tag $eeH$ and no-tag
$(ee)H$ events, it is worthwhile to examine the single-tag $(e)eH$
process  for completeness.
\begin{figure}[htb!]
  \begin{center}
    \includegraphics[width=7cm,height=5cm,angle=0]{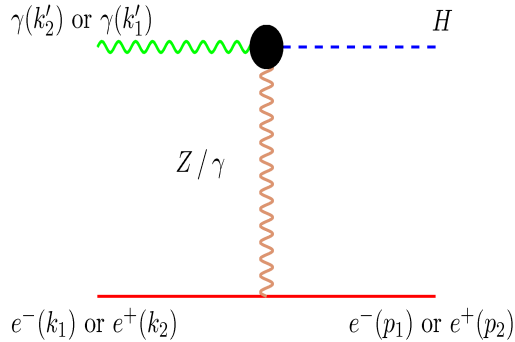}
    \caption{\small Feynman diagram of $e^{\pm}\gamma \rightarrow e^{\pm}H$}
    \label{STeeH_diagram}
  \end{center}
\end{figure}

\subsection{Cross section}
The momentum and the helicity assignments of the single-tag $(e)eH$
process is also the same as those in the double-tag $eeH$ process in
eq.(\ref{eeH_process}). We use the equivalent real photon approximation
for the  current from the untagged $e^{\pm}$. Since the helicity of the
equivalent real photon depends on the parent $e^{\pm}$ helicity, and
also the electroweak interactions of the tagged $e^{\pm}$ current are
sensitive to their polarization, we give the general expression of the
cross section with both $e^-$ and $e^+$ beam polarizations, $P$ and
$\bar{P}$, respectively;
\begin{eqnarray}
 \sigma^{(e)eH} &=& \sum_{\alpha,\beta}\,
  \left(\frac{1+\alpha \,P}{2}\right)\left( \frac{1+\beta\, \bar{P}}{2}\right) 
  \left[ \sum_{\lambda'_2} \,
   \hat{\sigma}_{\alpha\lambda'_2}^{e^-\gamma \rightarrow e^- H}\,\,
   D_{\gamma/e}^{\beta\lambda'_2}(z,Q^2) +
   \sum_{\lambda'_1}\,
   \hat{\sigma}_{\beta\lambda'_1}^{e^+\gamma \rightarrow e^+ H}\,\,
   D_{\gamma/e}^{\alpha\lambda'_1}(z,Q^2) \right]\, .\nonumber\\
  \label{STeeH_Xsec}
\end{eqnarray}
\noindent 
Here the polarization-dependent real photon distributions are as defined
in eq.(\ref{NTeeH_EPA2}), and the $e^{\pm}\gamma \rightarrow e^{\mp}H$
cross sections are
\begin{eqnarray}
 \hat{\sigma}_{\alpha\lambda'_i}^{e^{\pm}\,\,\gamma \rightarrow e^{\mp}H}
  = \frac{1}{2\,\,\hat{s}}\,\, \int \big\vert
  M_{\alpha\lambda'_i}^{e^{\pm}\gamma} \big\vert^2\,\, d\Phi_2 
  = \frac{1}{32\,\,\pi\,\,\hat{s}} \,\,
  \left( 1-\frac{\mhsq}{\hat{s}} \right)\,\,
  \int d\cos\theta^*_{e^{\pm}}\,\, 
  \big\vert M_{\alpha\lambda'_i}^{e^{\pm}\gamma}\big\vert^2 ,
  \label{STeeH_Xsec2}
\end{eqnarray}
where $\hat{s}=s\,\, z$, $\alpha/\,2$ is the $e^{\pm}$ helicity,
$\lambda'_i$ is the photon helicity, and $\theta^*_{e^{\pm}}$ is the
polar angle of the final tagged $e^{\pm}$ in the colliding
$e^{\pm}\gamma$ c.m.\ frame, measured from the incoming $e^{\pm}$
momentum direction. The integration region of $\cos\theta_{e^{\pm}}^*$
is constrained by the tagging criteria of eq.(\ref{Stag_cond})
\begin{eqnarray}
\big\vert\cos\theta_{e^{\pm}}\big\vert < 0.995 \equiv \cos\theta_{\rm min},
\end{eqnarray}
in the $e^+e^-$ collision c.m. frame. We find
\begin{eqnarray}
 -1 +\frac{z\,\, (1-\cos\theta_{\rm min})}
  {[z+(1-z)\,\, (1+\cos\theta_{\rm min})/\,2]} <
  \cos\theta^*_{e^{\pm}} < 
  1-\frac{1-\cos\theta_{\rm min}}
  {[z+(1-z)\,\, (1-\cos\theta_{\rm min})/\,2]}.
  \label{ST_etagcond_rest}
\end{eqnarray}

The interaction Lagrangian relevant for the 
$e^{\pm}\gamma \rightarrow e^{\pm}H$ process is
\begin{eqnarray}
 L_{\rm eff} = \frac{1}{v}\,\,\, \bgg \,\,\, A_{\mu\nu} \,\,A^{\mu\nu} + 
  \frac{2}{v} \,\, (\bg-\cg) \,\, H \,\,Z_{\mu\nu} \,\, A^{\mu\nu}\,\, .
\end{eqnarray}
The coupling $\bgg$ dictates the $\gamma$-exchange amplitude and the
combination $\bg-\cg$ dictates the $Z$-exchange amplitude in
fig.\ref{STeeH_diagram}. The helicity amplitudes for the $e^- \gamma
\rightarrow e^- H$ process are expressed as :
\begin{eqnarray}
 M_{\alpha\lambda'_2}^{e^-\gamma} &=& {j_{e^-}^{\mu}}_{(\alpha\alpha)} \,\,
  \left[ \frac{-e}{t_1} \,\,\Gamma_{\mu\nu}^{H\gamma\gamma} 
   +\frac{g_{\alpha}^{Zee}}{t_1-\mzsq} \,\,\Gamma_{\mu\nu}^{HZ\gamma}
  \right] \,\,\, \epsilon^{\nu}(k'_2,\,\,\lambda'_2), \nonumber\\
 &=& - \,\,\gz \,\,\sqrt{\hat{s}\,\, x_1} \,\,\frac{t_1-\mhsq}{\mz}\,\,
  \left[ \frac{-e}{t_1} \,\, \bgg 
   + \frac{g_{\alpha}^{Zee}}{t_1-\mzsq}\,\, (\bg-\cg) \right] \nonumber \\
 & & \hskip 3cm \sqrt{2}\,\,\sin\frac{\theta^*}{2}\,\,
  \left[ \delta_{\alpha\lambda'_2} 
   -\frac{\hat{s}-\mhsq}{t_1-\mhsq}\,\,\, \frac{1+\cos\theta^*}{2} \right],
\label{STeeH_ME}
\end{eqnarray}
\noindent 
where the $\Gamma_{\mu\nu}^{HVV}$ form factors are as in
eq.(\ref{HVVFormFactor}), ${j_{e^-}^{\mu}}_{(\alpha\alpha)}$ is the
massless $t$-channel current of eq.(\ref{tchancurrent1}), 
$t_1=(k_1-p_1)^2=-s\,\, x_1\,\,(1-\cos\theta^*)/\, 2$, $x_1$ is the
energy fraction of the tagged $e^-$, $\theta^*$ is the scattering angle
in the $e^-\gamma$ rest frame, $k'_2$ is the four momentum of the
equivalent real photon emitted from $e^+$. The amplitudes for 
$e^+\gamma \rightarrow e^+H$ are obtained from eq.(\ref{STeeH_ME})
simply by replacing $\alpha$, $g_{\alpha}^{Zee}$, $t_1=(k_1-p_1)^2$ and
$\lambda'_2$ by $\beta$, $g_{-\beta}^{Zee}$, $t_2=(k_2-p_2)^2$ and
$\lambda'_1$, respectively, where $\theta^*$ is now the scattering angle
between the initial and final $e^+$ in the $e^+\gamma$ rest frame.

It is instructive to show that the single-tag $(e)eH$ cross section
of eq.(\ref{STeeH_Xsec}) reduces to the no-tag $(ee)H$ cross section of
eq.(\ref{NTeeH_Xsec}) in the limit of $\big\vert t_1\big\vert \ll \mzsq$. 
In this limit, the $\gamma$-exchange contribution dominates the
amplitude, and by noting $\cos\theta^*\sim1$ and
$x_1=1-\mhsq/\,\hat{s}$, eq.(\ref{STeeH_Xsec2}) becomes
\begin{eqnarray}
 \hat{\sigma}_{\sigma\lambda}^{e^- \gamma \rightarrow e^- H}
 &\simeq& \frac{g_Z^2 \,\, e^2 \,\,{\bgg}^2 \,\,\mhsq}
 {8\,\,\pi\,\, \hat{s} \,\,\mzsq} \,\,\,
 \int \frac{d\cos\theta^*}{1-\cos\theta^*} 
 \, (1-x_1) \, \left[ \delta_{\sigma\lambda} +\frac{x_1}{1-x_1}
               \right]^2 , \nonumber\\
 &=& \frac{1}{\hat{s}} \,\,\,\frac{16\,\,\pi^2}{\mh} \,\,\,
  \Gamma_{\gamma\gamma} \,\,\int \frac{dt}{t} \,\, \frac{\alpha}{2\,\,\pi} \,\,
  \left[
   \frac{\delta_{\sigma,\lambda}+\delta_{\sigma,-\lambda}\,\,x_1^2}{1-x_1} 
  \right] , \nonumber\\
 &=& \frac{1}{\hat{s}} \,\,\,\frac{16\,\,\pi^2}{\mh} \,\, 
  \Gamma_{\gamma\gamma} \,\,
  D_{\gamma/\, e}^{\sigma\lambda}(1-x_1,Q^2).\label{egtogg}
\end{eqnarray}
Comparing the final expression of eq.(\ref{egtogg}) with
eq.(\ref{NTeeH_EPA2}a), we find that the term proportional to $z$, in
the  $\sigma\lambda=+$ distribution from the electron helicity flip
amplitudes, is not reproduced, since we have neglected the electron mass
of  the tagged $e^-$. Inserting eq.(\ref{egtogg})  in
eq.(\ref{STeeH_Xsec2}), we reproduce the cross section
(\ref{NTeeH_Xsec}) for the no-tag $(ee)H$ events.

\subsection{Error estimation for $\bgg$ and $\bg-\cg$}
To calculate the SM 1-loop contribution, we again adopt the
approximation of replacing the 1-loop vertices by localized effective
couplings,
\begin{eqnarray}
 \bgg = \bggSM + \Delta\bgg, \hskip 1cm \bg = \bgSM + \Delta\bg.
\label{bgCoup}
\end{eqnarray}
It is important to note that since the virtuality of the tagged
$e^{\pm}$ currents ($t$) can be larger than the weak boson mass scale
that dictates the spatial extension of the vertex, the approximation of
eq.(\ref{bgCoup}) is not as excellent as  that in eq.(\ref{bggCoup}) for
the no-tag $(ee)H$ events. In evaluating the SM contribution, we
therefore retain the $t$-dependence of the loop function \cite{Ginzburg},
\begin{eqnarray}
 \bggSM = \bggSM(t), && \bgSM = \bgSM(t),
  \label{bgCoupt}
\end{eqnarray}
and use the $t$-dependent \lq couplings\rq~ in the evaluation of the 
$e\gamma \rightarrow eH$ matrix elements. In other words, only the 
new physics contributions via $\Delta\bgg$ and
$\Delta\bg$ are assumed to be local in eq.(\ref{bgCoup}). With regard
to the $H\gamma\gamma$ couplings, $\bggSM(0)=-0.00187$ in
eq.(\ref{DTeeH_Xsec_bgg}) determines the $H\rightarrow \gamma\gamma$
width, while it reduces to $\bggSM(-\mzsq)=-0.00160$. Similarly, for the
$HZ\gamma$ coupling, $\bgSM(\mzsq)=-0.00322$ determines the 
$H\to Z\gamma$ width, while it reduces to $\bgSM(0)=-0.00276$ and
$\bgSM(-\,\mzsq)=-0.00243$, all for $\mh=120\GEV$. As a consequence of
the $W$ boson dominance in the loop, we find that the ratio 
$\bgSM(\mzsq)/\,\,\bggSM(0)\,\,\simeq1.7$, is essentially the ratio of
the $W$ boson gauge couplings $g_{ZWW}/g_{\gamma WW}\simeq1.8$. Since
the effective couplings are gauge-dependent when they are away from
their on-shell limit, $t=0$ for $\bggSM$ and $t=\mzsq$ for $\bgSM$,
respectively, the SM cross section based on the $t$-dependent effective
couplings should be regarded only as an order of magnitude
estimate. However, the box diagram contribution to the 
$e\gamma \to eH$ process, which is necessary to recover the gauge
invariance, is found to be negligibly small numerically
\cite{Ginzburg}.

Because the single-tag $(e)eH$ process has small cross sections as shown
in fig.\ref{rootsvariation} and fig.\ref{mhvariation}, we use only the
total cross section to constrain the $H\gamma\gamma$ and $HZ\gamma$
couplings.  In order to quantify the effects of the $e^-$ beam
polarization, we obtain the following parametrization for the single-tag
$e^-$ events for the $e^-$ helicity $\alpha$ and un-polarized
$e^{+{\footnotemark[2]}}$:
\footnotetext[2]{The coefficient $A_{\alpha}$ is determined by
evaluating the $\gamma$-exchange contribution only, such that the SM
cross section is reproduced with a constant $\bgg$, which may be
interpreted as the average of
$(\bgg)^2=\langle\bggSM(t)^2\rangle_{\gamma\gamma}$. Likewise, the
coefficient $C_{\alpha}$ is determined by requiring that the SM
contribution to the $\gamma$-$Z$ exchange interference is reproduced for
a constant value of $\bgg\,\, (\bg-\cg)$
=$\langle\bggSM(t)\rangle_{\gamma\gamma}\langle\bgSM(t)\rangle_{Z\gamma}$.
Finally, the coefficient $B_{\alpha}$ is fixed by demanding that the
total SM cross section is reproduced for the above average values.}
\begin{eqnarray}
 \sigma_{\alpha} &=& \frac{1}{2} \,\,\int dz \,\,\sum_{\lambda'_1} \,\,
  \sigma_{\alpha\lambda'_1}^{e^-\gamma \rightarrow e^-H}\,\,\,
  D_{\gamma/e^+}(z,Q^2) \nonumber\\
 &=& A_{\alpha}\,\,\,  (\bgg)^2
  +B_{\alpha}\,\,\, (\bg-\cg)^2 +C_{\alpha}\,\,\, \bgg\,\, (\bg-\cg)\, .
\end{eqnarray}
\label{STeeH_Xsec3}
\hskip -.14 cm
Here the average over $e^+$ helicities is replaced by the sum over the
photon helicity $\lambda'_1$. We find for $\sqrt{s}=500\GEV$,
\bsub
\begin{eqnarray}
\sigma_R &=& 15000\,\,(\bgg)^2  +\hphantom{0}890\,\,(\bg-\cg)^2
 -5500\,\,\bgg\,\, (\bg-\cg) = 0.025\,\,\,{\rm fb}, \\
 \sigma_L &=& 15000\,\,(\bgg)^2 +1200\,\, (\bg-\cg)^2 
  +6400\,\,\bgg\,\, (\bg-\cg) = 0.066\,\,\, {\rm fb},
\end{eqnarray}
\label{STeeH_ParaXsec500}
\esub
\hskip -.14cm 
and for $\sqrt{s}=1\TEV$,
\bsub
\begin{eqnarray}
 \sigma_R &=& 24000\,\, (\bgg)^2 +2900\,\,(\bg-\cg)^2 
  -15000\,\,\bgg\,\,(\bg-\cg) = 0.022\,\,\,{\rm fb}, \\
\sigma_L &=& 24000\,\,(\bgg)^2 +3900\,\, (\bg-\cg)^2 
 +17000\,\,\bgg\,\, (\bg-\cg) = 0.107\,\,\, {\rm fb}.
\end{eqnarray}
\label{STeeH_ParaXsec1000}
\esub
\hskip -1.4mm 
The large coefficients of $(\bgg)^2$ indicate that the $\gamma$-exchange
amplitude dominates the cross section even for the tagged events. The
effects of the $HZ\gamma$ couplings can be measured from the
interference between the $\gamma$-exchange and the $Z$-exchange
amplitudes, which is destructive for $e_R$ and constructive for $e_L$, 
reflecting the relative sign of their couplings, $g_L^{\gamma ee}$=
$g_R^{\gamma ee}=-\,e$ and 
$g_L^{Zee}=\gz\,\,(-\,1/2 \,+\sin^2\theta_{\rm W})$, 
$g_R^{Zee}=\gz\,\,\sin^2\theta_{\rm W}$. Therefore, we expect that
the beam polarization can be a powerful tool to distinguish the
$HZ\gamma$ coupling from the $H\gamma\gamma$ coupling.

It is also worth noting that the cross section after the $e$ tagging
condition of eq.(\ref{ST_etagcond_rest}) can become smaller at
high energies, as we find for $\sigma_R$ in eq.(\ref{STeeH_ParaXsec500}a)
and eq.(\ref{STeeH_ParaXsec1000}a).
\begin{figure}[htb!]
  \begin{center}
    \begin{minipage}[t]{0.45\textwidth}
      \includegraphics[width=7cm,height=8cm,angle=-90]
     {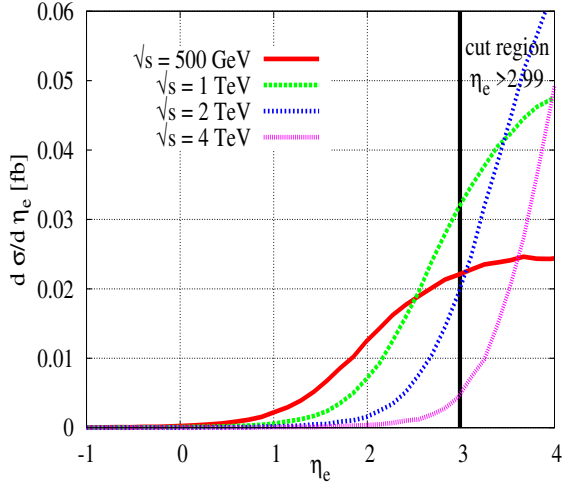} 
     \par {(a) Rapidity distributions for $\sigma_R$}
    \end{minipage}
    \hspace{5mm}
    \begin{minipage}[t]{0.45\textwidth}
      \includegraphics[width=7cm,height=8cm,angle=-90]
     {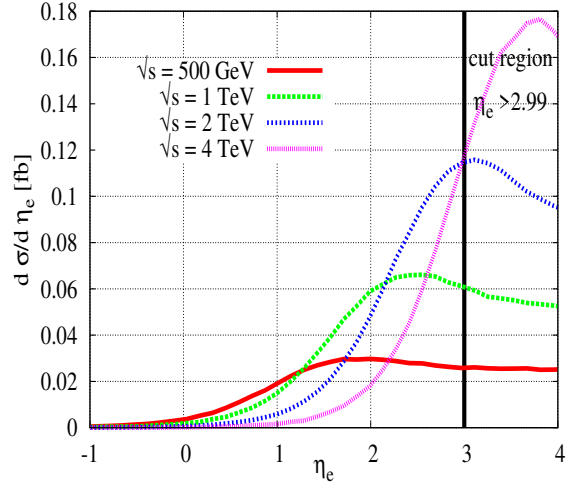} 
     \par {(b) Rapidity distributions for $\sigma_L$}
    \end{minipage}
    \caption{The rapidity distributions of tagged $e^-$ in the
   lab. frame for each initial helicity state and at each collision
   energy. The thick solid line shows the detection limit, above which
   $e^-$ escape detection.}
    \label{ST_RapDist}
  \end{center}
\end{figure}
\noindent 
In fig.\ref{ST_RapDist}, we show the rapidity distribution of tagged
$e^-$ in the laboratory frame. Although the total cross section increases
monotonically with $\sqrt{s}$, the distribution shifts  to large $\eta$ 
as $\sqrt{s}$ grows. It implies that more and more fraction of events
escape detection ($\cos\theta_e>0.995$ or $\eta_e>2.99$) at high
energies. The suppression of the cross section $\sigma_R$ for
($\eta_e<2.99$) as shown in fig.\ref{ST_RapDist}, can be attributed to
the cancellation between the $\gamma$- and $Z$- amplitudes for the
tagged $e_R$, with large scattering angle in the $e^- \gamma$ rest frame.

In terms of the above parameterizations eqs.(\ref{STeeH_ParaXsec500})
and (\ref{STeeH_ParaXsec1000}), we can determine the total cross section
for polarized $e^-$ beam and unpolarized $e^+$ beam as follows :
\bsub
\begin{eqnarray}
 \sigma^{e^-H}(P,\,\bar{P}=0) &=& \left(\frac{1+P}{2}\right)\,\,\,\sigma_R +
  \left(\frac{1-P}{2}\right)\,\,\,\sigma_L ,\\
 \sigma^{e^+H}(P,\,\bar{P}=0) &=&
 \frac{1}{2}\,\,\bigg(\sigma_R+\sigma_L\bigg). 
\end{eqnarray}
\label{STeeH_Xsec4}
\esub
\hskip -1.4mm 
At $\sqrt{s}= 500 $ GeV with $L=500\,\,\ifb$, we find
\bsub
\begin{eqnarray}
 \sigma^{e^-H}(P=+\,0.8,\,\bar{P}=0) &=& 0.030 \,\, 
  (\pm \,\,0.011) \,\,\, {\rm fb}, \\
 \sigma^{e^-H}(P=-\,0.8,\,\bar{P}=0) &=& 0.062 \,\, 
  (\pm \,\,0.016) \,\,\,{\rm fb}, \\
 \sigma^{e^+H}(P=\pm\,0.8,\,\bar{P}=0) &=& 0.046 \,\, 
  (\pm \,\, 0.0096) \,\,\, {\rm fb},
\end{eqnarray}
\label{STeeH_err500}
\esub
\hskip -1.4mm 
and at $\sqrt{s}=1$ TeV with $L=500\,\,\ifb$,  we find
\bsub
\begin{eqnarray}
 \sigma^{e^-H}(P=+\,0.8,\,\bar{P}=0) &=& 0.030 \,\,\, 
  (\pm 0.011) \,\,\, {\rm fb}, \\
 \sigma^{e^-H}(P=-\,0.8,\,\bar{P}=0) &=& 0.098 \,\,\, 
  (\pm 0.020) \,\,\, {\rm fb}, \\
 \sigma^{e^+H}(P=\pm\, 0.8,\,\bar{P}=0) &=& 0.064 \,\,\, 
  (\pm 0.011) \,\,\, {\rm fb}.
\end{eqnarray}
\label{STeeH_err1000}
\esub
\noindent 
\hskip -1.4mm 
The errors in the parentheses are for $L=250\ifb$ each for $P=0.8$ and
$P=-\,0.8$. 

From eqs.(\ref{STeeH_ParaXsec500}) and (\ref{STeeH_err500}), we find
\begin{eqnarray}
 \begin{array}{cl}
  \bg-\cg & = -\,0.00201 \pm\, 0.0012 \\
  \bgg    & = -\,0.00166 \pm \, 0.00016 \\
 \end{array}
 & &
 \left(
  \begin{array}{cc}
   1. & \\
   -\,.71 & 1.\\
  \end{array}
 \right)
 \label{STeeH_Res500} 
\end{eqnarray}
at $\sqrt{s}=500\GEV$. The mean values may be interpreted as the
average of the $t$-dependent SM couplings in the tagged events, which
turn out to be approximately $\bgSM(-\mzsq)$ and $\bggSM(-\mzsq)$, but
with slightly smaller magnitudes. The smallness of the magnitude of
$\langle\bgSM(t)\rangle$ reflects the larger mean value of the momentum
transfer $\langle \vert t\vert \rangle$ in the $\gamma$-$Z$ interference
than in the purely $\gamma$-exchange contribution. When we compare the
result eq.(\ref{STeeH_Res500}) with that of the no-tag $(ee)H$ events in
eq.(\ref{NTeeH_Res500}), we find that the no-tag events are about a
factor 2 more sensitive to the $H\gamma\gamma$ coupling.  On the other
hand, the sensitivity to the $HZ\gamma$ coupling, $\pm0.0012$, is
one order of magnitude worse than that of the $ZH$ process in
eq.(\ref{ZH_ResultPol500}) rescaled for $L=500\,\,\ifb$. Hence the
measurements of the single-tag $(e)eH$ events do not improve the
constraint on the $HZ\gamma$ coupling significantly.

For $\sqrt{s}=1\TEV$, we find
\begin{eqnarray}
 \begin{array}{cl}
  \bg-\cg & = -\,0.00180 \pm \,0.00063 \\
  \bgg    & = -\,0.00145 \pm \,0.00014 \\
 \end{array}
 & &
 \left(
  \begin{array}{cc}
   1. & \\
   -\,.77 & 1.\\
  \end{array}
 \right)
 \label{STeeH_Res1000} 
\end{eqnarray}
\noindent 
from eqs.(\ref{STeeH_ParaXsec1000}) and (\ref{STeeH_err1000}). The
magnitudes of the mean values decrease slightly from those of
eq.(\ref{STeeH_Res500}) at $\sqrt{s}=$ 500 GeV,  because of the higher
value of the typical momentum transfer, $\vert t\vert$,
at $\sqrt{s}$ = 1 TeV. The error of $\bgg$ is almost a factor of 3
larger than that of the no-tag $(ee)H$ events, eq.(\ref{NTeeH_Res1000}),
while that of $\cg$ is more than one order of magnitude larger than that
from $ZH$ events in  eq.(\ref{ZH_ResultPol1000}), rescaled for
$L=500\,\ifb$.

Although the process $e\gamma \to eH$ gives us clean measurements of the
$HZ\gamma$ couplings with $e^-$ beam polarization, the sensitivity is
rather low in $e^+e^-$ collisions because of the small and soft photon
flux from the bremsstrahlung. This process will become more important
once the photon linear collider option is realized \cite{Ginzburg}.

\section{Luminosity Uncertainty}
\label{luminos}
So far, the errors and their correlations based on the optimal
observables method for Higgs-gauge boson effective couplings are computed
by assuming the true luminosity $L$. However, the error in the
measurement of the $L$ can affect the measurements of some effective
couplings. We attempt here to study the impact of the luminosity
uncertainty on the precision measurements of the $HVV$ couplings.

In the presence of the luminosity uncertainty, the true luminosity $L$
can be estimated as
\begin{eqnarray}
 L = f \, \bar{L}, \hskip 1cm f = 1 \pm \Delta f,
\end{eqnarray}
where $\bar{L}$ is the measured mean value, and $\Delta f$ is its
1-$\sigma$ uncertainty. The $\chi^2$ function given in eq.(\ref{chidef})
of section \ref{OptPhaseSpace} should then be redefined as follows : 
\bsub
\begin{eqnarray}
 \chi^2 (c_i) &\rightarrow& \chi^2 (c_i \, ; \,f) \nonumber \\
 &=& \sum_{k=1}^{N} \biggl( \frac{N_{\rm exp}^k -N_{\rm th}^k(c_i)}
		  {\sqrt{N_{\rm exp}^k}} \biggr)^2 
  + \left( \frac{f-1}{\Delta f} \right)^2 , \label{chi_redef1} \\
 &=& \sum_{k=1}^{N} 
  \biggl( \frac{ L\,\,\Sigma_{\rm SM}(\Phi_k) \, \Delta 
   -\bar{L} \,\, \Bigl[ \Sigma_{\rm SM}(\Phi_k) 
		  +\displaystyle\sum_{i=1}^n \,\, 
		  c_i\,\,\Sigma_i(\Phi_k) \Bigr] \, \Delta} 
   {\sqrt{ L \,\,\Sigma_{\rm SM}(\Phi_k)\,\,\Delta}} \biggr)^2
  + \left( \frac{f-1}{\Delta f} \right)^2 , \hskip 0.5cm
  \label{chi_redef2} \\
 &=& \bar{L} \,\,\sum_{k=1}^{N} \,\, 
  \frac{ \Bigl[ (f-1) \,\, \Sigma_{\rm SM}(\Phi_k)
  +\sum_{i=1}^n\,\, c_i \,\,\Sigma_i(\Phi_k) \Bigr]^2}
  {\Sigma_{\rm SM}(\Phi_k)} \, \Delta
  + \left( \frac{f-1}{\Delta f} \right)^2 , \label{chi_redef3} \\
 &\stackrel{N\to\infty}{\to}& \sum_{i=0}^n \,\,\sum_{j=0}^n \,\,
  c_i\,\, \bigg(V^{-1}_f\bigg)_{ij} \,\,c_j 
  + \left( \frac{f-1}{\Delta f} \right)^2,
  \label{chi_redef4} 
\end{eqnarray}
\label{chi_redef}
\esub
\noindent
\hskip -1.4mm
where 
\begin{eqnarray}
\left( V^{-1}_f \right)_{ij} = \bar{L} \,\, \int 
 \frac{\Sigma_i(\Phi) \,\, \Sigma_j(\Phi)}{\Sigma_{\rm SM}(\Phi)} \,\,
 d\Phi,
\label{Vinv_LumiUn}
\end{eqnarray} 
is now $(n+1) \times (n+1)$ matrix with
\begin{eqnarray}
 c_0 = f-1, \hskip 1 cm  
  \Sigma_0(\Phi) = \Sigma_{\rm SM}(\Phi).
\end{eqnarray}

It is straightforward to integrate out the $c_0=f-1$ dependence and obtain
the probability distribution of the parameters $c_1$ to $c_n$ in the
presence of the luminosity uncertainty. In our study, we note that the
two couplings, $\aw$ and $\az$ have the weight functions that are
identical to the SM distribution.$^{\footnotemark[3]}$
\footnotetext[3]{For no-tag $(ee)H$ and single-tag $(e)eH$ processes,
the $H\gamma\gamma$ and $HZ\gamma$ couplings of the SM at one loop
order are used to calculate the SM cross-sections. There, however, the
statistical error is dominated, and the errors due to the luminosity
uncertainty can be safely neglected.}
Because of this, we can study the impacts of the luminosity uncertainty
algebraically by using the $\chi^2$ functions written in terms of $\aw$
and $\az$. We use
\begin{eqnarray}
 \Sigma_{\aw}(\Phi) = 2 \; \Sigma_{\rm SM}(\Phi),
\end{eqnarray}
for the $WW$-fusion process given in eq.(\ref{nunuh}), and 
\begin{eqnarray}
 \Sigma_{\az}(\Phi) = 2 \; \Sigma_{\rm SM}(\Phi),
\end{eqnarray}
for the $ZH$ production and double-tag $eeH$ processes given in
eq.(\ref{ZH_process}) and eq.(\ref{eeH_process}), respectively. We can
express the total $\chi^2$ function with the luminosity uncertainty
given in eq.(\ref{Vinv_LumiUn}) for a particular c.m. energy,
$\sqrt{s}$ as
\begin{eqnarray}
\chi^2(c_i;\,\, f)_{\sqrt{s}} &=&
 \chi^2\left(\aw\to\awp=\aw+\frac{f-1}{2}\right)^{HWW}_{\sqrt{s}} \nonumber \\
 &+& \chi^2\left(\az\to\azp=\az+\frac{f-1}{2}\right)^{HZZ}_{\sqrt{s}}
 + \left[ \frac{f-1}{(\Delta f)_{\sqrt{s}}} \right]^2.
\label{chisq_ReDef}
\end{eqnarray}
In the following analysis, we assume that the luminosity uncertainty is
common to all the processes at each collision energy $\sqrt{s}$.

Let us examine the effects at $\sqrt{s}=500\GEV$ in some detail. The
$\chi^2$ function for the $HWW$ process in eq.(\ref{w500}) can be
expressed in the form
\begin{eqnarray}
 \chi^2\bigg(\aw\to\awp\bigg)^{HWW}_{500} = \sum_{i,j}\,\,\, 
  c^\prime_{iWW} \,\,\, 
  \bigg[ (V^{HWW}_{500})^{-1} \bigg]_{ij} \,\,\, c'_{jWW},
\end{eqnarray}
where $\awp=\aw+(f-1)/\,\,2$, $c^\prime_{iWW}\equiv c_{iWW}$
$(i\ne1)$. Likewise, the $\chi^2$ function from the $ZH$ and the
double-tag $eeH$ processes in sections \ref{ZHProd} and \ref{ZZ} can be
expressed as
\begin{eqnarray}
 \chi^2\bigg(\az\to\azp\bigg)^{HZZ}_{500} = \sum_{k,l}\,\,\,  c'_k\,\,\, 
  \bigg[ (V^{HZZ}_{500})^{-1} \bigg]_{kl} \,\,\,c'_l,
\end{eqnarray}
where $\azp=\az+ (f-1)/\,\, 2$ and $c'_2,\,\,c'_3,\,\,c'_4,\,\,c'_5,\,\,c'_6$  
$\equiv \,\,\bz,\,\,\cz,\,\,\bg,\,\,\cg,\,\,\bgg $. Now, the luminosity
uncertainty in the $\chi^2$ function of eq.(\ref{chisq_ReDef}) at
$\sqrt{s}=500\GEV$ can easily be factored out as
\begin{eqnarray}
 \chi^2\big(c_i;\,\,f\big)_{500} &=& 
  \left[ \frac{f-1}{(\Delta f)_{500}^{\rm eff}} 
   + (\Delta f)_{500}^{\rm eff}\,\, R \right]^2 
  + \widetilde{\chi}^2(c_i)_{500},
  \label{chisq_500-2}
\end{eqnarray}
where
\bsub
\begin{eqnarray}
 \frac{1}{[(\Delta f)_{500}^{\rm eff}]^2} &=& 
  \frac{1}{[(\Delta f)_{500}]^2} 
  + \frac{1}{4}[(V^{HWW}_{500})^{-1}]_{11}
  + \frac{1}{4}[(V^{HZZ}_{500})^{-1}]_{11}, \\
 R &=& \frac{1}{2} 
  \left( \sum_{k=1}^3 c_{kWW} [(V^{HWW}_{500})^{-1}]_{1k}
   + \sum_{k=1}^6 c_k [(V^{HZZ}_{500})^{-1}]_{1k} \right),
\end{eqnarray}
\esub
and the reduced $\chi^2$ function is
\begin{eqnarray}
 \widetilde{\chi}^2(c_i)_{500} = \chi^2(c_i)_{500} 
  - [(\Delta f)_{500}^{\rm eff}]^2 R^2.
  \label{Reduced_chi}
\end{eqnarray}
We can use the  reduced $\chi^2$ function to study the constraints on
the non-standard couplings in the presence of the luminosity
uncertainty. It should be noted that because of the last term of
eq.(\ref{Reduced_chi}) there appear correlations between the $HWW$
couplings and the $HZZ/HZ\gamma/H\gamma\gamma$ coupling
measurements. Therefore, the effects of the luminosity uncertainty is
important when we study the constraints on the coefficients of the
dimension-6 operators, eq.(\ref{effcoup_Operator}), since some of them
contribute to both couplings.

The reduced $\chi^2$ function of eq.(\ref{Reduced_chi}) gives a
$9\times9$ covariance matrix. The results for $\sqrt{s}=500\GEV$,
$L=500\ifb$ and $\Delta f=0.01$ can be expressed as,
\begin{eqnarray}
\begin{array}{ll}
 \aw  & = \pm .045 \\
 \bw  & = \pm .046 \\
 \dw  & = \pm .013 \\
 \az  & = \pm .013 \\
 \bz  & = \pm .00059 \\
 \cz  & = \pm .00044 \\
 \bg  & = \pm .00037 \\
 \cg  & = \pm .000080 \\
 \bgg & = \pm .0022 \\
\end{array}
& &
\left(
\begin{array}{cccccccccc}
  1 \\
  .990 &  1 \\
  .991 & .995 & 1 \\
  .043 &   0.  & 0. &  1 \\
  .0   &   0.  & 0. & -.63  & 1 \\
  .0   &   0.  & 0. & -.81  & .033  &  1 \\
  .0   &   0.  & 0. & -.006 & -.063 & .043  &  1 \\
  .0   &   0.  & 0. & .0    &  .050 & -.046 & -.83  & 1 \\
  .0   &   0.  & 0. & .22   & -.27  & -.28 & -.040 & .034 & 1 \\
\end{array}
\right).
\label{CombResCoup_500_df}
\end{eqnarray}
When the above results are compared with those without the luminosity
uncertainty in eq.(\ref{WWfusion_500-2}) for the $HWW$ couplings and the
corresponding expression for the $HZZ/HZ\gamma/H\gamma\gamma$ couplings,
after rescaling of the errors by $1/\sqrt{5}\simeq 0.447$, we find the
followings. The errors of $\aw$ and $\az$ are slightly larger than
$1/\sqrt{5}$ of the corresponding statistical errors, while all the
other errors are not affected much. The correlations between $\aw$ and
$\bw$ or $\dw$ are reduced slightly in magnitude from
eq.(\ref{WWfusion_500-2}), and those correlations between $\az$ and the
other $HZZ$, $HZ\gamma$, $H\gamma\gamma$ couplings are also reduced
slightly in magnitude. Finally, the off-diagonal sub-correlation matrix in
eq.(\ref{CombResCoup_500_df}) is almost vacant except for the
$(\aw,\az)$ component, which shows positive correlation because the
luminosity uncertainty affects both $\aw$ and $\az$ couplings in the
same way. All the effects of the luminosity uncertainty
are rather small because the statistical errors of $\aw$ and $\az$ are
$\pm 0.045$ and $\pm 0.013$, respectively, which are significantly
larger than the error from the postulated luminosity uncertainty, 
$\Delta f/2=0.005$. When the errors due to the luminosity uncertainty
become dominant, the errors of $\aw$ and $\az$ will stop decreasing with
the luminosity, and the correlation between $\aw$ and $\az$ will grow.

It is instructive to study the impact of the luminosity uncertainty
analytically in a very simplified example where only the two couplings,
$\aw$ and $\az$, are retained in the amplitudes at
$\sqrt{s}=500\GEV$. In this limit, the reduced $\chi^2$ function of
eq.(\ref{Reduced_chi}) is simply
\begin{eqnarray}
 \widetilde{\chi}^2(c_i)_{500} &=&
  \awsq \,\, [ \big(V^{HWW}_{500}\big)^{-1} ]_{11} 
  +\azsq \,\, [ \big(V^{HZZ}_{500}\big)^{-1} ]_{11} \nonumber \\
 & & -\frac{1}{4}\,\, \left[\big(\Delta f\big)_{500}^{\rm eff}\right]^2\,\,
  \left( \aw\,\, [ \big(V^{HWW}_{500}\big)^{-1} ]_{11} 
   +\az\,\, [ \big(V^{HZZ}_{500}\big)^{-1} ]_{11} \right)^2, \nonumber\\ 
 &=& \left( \frac{\aw}{\Delta \aw} \right)^2 
  + \left( \frac{\az}{\Delta \az} \right)^2 
  - \frac{1}{4}\,\,\left[\big(\Delta f\big)_{500}^{\rm eff}\right]^2\,\,
  \left[ \frac{\aw}{(\Delta \aw)^2} + \frac{\az}{(\Delta \az)^2}
  \right]^2, \nonumber\\
\label{Reduced_chi_aWaZ}
\end{eqnarray}
\noindent 
where the combined error of $\big(\Delta f\big)_{500}^{\rm eff}$ is
\begin{eqnarray}
 [ \big(\Delta f\big)_{500}^{\rm eff} ]^{-2} = 
  [ \big(\Delta f\big)_{500} ]^{-2} 
  + \frac{1}{4}\,\,\big(\Delta \aw\big)^{-2} 
  + \frac{1}{4}\,\,\big(\Delta \az\big)^{-2}. 
\end{eqnarray}
As noted above, the second term in eq.(\ref{Reduced_chi_aWaZ})
generates the correlation between the errors of $\aw$ and $\az$. For
example, if we take the limit where the statistical errors are much
smaller than the luminosity uncertainty, 
$(\Delta \aw)^2, (\Delta \az)^2 \ll (\Delta f)^2$, the above
reduced $\chi^2$ function can be expressed as
\begin{eqnarray}
 \widetilde{\chi}^2(c_i)_{500} 
  &\stackrel{(\Delta c_{1VV})^2 \ll \Delta f^2}{\longrightarrow}&
  \nonumber \\
 & & \hskip -4.5cm
  \frac{(\aw-\az)^2}{\big(\Delta \aw\big)^2+\big(\Delta \az\big)^2}
  + \frac{4}{({\Delta f})^2} \,
   \frac{ [ \big(\Delta \az\big)^2 \,\aw +\big(\Delta \aw\big)^2 \,\az ]^2 }
   { [ \big(\Delta \aw\big)^2 + \big(\Delta \az\big)^2 ]^2 } 
   + {\cal O}\left( \frac{1}{(\Delta f)^4} \right).
 \label{Reduced_chi_aWaZ_limit}
\end{eqnarray}
In the leading term, only the combination $(\aw-\az)$ is constrained, and the
next-to-leading term proportional to $1/\,\,\big(\Delta f\big)^2$
constrain the combination 
$\big(\Delta \az\big)^2\,\,\aw +\big(\Delta \aw\big)^2\,\, \az$. In the
limit of large statistics, the first term dominates the $\chi^2$
function and the correlation approaches the unity.

In fig.\ref{Fig:aWaZError}, we show the errors of $\aw$ and $\az$
couplings as a function of the integrated luminosity $L$ for the
luminosity uncertainty $\Delta f=0.01$ at $\sqrt{s}=500\GEV$.
\begin{figure}[htb!]
 \begin{center}
  \includegraphics[width=10cm, height=14cm, angle=-90]{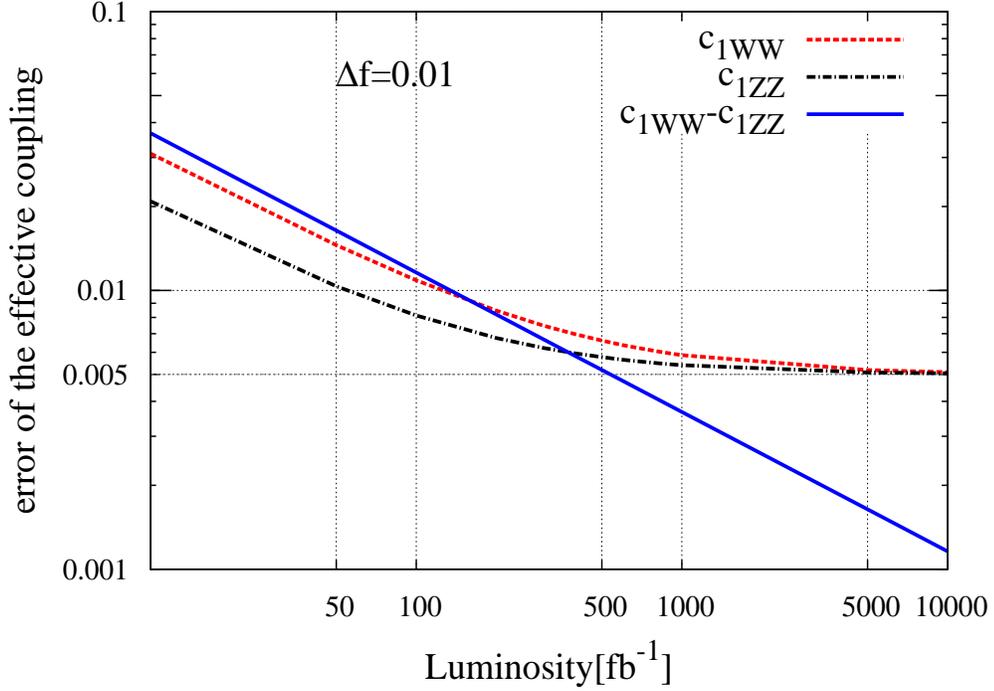}
  \caption{Error of $\aw$ and $\az$ vs integrated Luminosity at
  $\sqrt{s}=500\GEV$ when the luminosity uncertainty is $\Delta f=0.01$.}
  \label{Fig:aWaZError}
 \end{center}
\end{figure}
The solid line shows the error of the combination $\aw-\az$ which
decreases as $1/\,\,\sqrt{L}$ asymptotically, since the luminosity
uncertainty is canceled out by taking the difference. On the other
hand, the dashed line for $\aw$, and the dash-dot line for $\az$,
approaches the same value $\big(\Delta f\big)/\,\, 2$, showing that the
individual errors will be dominated by the luminosity uncertainty. More
explicitly, the errors of the couplings, $\aw$ and $\az$, can be
expressed as
\bsub
\begin{eqnarray}
 \aw = \pm \,\,\sqrt{ \left( \frac{\Delta f}{2} \right)^2 
  +\big(\Delta \aw\big)^2 },
  \hskip 0.5cm
 \az = \pm\,\, \sqrt{ \left( \frac{\Delta f}{2} \right)^2 
  +\big(\Delta \az\big)^2 },
\end{eqnarray}
with the correlation
\begin{eqnarray}
 \rho = 
  \left[ 
   \left( 1 +4\,\,
    \frac{\big(\Delta \aw\big)^2}{\big(\Delta f\big)^2} \right) \,\, 
   \left( 1 +4\,\,\frac{\big(\Delta \az\big)^2}{\big(\Delta f\big)^2} \right)
  \right]^{-\,\frac{1}{2}}.
\end{eqnarray}
\esub

\section{Conclusion and Discussion}
\label{conclu}
The imprint of the dynamics of the symmetry breaking physics is
inherent in the interactions of the Higgs boson and the gauge bosons. 
In this article, we attempt to evaluate the potential of the
future linear $e^+e^-$ colliders, such as the ILC, in probing the
dynamics of all the CP-even and gauge-invariant
dimension six operators of the Standard Model (SM) fields that affect the
Higgs-gauge-boson couplings when there is one SM-like light
Higgs-boson. For this purpose, we study all the processes that are
sensitive to the $HVV$ couplings: the $\nu_e \bar{\nu}_e H$ production
via $WW$-fusion is sensitive to $HWW$ couplings, the $ZH$ production
process is sensitive to $HZZ$ and $HZ\gamma$ couplings, the double-tag
$eeH$ production process via $t$-channel $Z$ and $\gamma$ exchange is
also sensitive to $HZZ$, $HZ\gamma$  and $H\gamma\gamma$ coupling, 
the no-tag $(ee)H$ process measures the $H\gamma\gamma$ coupling, and
the single-tag $(e)eH$ process measures  $HZ\gamma$ and $H\gamma\gamma$
couplings. In order to quantify the resolving power of each process,  we
allow all the effective $HVV$ couplings to vary freely in the fit and
adopt the optimal observables method to constrain them in each process,
at a few selected collision energies ($\sqrt{s}=$250, 350, 500,
1000$\GEV$), and with or without $e^-$ beam polarization.

\subsection{Summary of the constraints on the $HVV$ couplings}
Here we summarize our results for the effective $HVV$ couplings at the ILC. 
\begin{itemize}
 \item All our results have been presented for nominal integrated
       luminosity of $L_0=100\ifb$, except for the no-tag $(ee)H$ and
       single-tag $(e)eH$ events, where we give our error estimates for
       $L=500\ifb$ at $\sqrt{s}=500\GEV$ and $1\TEV$ only.
 \item Most of our results have been given for $e^-$ beam polarization
       of $|P|=80\%$ with no $e^+$ beam polarization $|\bar{P}|$=0,
       where exactly half of the total integrated luminosity $L$ is
       delivered with $P=|P|$ and $P=-|P|$.
 \item Results for $|P|=|\bar{P}|=0$ have also been calculated for all
       the cases~\cite{YMthesis}, but presented only for the $ZH$
       production process  at $\sqrt{s}=250\GEV$; see
       eq.(\ref{ZH_ResultUnPol250}).
 \item Our results for the $HWW$ effective couplings are shown in
       eq.(\ref{w250stcomb}) for $\sqrt{s}=250\GEV$,
       eq.(\ref{w350stcomb}) for $\sqrt{s}=350\GEV$, eq.(\ref{w500}) for
       $\sqrt{s}=500\GEV$ and eq.(\ref{w1tev}) for $\sqrt{s}=1\TEV$. 
 \item The results for the $HZZ$ and $HZ\gamma$ couplings in $ZH$
       production process with $\vert P\vert$ = 80\%  are shown in
       eq.(\ref{ZH_ResultPol250}) for 
       $\sqrt{s}=250\GEV$, eq.(\ref{ZH_ResultPol350}) for
       $\sqrt{s}=350\GEV$, eq.(\ref{ZH_ResultPol500}) for
       $\sqrt{s}=500\GEV$, and eq.(\ref{ZH_ResultPol1000}) for
       $\sqrt{s}=1\TEV$.
 \item The results for the $HZZ$, $HZ\gamma$ and $H\gamma\gamma$
       couplings in double-tag $eeH$ process from $t$-channel vector
       boson fusion with $\vert P\vert$ = 80\% are shown in
       eq.(\ref{ZZeigencomb250}) for $\sqrt{s}=250\GEV$,
       eq.(\ref{ZZeigencomb350}) for $\sqrt{s}=350\GEV$,  
       eq.(\ref{ZZeigen500}) for $\sqrt{s}=500\GEV$, and 
       eq.(\ref{ZZeigen1000}) for $\sqrt{s}=1\TEV$.
 \item The $H\gamma\gamma$ coupling results from the no-tag $(ee)H$
       process are given for $L=500\ifb$ in eq.(\ref{NTeeH_Res500}) at
       $\sqrt{s}=500\GEV$, and in eq.(\ref{NTeeH_Res1000}) at
       $\sqrt{s}=1\TEV$. 
 \item The $HZ\gamma$ and $H\gamma\gamma$ coupling results from the
       single-tag $(e)eH$ process for $L=500\ifb$ with $|P|$ = 80\%
       are given in eq.(\ref{STeeH_Res500}) at $\sqrt{s}=500\GEV$, and
       in eq.(\ref{STeeH_Res1000}) at $\sqrt{s}=1\TEV$.
 \item The impact of the luminosity uncertainty, $L=f\bar{L}$ with
       $f=1\pm\Delta f$, should be taken into account according to
       eq.(\ref{chisq_ReDef}), where the coefficients $\aw$ and $\az$ in
       the $\chi^2$ function at each energy is replaced by
       $c'_{1VV}=c_{1VV}^{}+(f-1)/\Delta f$ and by adding the term
       $(f-1)^2/(\Delta f)^2$ at each energy. After squaring out the
       $(f-1)$ dependence, we obtain the reduced $\chi^2$ function
       $\widetilde{\chi}^2(c_i)$ at each energy.
\end{itemize}
The combined analysis in this section are performed by gathering all the
above results, by adding all the $\chi^2$ functions with appropriate
weights.

In particular, we report the following two cases;
\bsub
\begin{eqnarray}
 \chi^2(c_i)_{\rm I} &=& 
  \frac{L_{250}}{L_0}\,\,\,\, \widetilde{\chi}^2(c_i)_{250} 
  +\frac{L_{350}}{L_0} \,\,\,\,\widetilde{\chi}^2(c_i)_{350} 
  +\frac{L_{500}}{L_0} \,\,\,\,\widetilde{\chi}^2(c_i)_{500}
  \label{Comb_chisqI} \\
 \chi^2(c_i)_{\rm I+II} &=& \chi^2(c_i)_{\rm I} 
  +\frac{L_{1000}}{L_0} \,\,\,\,\widetilde{\chi}^2(c_i)_{1000},
  \label{Comb_chisqII}
\end{eqnarray}
\label{Comb_chisq}
\esub
\hskip -1.4 mm
where $\chi^2(c_i)_{\rm I}$ gives the combined results of the ILC phase
I with the maximum energy of $\sqrt{s}=500\GEV$ and 
$\chi^2(c_i)_{\rm I+II}$ gives that of combining the results from ILC-I
and ILC-II at $\sqrt{s}=1\TEV$. The reduced $\chi^2$ function at each
energy, $\widetilde{\chi}^2(c_i)_{\sqrt{s}}$, are obtained for the
luminosity uncertainty of $(\Delta f)_{\sqrt{s}}$ at each energy.

Since we present the individual covariance matrix separately in each
process at $\sqrt{s}=250\GEV$, $350\GEV$, $500\GEV$ and $1\TEV$, in
terms of its eigenvectors and square root of eigenvalues, we can
estimate the constraints for an arbitrary integrated luminosity with an
arbitrary luminosity uncertainty $\Delta f$ at each energy. As an
example, we show the results for
\begin{eqnarray}
 L_{250} = L_{350} = 100\ifb, \hskip 1cm
 L_{500} = L_{1000} = 500\ifb,
\label{LumAssign}
\end{eqnarray}
\noindent 
when the luminosity uncertainty is $\Delta f=1\%$ at all energies.

Using the $\chi^2$ function defined in eq.(\ref{Comb_chisqI}), we
evaluate the errors and correlations of the effective couplings for
ILC-I, with the integrated luminosities given in eq.(\ref{LumAssign})
and obtain
\begin{eqnarray}
\begin{array}{ll}
 \aw  & = \pm \,\, .022 \\
 \bw  & = \pm \,\, .023 \\
 \dw  & = \pm \,\, .0065 \\
 \az  & = \pm \,\, .0067 \\
 \bz  & = \pm \,\, .00048 \\
 \cz  & = \pm \,\, .00021 \\
 \Delta \bg  & = \pm \,\, .00030 \\
 \cg  & = \pm \,\, .000073 \\
 \Delta \bgg & = \pm \,\, .000075 \\
\end{array}
& &
\left(
\begin{array}{cccccccccc}
  1 \\
  .96 &  1 \\
  .96 & .98  & 1 \\
  .17 &  .0  & .0 &  1 \\
 -.0  &  .0  & .0 & -.49  &   1 \\
  .0  &  .0  & .0 & -.26  & -.16  & 1 \\
 -.0  &  .0  & .0 &  .0   & -.067 & .072   &  1 \\
  .0  &  .0  & .0 & -.001 &  .049 & -.090  & -.81   & 1 \\
  .0  &  .0  & .0 & .004  & -.006 & -.002 & -.001 & .001 & 1 \\
\end{array}
\right).
\label{CombResCoup_ILC-I}
\end{eqnarray}
When we compare the above combined results from the three ILC-I energies
with those of eq.(\ref{CombResCoup_500_df}) at $\sqrt{s}=500\GEV$, we
observe the followings. As for the $HWW$ couplings, we observe reduction
of the errors of all the 3 couplings and that of correlations. This is
essentially because contributions have different dependence on
$\sqrt{s}$. As for the $HZZ/HZ\gamma/H\gamma\gamma$ couplings, the
errors of $\bz$, $\cz$, $\bg$ and $\cg$ are essentially determined by
the $\sqrt{s}=500\GEV$ experiment, while that of $\az$ is reduced
significantly by including the lower energy data, because the $ZH$
production cross section is larger at lower energies, see
fig.\ref{rootsvariation}. The error of $\bgg$ is essentially determined
by the no-tag $(ee)H$ events.We find that the single-tag $(e)eH$ events,
eq.(\ref{STeeH_Res500}), do contribute significantly to improve the
$HZ\gamma$ couplings measurement.

Only the errors of $\aw$ and $\az$ and the corresponding rows of the
correlation matrix are affected by the luminosity uncertainty 
$\Delta f=0.01$ at each energies, assumed in the fit. By combining the
results of eqs.(\ref{w250}), (\ref{w350}) and (\ref{w500}) we obtain the
combined error of $\aw$ without the luminosity uncertainty to be 0.025,
which is still 5 times larger than the error due to the luminosity
uncertainty, $\Delta f/2=0.005$ at each energy. Therefore, the error of
$\aw$ does not increase significantly with the inclusion of the
luminosity error. On the other hand, by combining the results at
$\sqrt{s}=$ 250, 350 and 500 GeV, we find that the combined error of
$\az$ without the luminosity uncertainty is 0.0045, which is comparable
to the error due to the luminosity uncertainty. We therefore find almost
30\% larger error for $\az$ in eq.(\ref{CombResCoup_ILC-I}). The
correlation between $\aw$ and $\az$ is now 0.17, which is still small
because the statistical error of $\aw$ is much larger than the error due
to the luminosity uncertainty. 

After combining the ILC-I and ILC-II results with the integrated
luminosities of eq.(\ref{LumAssign}), we find
\begin{eqnarray}
\begin{array}{ll}
 \aw  & = \pm \,\, .0089 \\
 \bw  & = \pm \,\, .0077 \\
 \dw  & = \pm \,\, .0015 \\
 \az  & = \pm \,\, .0058 \\
 \bz  & = \pm \,\, .00032 \\
 \cz  & = \pm \,\, .000063 \\
 \Delta \bg  & = \pm \,\, .00018 \\
 \cg  & = \pm \,\, .000023 \\
 \Delta \bgg & = \pm \,\, .000039 \\
\end{array}
& &
\left(
\begin{array}{cccccccccc}
 1 \\
 .80  &  1 \\
 .80  & .96 &  1 \\
 .49  &  .0 & .0 &  1 \\
 -.0  &  .0 & .0 & -.36  & 1 \\
 -.0  &  .0 & .0 & -.10  & -.23  & 1 \\
 -.0  &  .0 & .0 & -.003 & -.061 & .055  & 1 \\
  .0  &  .0 & .0 &  .0   &  .034 & -.096 & -.60  & 1 \\
  .0  &  .0 & .0 & .003  & -.007 & -.0   & -.002 & .001 & 1 \\
\end{array}
\right).
\label{CombResCoup_ILC-II}
\end{eqnarray}
The $HWW$ couplings are measured much more accurately than the ILC-I
alone case, mainly because of the large $WW$-fusion cross section at
$\sqrt{s}=1\TEV$, see fig.\ref{rootsvariation}. The correlations between
the error of $\aw$ and those of $\bw$ and $\dw$ are reduced because of
the strong energy dependence of the contributions from the higher
dimensional operators, as discussed in section \ref{WWFusion}. In
contrast, the reduction of the error of $\az$ is marginal and the
correlation between $\aw$ and $\az$ grows to 0.49, reflecting the
dominance of the luminosity uncertainty $\Delta f/2=0.005$. The errors
of $\cz$ and $\cg$ are a factor of 3 smaller than the ILC-I 
results in eq.(\ref{CombResCoup_ILC-I}),
mainly because of the strong energy dependence of their contributions;
as discussed in section \ref{ZZ}. The error of $\bgg$ is roughly  half of the
ILC-I result in eq.(\ref{CombResCoup_ILC-I}), partly because of the 2.3
times larger cross section of the no-tag $(ee)H$ events, see
fig.\ref{rootsvariation} and eqs.(\ref{xsecsignal}) and
(\ref{NTeeH_Result}), and partly because of the contributions from the
double tag $ee\,H$ events.

\subsection{Comparison with other papers}
First, we would like to compare our results with those of the reference
 \cite{kniehl}, as the present work is envisaged as an extension of this
piece of work. The authors of ref.\cite{kniehl} analyzed the $ZH$
production process for $m_H$ = 120 GeV both with unpolarized and 90\%
polarized $e^-$ beam in $e^+\, e^-$ collisions, using all  the $Z$-boson
decay modes.  The study was based on the optimal observables  method by
allowing all the couplings to vary simultaneously.  Although the
formalism presented in section \ref{ZHProd}  of this report is
significantly more compact than that of   ref.\cite{kniehl}, we
reproduce all the  errors and their correlations for both  unpolarized
and 90\% polarized $e^-$ beams  at $\sqrt{s}=250\GEV$ as given in
eqs. (5.4) and (5.9), respectively, as well as eq.(5.13) of
ref.\cite{kniehl}, for $\sqrt{s}=500\GEV$ with 90\% polarized $e^-$
beam. Note that the authors in ref.\cite{kniehl} considered the
integrated luminosity $10\ifb$, and therefore  the errors in
eqs. (5.4), (5.9)  and (5.13) should  be multiplied by $1/\,\sqrt{10}$
for comparison with our results. It is worth noting here that the
$\tau$ lepton polarization and the $b$ jet charge identification with
the efficiency 40\% and 20\%, respectively, considered in
ref.\cite{kniehl} has little impacts on the final results once the
$e^-$  beam polarization is available. The reduction of the beam
polarization from 90\% considered in  ref.\cite{kniehl} to 80\% in this
report does lead to a slight increase in the error of the $HZ\gamma$
couplings by about 13\%.
 
Next we compare our  results with  tables II  and III of
 ref.\cite{debchou},  where the authors  estimated the $3\, \sigma$
bounds on $\az$, $\bz $, $\aw$ and $\bw $  ( $\Delta a_Z$, $\Delta b_Z
/2$, $\Delta a_W$ and $\Delta b_W/2$, respectively, in their notation).
This analysis was performed  for unpolarized $e^+\, e^-$ collisions at
$\sqrt{s}=500\GEV$, with  $L=500\ifb$ and $\mh=120\GEV$.
It is worth   noting  that to derive these bounds  they vary one
coupling at a time and hence the inclusion of the  1\% systematic error
(accruing from luminosity uncertainty etc.) dominates their  fluctuation
estimation for all the observed cross sections.   Their results can be
easily reproduced from ours simply  by setting all  the couplings to
zero except for the one whose error is  estimated, in the   respective
$\chi^2$ function for the specified process with an appropriate
integrated luminosity, which should be corrected for the Higgs boson
decay branching fraction.  The systematic error of 1\% is then added to
the statistical error  in quadrature. For instance, the $3\,\sigma$
limit $\big\vert \az\big\vert \le 0.034$ \cite{debchou} is found from
the double-tag $eeH$ process excluding the $Z\rightarrow e^+\,e^-$
events. To compare, we set all the other couplings to zero  in the
$\chi^2$ function and find $\az =\pm$ 0.0092 for $L=500\ifb$ with 
$B\big(H\rightarrow b\,\bar b\big)$ = 0.9. Now, adding the systematic
error of 1\% in quadrature we find $ \az=\pm\,\, 0.0105$ and the 3
$\sigma $ limit is $\big\vert\az\big\vert \le$ 0.032. The limit
$\big\vert \bz  \big\vert \le \, 0.0022$ \cite{debchou} is obtained
from the observed $ZH$ production cross section, where $Z$ decays to
muon pair and light quarks (excluding $b$ quark) pair have been
considered. Under similar conditions we find the statistical error
of $\bz=\pm\,\, .00059$. Since the luminosity error does not affect
the measurement of $\bz$, this gives the $3\,\sigma$ limit of
$\big\vert \bz \big\vert <0.0018$, which shows an improvement by a
factor of 1.3. We find that this difference is due to the  use of
the optimal observable, {\it i.e.}, the differential distribution which
is linear in $\bz$; the quoted limit in \cite{debchou} is reproduced
if we use only the $\bz$ effects on the total cross section. Likewise,
we reproduce the bound $\big\vert \aw\big\vert $, while we find a
factor of two better bound on $\big\vert \bw\big\vert$, which can be
attributed to our use of optimal weight function.  

We also compare our results on $\az$ and $\aw$ couplings with those of
the ref.\cite{TaoHan}. Since they studied possible constraint on the
operator ${\cal O}_{\phi2}$ (${\cal O}_1$ in their notation), we present
our comparison result in the next subsection; see the second footnote.

\subsection{Constraints on the dimension six operators}
The constraints on the effective $HVV$ couplings given in
eqs.(\ref{CombResCoup_ILC-I}) and (\ref{CombResCoup_ILC-II}) should be
expressed as those of  8 dimension-six operators of
eq.({\ref{Operators}) in order to compare the power of ILC precision
measurements with that of the other experiments. All the effective $HVV$
couplings are linear combinations of the coefficients $f_i/\Lambda^2$'s
of these operators, as given in eq.(\ref{effcoup_Operator}). It is clear
from eq.(\ref{effcoup_Operator}) that the coefficients $\fphione,
\fphitwo, \fphifour$ cannot be determined uniquely from the $HVV$
coupling measurements alone$^{\footnotemark[4]}$. 
\footnotetext[4]{For instance, the measurement of the triple
Higgs-boson coupling is necessary to constrain all the three operators.}
We therefore present constraints on two combinations of the three
coefficients, $\fphione$ and $3\fphifour-2\fphitwo$. Our results are
hence for the 7 coefficients; $\fphione$, $\fBW$, $\fW$, $\fB$, $\fWW$,
$\fBB$ and $3\fphifour-2\fphitwo$. 

The combined ILC-I results of eq.(\ref{CombResCoup_ILC-I}) lead to the
following constraints on the dimension-6 operators,
\begin{eqnarray}
\begin{array}{ll}
 \left( \frac{1\TEV}{\Lambda} \right)^2  \,\, \fphione               & = \pm \, .091\\
 \left( \frac{1\TEV}{\Lambda} \right)^2  \,\, \fBW                   & = \pm \, .35 \\
 \left( \frac{1\TEV}{\Lambda} \right)^2  \,\, \fW                    & = \pm \, .051 \\
 \left( \frac{1\TEV}{\Lambda} \right)^2  \,\, \fB                    & = \pm \, .084 \\
 \left( \frac{1\TEV}{\Lambda} \right)^2  \,\, \fWW                   & = \pm \, .22 \\
 \left( \frac{1\TEV}{\Lambda} \right)^2  \,\, \fBB                   & = \pm \, .56 \\
 \left( \frac{1\TEV}{\Lambda} \right)^2  \,\, (3\fphifour-2\fphitwo) & = \pm \, .36 \\
\end{array}
& &
\left(
\begin{array}{ccccccc}
  1 \\
 -.40  &  1 \\
  .37  &  .23  &  1 \\
  .23  & -.49  &  .35 &  1 \\
  .59  & -.91  & -.20 &  .26 & 1 \\
 -.49  &  .98  &  .22 & -.41 & -.96 & 1 \\
 -.052 & -.017 & .060 & .010 & .053 & -.032 & 1 \\
\end{array}
\right) 
\label{ResOpe_500_df}
\end{eqnarray}
We find that the three coefficients, $\fphione$, $\fW$ and $\fB$ can be
constrained rather accurately, with around 5\% accuracy for
$\Lambda=1\TEV$, and rather independently of the other operators. The
coefficient  $\fphione$ is measured accurately, because it contributes
with the opposite sign to $\aw$ and $\az$, see eq.(\ref{ninecoupstart})
and eq.(\ref{aWeq}). In other words, $\fphione$ is a measure of the
difference between the $HZZ$ and the $HWW$ coupling strengths.
On the other hand, the error of the $3\fphitwo-2\fphifour$ is
rather large, since it measures the overall strengths of the $HVV$
couplings. In fact, we find that it is only the error of
$3\fphitwo-2\fphifour$ which is affected significantly by the luminosity
uncertainty of $\Delta f=0.01$.  When we set $\Delta f=0$, its error is
reduced to 0.14$^{\footnotemark[5]}$,
\footnotetext[5]{ The authors of ref.\cite{TaoHan} studied possible
constraints on the dimension six operator ${\cal O}_{\phi2}$ 
(${\cal O}_1$ in their notation) which affects the $HZZ$ and $HWW$
couplings by considering $ZH$ production, $WW$-fusion and double-tag
$eeH$ processes. They found the uncertainty of .005 for the coulpling
$2\az=2\aw$ ( $a_1$ in their notation) at $\sqrt{s}=500\GEV$ for
$L=1~{\rm ab}^{-1}$ with $B(H\to b\,\bar b)=0.9$ and 80\% $b$-tagging
efficiency. Our result of $3\fphifour-2\fphitwo =\pm .14$ corresponds
to $2\,\az=2\aw \approx \pm .0042$ even though the total integrated
luminosity of our analysis is 700$\ifb$. This improvement is mainly due
to the optimal observable method, but the luminosity uncertainty will
limit our measurement.}
but none of the errors of the other operators and their
correlations are affected significantly. In addition, the three
coefficients, $\fBW$, $\fWW$ and $\fBB$ are poorly constrained while
their errors are strongly correlated, with the correlation matrix
elements of $-0.91$, $0.98$, $-0.96$, suggesting the presence of their
linear combination, which can be measured accurately. We therefore
present constraints on the 6 operators, after integrating out the
contributions from $3\fphitwo-2\fphifour$  eq.(\ref{ResOpe_500_df}), in
terms of the eigenvectors and their errors :
\bsub
\begin{eqnarray}
 -.096 \fphione -.29 \fBW -.26 \fW +.31 \fB +.73 \fWW +.47 \fBB
  & =&  \pm \, .026 \label{EigenOPRes-I_dL66-1} \\
 -.35 \fphione -.18 \fBW +.84 \fW -.28 \fB +.21 \fWW +.13 \fBB 
  & =&  \pm \, .029 \label{EigenOPRes-I_dL66-2} \\
 -.49 \fphione +.63\fBW +.077\fW +.47\fB +.22\fWW -.32\fBB 
  & =& \pm \, .039 \label{EigenOPRes-I_dL66-3} \\
 .59\fphione +.020\fBW +.47\fW  +.65\fB  -.078\fWW +.062\fBB 
  & =& \pm \, .088 \label{EigenOPRes-I_dL66-4} \\
 .52\fphione  +.49\fBW +.057\fW -.43\fB  +.53\fWW  -.093\fBB 
  & =& \pm \, .11 \label{EigenOPRes-I_dL66-5} \\
 -.064\fphione  +.50\fBW  +.016\fW -.052\fB -.31\fWW +.81\fBB
  & =& \pm \, .69 \label{EigenOPRes-I_dL66-6}
\end{eqnarray}
\label{EigenOPRes_500_dL_6by6}
\esub
\hskip -1.4 mm
As anticipated, we find 3 combinations of the 6 coefficients, whose
errors are smaller than 5\% for $\Lambda=1\TEV$. The worst constrained
combination of eq.(\ref{EigenOPRes-I_dL66-6}) has a much larger error of
68\%, showing the poorly constrained combination of the three
coefficients, $\fBW$, $\fWW$ and $\fBB$, which leads to their large
errors and the strong correlations among themselves in
eq.(\ref{ResOpe_500_df}). It is worth noting here that the eigenvector
of the most accurately measured combination in
eq.(\ref{EigenOPRes-I_dL66-1}) has a significant contribution from the
constraint on $\bgg$, which is proportional to $\fBW-\fWW-\fBB$ in
eq.(\ref{bggeq}). In fact, if we drop the no-tag $(ee)H$ events from the
analysis, the eigenvector with smallest error becomes essentially that
of eq.(\ref{EigenOPRes-I_dL66-2}) with a dominant $\fW$ term,
which contribute to the  $\cz$, $\cg$ and $\dw$ couplings
in eqs.(\ref{czeq}-\ref{ninecoupend}). Also as expected, none of the
results of eq.(\ref{EigenOPRes_500_dL_6by6}), neither the eigenvectors
nor errors, are affected significantly by the luminosity uncertainty of
$\Delta f=0.01$.

It is worth reporting here the importance of the $e^-$ beam polarization
to obtain the previous results. By setting $\big\vert P\big\vert=0$, we
find for the same ILC-I integrated luminosities and their errors of
$\Delta f=0.01$, 
\begin{eqnarray}
\begin{array}{ll}
 \left( \frac{1\TEV}{\Lambda} \right)^2  \,\, \fphione               & = \pm \, 0.17\\
 \left( \frac{1\TEV}{\Lambda} \right)^2  \,\, \fBW                   & = \pm \, 1.6 \\
 \left( \frac{1\TEV}{\Lambda} \right)^2  \,\, \fW                    & = \pm \, 0.22 \\
 \left( \frac{1\TEV}{\Lambda} \right)^2  \,\, \fB                    & = \pm \, 1.2 \\
 \left( \frac{1\TEV}{\Lambda} \right)^2  \,\, \fWW                   & = \pm \, 0.61 \\
 \left( \frac{1\TEV}{\Lambda} \right)^2  \,\, \fBB                   & = \pm \, 2.1 \\
 \left( \frac{1\TEV}{\Lambda} \right)^2  \,\, (3\fphifour-2\fphitwo) & = \pm \, 0.38 \\
\end{array}
& &
\left(
\begin{array}{ccccccc}
  1 \\
  .12  &  1 \\
  .45  & -.051 &  1 \\
 -.62  & -.089 & -.91  &  1 \\
  .41  & -.83  & .21   & -.24  & 1 \\
 -.031 &  .986 & -.099 &  .005 & -.91 & 1 \\
  .13  &  .30  & -.078 & -.009 & -.14 & .26 & 1 \\
\end{array}
\right)
\label{ResOpe_500_Unpdf} 
\end{eqnarray}
It is striking to find that all the errors except that of
$3\fphitwo-2\fphifour$ are larger by more than a factor of 3 to 9 for
the same luminosity. This is essentially because of the incapability to
resolve the non-standard $HZZ$ and $HZ\gamma$ couplings in the absence
of beam polarization. On the other hand, we notice extremely strong
correlations  among the errors of $\fBW$, $\fWW$ and $\fBB$, and
moderately strong correlation of -0.91 between the errors of $f_W$ and
$f_B$. We therefore give the eigenvectors and their errors for the three
most accurately measured combinations after $3\fphifour-2\fphitwo$ is
integrated out ;
\bsub
\begin{eqnarray}
 -\,.24\,\fphione -\,.43\,\fBW +\,.27\,\fW +.052\,\fB +.\, 66\,\fWW 
  +\,.50\,\fBB &=& \,\pm \, .028 \label{EigenOPRes-I_UnpdL66-1} \\
 -\,.53\,\fphione +\,.44\,\fBW +\,.64\,\fW +\,.12\,\fB +\,.052\,\fWW 
  -\,.30\,\fBB &=& \,\pm \, .035 \label{EigenOPRes-I_UnpdL66-2} \\
 .72\,\fphione  -\,.092\,\fBW +\,.66\,\fW +\,.15\,\fB -\,.12\,\fWW   
  +\,.045\,\fBB &=& \,\pm \, .083 \label{EigenOPRes-I_UnpdL66-3}
\end{eqnarray}
\label{EigenOPRes_500_UnpdL_6by6}
\esub
\hskip -1.4mm
It is remarkable that the error of the most accurately measured combination 
in eq.(\ref{EigenOPRes-I_UnpdL66-1}) is not  much different from that of
eq.(\ref{EigenOPRes-I_dL66-1}). The reason is partly because both of
them  receive dominant contribution from the no-tag $(ee)H$ events that
measure the $H\gamma\gamma$ coupling, which does not depend on  the beam
polarization. The second and the third combination of
eq.(\ref{EigenOPRes-I_UnpdL66-2}) and eq.(\ref{EigenOPRes-I_UnpdL66-3})
have dominant contributions from the $\fW$ and $\fphione$.  Except for the
these three combinations, all the other eigenvectors have errors larger
than 0.1 for $\Lambda=1\TEV$.

Finally, our results for the combined ILC-I and ILC-II analysis,
eq.(\ref{CombResCoup_ILC-II}) gives
\begin{eqnarray}
\begin{array}{ll}
 \left( \frac{1\TEV}{\Lambda} \right)^2  \,\, \fphione               & = \pm \, .056 \\
 \left( \frac{1\TEV}{\Lambda} \right)^2  \,\, \fBW                   & = \pm \, .22 \\
 \left( \frac{1\TEV}{\Lambda} \right)^2  \,\, \fW                    & = \pm \, .015 \\
 \left( \frac{1\TEV}{\Lambda} \right)^2  \,\, \fB                    & = \pm \, .026 \\
 \left( \frac{1\TEV}{\Lambda} \right)^2  \,\, \fWW                   & = \pm \, .14 \\
 \left( \frac{1\TEV}{\Lambda} \right)^2  \,\, \fBB                   & = \pm \, .34 \\
 \left( \frac{1\TEV}{\Lambda} \right)^2  \,\, (3\fphifour-2\fphitwo) & = \pm \, .34 \\
\end{array}
& &
\left(
\begin{array}{ccccccc}
  1 \\
 -.42 &  1 \\
  .22 &  .21  &  1 \\
  .12 & -.32  &  .32 &  1 \\
  .61 & -.91  & -.21 &  .13 & 1 \\
 -.51 &  .98  &  .21 & -.25 & -.96 & 1 \\
 .005 & -.005 & .029 & .002 & .028 & -.014 & 1 \\
\end{array}
\right) 
\label{ResOpe_1000_df} 
\end{eqnarray}
When compared with the ILC-I only results of eq.(\ref{ResOpe_500_df}),
we find that the errors of $\fW$ and $\fB$ are reduced to 1/3, those of
$\fphione$, $\fBW$, $\fWW$ and $\fBB$ are reduced to 2/3, while that
of $3\fphitwo-2\fphifour$ remains the same. The reduction of the errors
in $\fB$ and $\fW$ is a result of the strong constraints on the $\dw$,
$\cz$ and $\cg$ couplings at high energies. The strong correlations
among the errors of $\fBW$, $\fWW$ and $\fBB$ remain unchanged,
suggesting the persistent importance of the $H\gamma\gamma$ measurement
via no-tag $(ee)H$ events. The error of the combination
$3\fphitwo-2\fphifour$ does not change, because it is dominated by the
luminosity uncertainty. If we set $\Delta f=0$, it reduces to $\pm 0.070$.

As in the ILC-I only case, we obtain the eigenvectors and their errors
after integrating out the uncertainty in $3\fphitwo-2\fphifour$, and
hence also over the luminosity error. We find,
\bsub
\begin{eqnarray}
 -.12\,\fphione +\,.001\,\fBW +\,.95\,\fW -\,.29\,\fB -\, .005\,\fWW 
  -\,.028\,\fBB &=& \pm \, .012 \label{EigenOPRes-II_dL66-1} \\
 -.085\,\fphione -\,.48\,\fBW +\,.057\,\fW +\,.16\,\fB +\,.66\,\fWW  
  +\,.54\,\fBB &=& \pm \, .015 \label{EigenOPRes-II_dL66-2} \\
 -\,.27\,\fphione +\,.36\,\fBW +\,.21\,\fW +\,.84\,\fB +\,.15\,\fWW  
  -\,.17\,\fBB &=& \pm \, .021 \label{EigenOPRes-II_dL66-3} \\
 .70\,\fphione -\,.35\,\fBW +\,.22\,\fW +\,.42\,\fB -\,.39\,\fWW  
  +\,.13\,\fBB &=& \pm \, .038 \label{EigenOPRes-II_dL66-4} \\
 .64\,\fphione +\,.53\,\fBW +\,.043\,\fW -\,.14\,\fB +\,.54\,\fWW  
  -\,.063\,\fBB &=& \pm \, .062 \label{EigenOPRes-II_dL66-5} \\
 -.066\,\fphione +\,.50\,\fBW +\,.008\,\fW -\,.016\,\fB -\,.31\,\fWW  
  +\,.81\,\fBB &=& \pm \, .43 \,.\label{EigenOPRes-II_dL66-6}
\end{eqnarray}
\label{EigenOPRes_1000_dL_6by6}
\esub
\hskip -1.4 mm
We now find that 5 combinations out of 6 coefficients are constrained
better than 5\% for $\Lambda=1\TEV$. We noticed that the most accurately
measured combination in eq.(\ref{EigenOPRes-II_dL66-1}) is now dominated
by $\fW$, reflecting the strong $\sqrt{s}$ dependence of the $\dw$,
$\cz$ and $\cg$ couplings; see eq.(\ref{czeq}) to
eq.(\ref{ninecoupend}). The reduction of the error from that of the
corresponding combination in eq.(\ref{EigenOPRes-I_dL66-2}) is 60\%. The
second best constrained combination, eq.(\ref{EigenOPRes-II_dL66-2}),
can be identified as the $\bgg$ combination, whose error is reduced by 40\%
from the ILC-I result of eq.(\ref{EigenOPRes-I_dL66-1}). The coefficient
of $\fB$ dominates the third accurately measured combination,
eq.(\ref{EigenOPRes-II_dL66-3}). The worst measured combination in
eq.(\ref{EigenOPRes-II_dL66-6}) is exactly the same as that of ILC-I only
result in eq.(\ref{EigenOPRes-I_dL66-6}), while its error is reduced to
about 2/3 reflecting a factor of 3 larger cross sections of the
dominant $WW$- and $ZZ$-fusion processes at $\sqrt{s}=1\TEV$ as compared
to those at $\sqrt{s}=500\GEV$, see fig.\ref{rootsvariation}.

\subsection{Comparison with the precision electroweak measurements}
A clear advantage of using the higher dimensional operators to
parametrize possible new physics contribution is that we can compare the
sensitivity and complementarity of any experiments, whether at high
energies or low energies, in a model-independent manner.

Although the $HVV$ couplings can also be measured at the LHC in the
Higgs-strahlung processes ($WH$ and $ZH$ production) and in the weak
boson fusion processes, the expected sensitivity to the higher
dimensional operators \cite{zhang,eboli} is not competitive with that
expected at the ILC. On the other hand, the sensitivity of the precision
measurements of the $Z$-boson and the $W$-boson properties on the higher
dimensional operators will remain competitive even in the ILC era. In
this last subsection, we therefore compare our results with those of
the present and future precision electroweak measurements. Although the
results from LEP and SLC experiments have been
finalized \cite{Z-Pole_Param}, both the mean values and the errors of the
coefficients of the two operators, ${\cal O}_{\phi 1}$ and ${\cal O}_{BW}$ 
in eq.(\ref{Operators}), will depend not only on $\mh$ but also strongly
on the continuously improving measurements of $\mt$ and $\mw$, and to a
lesser extent on $\alpha_s(\mz)_{\overline{\rm MS}}$ and
$\alpha(\mzsq)$. We therefore present details of the dependences of the
precision observables on these parameters.

It is well known that the two operators ${\cal O}_{\phi 1}$ and 
${\cal O}_{BW}$ in eq.(\ref{Operators}) contribute to the $Z$ and
$W$-boson properties~\cite{Grinstein,Oblique2} via the oblique
parameters $S$ and $T$ \cite{Oblique1,Oblique3,Oblique4} 
\bsub
\begin{eqnarray}
 (\Delta S)_{\rm NP} &=& 
  -4\,\, \pi\,\, \frac{v^2}{\Lambda^2}\,\, \,\,  \fBW, \\
 (\Delta T)_{\rm NP} &=& 
  -\frac{1}{2\alpha} \,\, \,\, \frac{v^2}{\Lambda^2}\,\, \,\,  \fphione.
\end{eqnarray}
\label{ST_OPE}
\esub
\hskip -1.7 mm
Here $(\Delta S)_{\rm NP}$ and $(\Delta T)_{\rm NP}$ are the new physics
contributions to the $S$ and $T$ parameters, respectively. All the $Z$
boson parameters are parametrized in terms of the two parameters,
$\Delta S_Z$ and $\Delta T_Z$ \cite{SzTz}, which are related to the $S$
and $T$ parameters as
\bsub
\begin{eqnarray}
 \Delta S_Z &=& \Delta S +\Delta R_Z, \\
 \Delta T_Z &=& \Delta T +1.49\,\,  \Delta R_Z,
\end{eqnarray}
\esub
where $\Delta R_Z$ denotes the difference in the effective $Z$-boson
coupling, $\bar{g}_Z^2(q^2)$ \cite{Hagiwara:1998}, between $q^2=0$ (where
$S$ and $T$ parameters are defined) and $q^2=\mzsq$ (where the coupling
is measured precisely at LEP and SLC). 
By using the 13 data set of Z-pole parameters \cite{Z-Pole_Param}, we
obtain the following fit in terms of 4 parameters, $\Delta S_Z$, 
$\Delta T_Z$, $m_t$ and $\alpha_s(\mz)_{\overline{\rm MS}}$ ;
\bsub
\begin{eqnarray}
 & &
 \begin{array}{lc}
  \Delta S_Z = 0.037 +0.0045\,\, x_t -0.037\,\, x_s & \,\,\pm \,\, 0.105 \\
  \Delta T_Z = 0.043 +0.0084\,\, x_t -0.065\,\, x_s & \,\,\pm \,\, 0.136 \\
 \end{array}
 \left(
  \begin{array}{cc}
   1. & \\
   0.90 & 1.\\
  \end{array}
 \right), \\
 & & \hskip 1.4mm
  \chi^2_{\rm min} = 15.5 
  + \left( \frac{x_t + x_s + 2.8 }{4.3} \right)^2
  + \left( \frac{x_s + 0.17}{0.79} \right)^2 .
\end{eqnarray}
\label{SzTz_Const}
\esub
\noindent
\hskip -1.4mm
Here $x_t=(m_t-172\GEV)/\,\, 3\GEV$ and $x_s=(\alpha_s-0.118)/\,\,0.003$. 
The direct $m_t$ dependence of the fit comes from the $Zb\bar{b}$ vertex
correction, and the $\alpha_s$ dependence comes from 
$\Gamma(Z\to {\rm hadrons})$.

In the above fit, $\Delta S$, $\Delta T$ and $\Delta R_Z$ are measured
from their reference values in the SM at $\mh=100\GEV$, $\mt=172\GEV$,
$\alpha_s(\mz)_{\overline{\rm MS}}=0.118$ and  
$\Delta \alpha_{\rm had}^5=0.0277$. Because the values of $\mh$, $m_t$,
$\alpha_s$ and $\Delta \alpha_{\rm had}^5$ will be measeured precisely
in the future, we parametrized their dependence
as~\cite{SzTz,NextEWSUSY},
\bsub
\begin{eqnarray}
 (\Delta S)_{\rm SM} &=&  0.0963\,\, x_h -0.0224\,\, x_h^2 
  +0.0026\,\,x_h^3 -0.0014\,\, x_t -0.033\,\, x_a, \\
 (\Delta T)_{\rm SM} &=& -0.0432\,\, x_h -0.0539\,\, x_h^2 
  +0.0096\,\, x_h^3 +0.0367\,\, x_t \nonumber \\
 & & -0.0007\,\, x_h\,\, x_t -0.0033\,\, x_s \\
 (\Delta R_Z)_{\rm SM} 
 &=& 0.00838 (1-e^{-2\,x_h}).
\end{eqnarray}
\label{STRz_para}
\esub
\hskip -1.2 mm
where, $x_h=\ln(\mh/\,\, 100\GEV)$ and 
$x_a=(\Delta \alpha_{\rm had}^5-0.0277)/\,\, 0.0003$.

In addition to the $Z$-boson parameters, the $W$-boson mass is also
sensitive to the operators $\fphione$ and $\fBW$. Their dependences can
be parametrized as
\begin{eqnarray}
 \mw{\rm [GeV]} &=& 80.318 -0.288 \,\, 
  \big(\Delta S\big) +0.418\,\, \big(\Delta T\big) 
  +0.337\,\, \big(\Delta U\big) -0.0055 \,\, x_a,
\end{eqnarray}
where the SM contribution to the $U$ parameter is
\begin{eqnarray}
 (\Delta U)_{\rm SM} &=& -\,\, 0.2974 \,\, x_h -\,\, 0.0260 \,\, x_h^2 
  +\,\, 0.0772 \,\, x_t +\,\, 0.0004 \,\, x_t^2, 
\end{eqnarray}
We note that none of the dimension-6 operators in eq.(\ref{Operators})
contribute to $\Delta U$ and $\Delta R_Z$;
\begin{eqnarray}
 (\Delta U)_{\rm NP} = (\Delta R_Z)_{\rm NP} = 0.
\end{eqnarray}

By using the fit eq.(\ref{SzTz_Const}) of the LEP and SLC
results \cite{Z-Pole_Param} on the $Z$ parameters, and the latest
estimates of $\mw$, $\mt$, $\alpha_s$ \cite{pdg2006} and 
$\Delta \alpha_{\rm had}^5$ \cite{alpha_h5}
\bsub
\begin{eqnarray}
 \mw{\rm [GeV]} &=& 80.403 \,\,\pm \,\, 0.029, \\
 \mt{\rm [GeV]} &=& 172.5 \,\,\pm \,\, 2.3, \\
 \alpha_s(\mz)_{\overline{\rm MS}} &=& 0.118 \,\,\pm \,\, 0.002, \\
 \Delta \alpha_{\rm had}^5 &=& 0.02768 \,\,\pm \,\, 0.00022,
\end{eqnarray}
\label{PMesConsts}
\esub
\noindent
\hskip -1.4 mm
we find the constraints 
\begin{eqnarray}
 \begin{array}{c}
  \left(\frac {1\, {\rm TeV}} { \Lambda }\right)^2 \,\, \fphione
   = -\,\, 0.048 \,\,\pm 0.028 \\
  \left(\frac {1\, {\rm TeV}} { \Lambda }\right)^2 \,\, \fBW
   = -\,\, 0.093 \,\,\pm 0.14 \\
 \end{array}
 & &
 \left(
  \begin{array}{cc}
   1. & \\
   0.83  & 1.\\
  \end{array}
 \right),
\label{STconst} 
\end{eqnarray}
with $\chi^2_{\rm min}$/d.o.f. = $20.5/12$, for $\mh=120\GEV$
($x_h=0.1823$). By comparing eq.(\ref{STconst}) with
eq.(\ref{ResOpe_500_df}) for ILC-I and eq.(\ref{ResOpe_1000_df}) for the
combined ILC-I and ILC-II analysis, we find that the low energy data
constrain $\fphione$ and $\fBW$ better than the ILC. On the other hand,
the two combination of the operators that are constrained by the low
energy data
\bsub
\begin{eqnarray}
 .987 \,\, \fphione - \,\, .16\,\, \fBW &=& 
  -\,\, 0.032 \,\,\pm \,\, 0.015, \\
 .16 \,\, \fphione + \,\, .987\,\, \fBW &=& 
  -\,\, 0.10 \,\,\pm \,\, 0.14.
\end{eqnarray}
\label{STconst_Eigen}
\esub
\noindent
\hskip -1.4mm
are quite orthogonal to the most precisely measured combinations at
ILC-I, eq.(\ref{ResOpe_500_df}), and those in the combined ILC-I and
ILC-II analysis, eq.(\ref{ResOpe_1000_df}). For instance, if we combine
eq.(\ref{ResOpe_1000_df}) and eq.(\ref{STconst}), we find
\begin{eqnarray}
\begin{array}{lll}
 \left( \frac{1\TEV}{\Lambda} \right)^2  \,\, \fphione               
  & = -\,\, 0.048 & \,\,\pm \,\,.020 \\
 \left( \frac{1\TEV}{\Lambda} \right)^2  \,\, \fBW                   
  & = -\,\, 0.093  & \,\,\pm \,\, .10 \\
 \left( \frac{1\TEV}{\Lambda} \right)^2  \,\, \fW                    
  & =        & \,\,\pm \,\,.015 \\
 \left( \frac{1\TEV}{\Lambda} \right)^2  \,\, \fB                    
  & =        & \,\,\pm \,\, .025 \\
 \left( \frac{1\TEV}{\Lambda} \right)^2  \,\, \fWW                   
  & =        & \,\,\pm \,\, .065 \\
 \left( \frac{1\TEV}{\Lambda} \right)^2  \,\, \fBB                   
  & =        & \,\,\pm \,\,.15 \\
\end{array}
& &
\left(
\begin{array}{cccccc}
  1 \\
  .68  &  1 \\
  .27  &  .28  &  1 \\
 -.12  & -.17  & .38  &  1 \\
 -.34  & -.65  & -.40 & -.25  & 1 \\
  .59  &  .93  & .36  & -.009  & -.86 & 1 \\
\end{array}
\right) 
\label{ResOpe_ILCII+EWPM} 
\end{eqnarray}
and now all the 6 operator coefficients are constrained rather
independently, except for $\fBB$ which is still correlated with $\fBW$
and $\fWW$. It is remarkable that even the errors of $\fphione$ and
$\fBW$ are reduced significantly ($\sim 30\%$) by the ILC data. Because
of the strong correlation among the errors of $\fBW$, $\fWW$ and $\fBB$
in eq.(\ref{ResOpe_1000_df}), the addition of the low-energy data
(\ref{STconst}) leads to reduction of $\fWW$ and $\fBB$ errors.

We note here that the  result  given in eq.(\ref{ResOpe_ILCII+EWPM})
does not take into account improvements in the measurements of $\mw$
and $\mt$ which should certainly take place at the ILC-I, and also
possible improvements in the measurements of 
$\alpha_s(\mz)_{\overline{\rm MS}}$ and $\Delta \alpha_{\rm had}^5$ are
expected. If we replace the present constraints of eq.(\ref{PMesConsts})
by
\bsub
\begin{eqnarray}
 \mw{\rm [GeV]} &=& 80.403 \,\,\pm \,\, 0.010, \\
 \mt{\rm [GeV]} &=& 172.50 \,\,\pm 0.10, \\
 \alpha_s(\mz)_{\overline{\rm MS}} &=& 0.1180 \,\,\pm \,\, 0.0010, \\
 \Delta \alpha_{\rm had}^5 &=& 0.02768 \,\,\pm \,\, 0.00010,
\end{eqnarray}
\label{OptMesConsts}
\esub
\noindent
\hskip -1.4mm
without changing their mean values, the low energy constraints will
become
\begin{eqnarray}
 \begin{array}{lll}
  \left(\frac {1\, {\rm TeV}} { \Lambda }\right)^2 \Delta f_{\phi 1}
   & = -\,\, 0.082 & \,\,\pm \,\, 0.018 \\
  \left(\frac {1\, {\rm TeV}} { \Lambda }\right)^2 \Delta f_{BW}
   & = -\,\, 0.21  & \,\,\pm \,\, 0.12 \\
 \end{array}
 & &
 \left(
  \begin{array}{cc}
   1. & \\
   0.95  & 1.\\
  \end{array}
 \right).
\label{STconst_new} 
\end{eqnarray}
Most importantly, the error of $\fphione$ is reduced to about a half,
because $\fphione$ has rather strong dependence on $m_t$. 

In addition, if there are new measurements at GigaZ, we can measure the
effective weak-mixing angle much more accurately. For instance, an
estimate in ref.~\cite{GigaZ} gives 
\begin{eqnarray}
 \sin^2\theta^{\rm eff}_{\rm W} &=& 0.23153 \,\,\pm 0.000013.
\label{GigaZMesConsts}
\end{eqnarray}
\noindent 
The effective mixing angle can also be parametrized
as~\cite{SzTz,NextEWSUSY}
\begin{eqnarray}
 \sin^2\theta^{\rm eff}_{\rm W} = 0.23148 +0.00359(\Delta S_Z)
  -0.00241(\Delta T_Z) +0.00011x_{\alpha},
  \label{sintheff_th}
\end{eqnarray}
and the constraints on $\fphione$ and $\fBW$ will become
\begin{eqnarray}
 \begin{array}{lll}
  \left(\frac {1\, {\rm TeV}} { \Lambda }\right)^2\,\,\,\, \Delta f_{\phi 1}
   & = -\,\,0.097 & \,\,\pm \,\,0.011 \\
  \left(\frac {1\, {\rm TeV}} { \Lambda }\right)^2\,\,\,\, \Delta f_{BW}
   & = -\,\,0.32  & \,\,\pm \,\,0.062 \\
 \end{array}
 & &
 \left(
  \begin{array}{cc}
   1. & \\
   0.95  & 1.\\
  \end{array}
 \right),
\label{STconst_GigaZ} 
\end{eqnarray}
with $\chi^2$/d.o.f = 30.9/13$^{\footnotemark[6]}$
\footnotetext[6]{Neither the low probability of the fit nor the
13$\sigma$ evidence for ($\Delta\fphione$,$\Delta\fBW$) in
eq.(\ref{STconst_GigaZ}) is our concern, since they are artifacts of
our keeping the present mean values eq.(\ref{OptMesConsts}) and
eq.(\ref{GigaZMesConsts}) when reducing their errors.}.
Both errors of $\fphione$ and $\fBW$ are reduce by a factor 2.5 and 2.3,
respectively, from eq.(\ref{STconst}). The eigenvectors and its errors
are
\bsub
\begin{eqnarray}
 .98 \,\, \fphione - \,\,.17\,\,\fBW &=& -\,\,0.041 \,\, \pm \,\,0.0034, \\
 .17 \,\, \fphione + \,\,.98\,\,\fBW &=& -\,\,0.33  \,\, \pm \,\,0.062.
\end{eqnarray}
\esub
Although the most accurately measured combination of the dimension-6
operator does not change from eq.(\ref{STconst_Eigen}), the error is
reduced by a factor of 4.4. We note, however, that the reduction of the
errors are limited by the error of $\Delta \alpha_{\rm had}^5$ assumed in
eq.(\ref{OptMesConsts}d), whose contribution to the uncertainty of the
$x_{\alpha}$ term in eq.(\ref{sintheff_th}) is 3 times larger than the
error in eq.(\ref{GigaZMesConsts}).

If we combine eq.(\ref{STconst_GigaZ}) with ILC-I + ILC-II result of 
eq.(\ref{ResOpe_ILCII+EWPM}), we find
\begin{eqnarray}
\begin{array}{lll}
 \left( \frac{1\TEV}{\Lambda} \right)^2  \,\, \fphione               
  & = -\,\,0.097 & \,\,\pm \,\,0.010 \\
 \left( \frac{1\TEV}{\Lambda} \right)^2  \,\, \fBW                   
  & = -\,\,0.32 & \,\,\pm \,\,0.056 \\
 \left( \frac{1\TEV}{\Lambda} \right)^2  \,\, \fW                    
  & =        & \,\,\pm \,\,0.014 \\
 \left( \frac{1\TEV}{\Lambda} \right)^2  \,\, \fB                    
  & =        & \,\,\pm \,\,0.025 \\
 \left( \frac{1\TEV}{\Lambda} \right)^2  \,\, \fWW                   
  & =        & \,\,\pm \,\,0.053 \\
 \left( \frac{1\TEV}{\Lambda} \right)^2  \,\, \fBB                   
  & =        & \,\,\pm \,\,0.096 \\
\end{array}
& &
\left(
\begin{array}{cccccc}
  1 \\
  .94  &  1 \\
  .17  &  .18  &  1 \\
 -.088 & -.093  & .43  &  1 \\
 -.38  & -.42  & -.36 & -.40  & 1 \\
  .75  &  .80  & .31  & .16   & -.81 & 1 \\
\end{array}
\right) .
\label{ResOpe_ILCII+GigaZ} 
\end{eqnarray}
Thanks to the precise measurements of $\fphione$ and $\fBW$, the errors
of $\fBB$ and $\fWW$ are reduced. The error of $\fW$ and $\fB$ are not
affected much by the improved measurements of the weak-boson parameters,
and the constraints of the type (\ref{EigenOPRes-II_dL66-1}) $\sim$ 
(\ref{EigenOPRes-II_dL66-3}) will still give additional informations on
new physics from the $HVV$ coupling measurements. 

\vskip 0.5cm

We hope that our report will be useful in studying the physics potential
of the ILC project.

\vskip 1 cm
\noindent

{\bf Acknowledgment}
We thank Satoshi Ishihara, Jun-ichi Kamoshita and Yosuke Uehara for
participating in stimulating discussions in the early stage of this
study. This work is supported in part by the JSPS Core University
Program, and Grant-in-Aid for Scientific Research (\#17540281 and
\#18340060) from MEXT, Japan. S.D. would like to thank the Science and
Engineering Research Council, Department of Science and Technology, the
Government of India  for partial financial support (SR/S2/HEP-12/2006).

\appendix
\setcounter{equation}{0}
\section{The 3-body phase space }
\label{app:phasespace}

We parametrize the four momentum of the $e^+e^- \to f\bar{f}H$ process,
see eq.(\ref{ffbarh}), in the laboratory frame as follows
\bsub
\begin{eqnarray}
 k_{1}^{\mu} &=& \frac{\sqrt{s}}{2} \,\, \bigg(1,\, 0,\,0,\,1\bigg), \\
 k_{2}^{\mu} &=& \frac{\sqrt{s}}{2} \,\, \bigg(1,\, 0,\,0,\,-1\bigg), \\
 p_{1}^{\mu} &=& \frac{\sqrt{s}}{2} x_1
  \,\, \bigg(1, \,\,\sin\theta_1 \,\cos\phi_1,\,\, \sin\theta_1
  \,\sin\phi_1,\,\, \cos\theta_1\bigg), \\
 p_{2}^{\mu} &=& \frac{\sqrt{s}}{2} x_2 
  \,\, \bigg(1,\,\, \sin\theta_2 \,\cos\phi_2,\,\, \sin\theta_2
  \,\sin\phi_2,\,\, \cos\theta_2\bigg), \\
 p_{\rm H}^{\mu} &=& k_1^{\mu} + k_2^{\mu} - p_1^{\mu} - p_2^{\mu}, \\
 &=& x_H \,\, \frac{\sqrt{s}}{2} \, \,
  \biggl( 1,\, \,\beta_H \,\sin\thetaH,\,\, 0,\,\, \beta_H \,\cos\thetaH
  \biggr), 
  \end{eqnarray}
\label{oldlabmom}
\esub
where we ignore the masses of $e^{\pm}$, $f$ and $\bar{f}$, and the
Higgs-boson is produced in the xz-plane
\begin{eqnarray}
 x_1 \sin \theta_1 \sin \phi_1 + x_2 \cos \theta_2 \cos \phi_2 = 0.
  \label{Higgs_xz}
\end{eqnarray}
The Higgs-boson energy fraction and its velocity are
\begin{eqnarray}
 x_H = \frac{2E_H}{\sqrt{s}} = 2 -x_1 -x_2, \hskip 0.5cm 
  \beta_H = \sqrt{1-4\mhsq\,/\,sx_H^2}.
\end{eqnarray}
We parametrize the $f$ and $\bar{f}$ four momenta also in the rest frame of
the $f \bar{f}$ system,
\begin{eqnarray}
 {p_1^*}^\mu &=& \frac{m_{f \bar{f}}}{2}\,
  \biggl( 1,\, \sin\theta^* \cos\phi^*,\,  \sin\theta^* \sin\phi^*,\, 
  \cos\theta^* \biggr), \nonumber \\
 {p_2^*}^\mu &=& \frac{m_{f \bar{f}}}{2}\,
  \biggl( 1,\, -\sin\theta^* \cos\phi^*,\, -\sin\theta^* \sin\phi^*,\,
  -\cos\theta^* \biggr),
\end{eqnarray}
where $m^2_{f\bar{f}}=(p_1+p_2)^2$, $\theta^*$ and $\phi^*$ are the
polar and azimuthal angle, respectively, with respect to the momentum
direction of the $f\bar{f}$ system in the laboratory frame. After making
an appropriate boost to the laboratory frame with
\begin{eqnarray}
 \beta = \frac{x_H \beta_H}{2-x_H}, \hskip 0.5cm
  \gamma = \frac{1}{\sqrt{1-\beta^2}},
\end{eqnarray}
and an appropriate rotation about the y-axis by $\thetaH$, we find
$p_1^\mu$ and $p_2^\mu$ in the laboratory frame
\begin{eqnarray}
{p_1}^\mu &=& \frac{m_{f \bar{f}}}{2} 
 \biggl( \gamma (1+\beta c_{\theta^*}),\, 
 c_{\thetaH} s_{\theta^*} c_{\phi^*} 
 +s_{\thetaH} \gamma(\beta +c_{\theta^*}),\, 
 s_{\theta^*} s_{\phi^*},\,
 c_{\thetaH} \gamma(\beta +c_{\theta^*}) 
 -s_{\thetaH} s_{\theta^*} c_{\phi^*} \biggr), \nonumber \\
 {p_2}^\mu &=& \frac{m_{f \bar{f}}}{2}
  \biggl( \gamma (1-\beta c_{\theta^*}),\, 
  -c_{\thetaH} s_{\theta^*} c_{\phi^*}
  +s_{\thetaH} \gamma(\beta -c_{\theta^*}),\, 
  -s_{\theta^*} s_{\phi^*},\,
  c_{\thetaH} \gamma(\beta -c_{\theta^*})
  +s_{\thetaH} s_{\theta^*} c_{\phi^*} \biggr). \nonumber \\
 \label{newlabmom}
\end{eqnarray}
Here we introduce a short-hand, $s_\theta \equiv \sin \theta$ and
$c_\theta \equiv \cos \theta$. Comparing eq.(\ref{oldlabmom}) with
eq.(\ref{newlabmom}), we find 
\bsub
\begin{eqnarray}
 x_1 &=& \frac{m_{f \bar{f}}}{\sqrt{s}}\, \gamma\, 
  (1+\beta c_{\theta^\star}), \hskip 2.75cm 
 x_2 = \frac{m_{f \bar{f}}}{\sqrt{s}}\, \gamma\, 
  (1-\beta c_{\theta^\star}), \\
 \cos\theta_1 &=&
  \frac{ c_{\thetaH} \gamma (\beta +c_{\theta^*}) 
   -s_{\thetaH} s_{\theta^*} c_{\phi^*} }
   { \gamma (1 +\beta c_{\theta^*}) }, \hskip 0.5cm
 \cos\theta_2 =
  \frac{ c_{\thetaH} \gamma (\beta -c_{\theta^*})
   +s_{\thetaH} s_{\theta^*} c_{\phi^*} }
   { \gamma (1 -\beta c_{\theta^*}) },\\
 \sin\theta_1 \sin\phi_1 &=& \frac{ s_{\theta^*} s_{\phi^*} }
   {\gamma (1+\beta c_{\theta^*})}, \hskip 2.0cm 
 \sin\theta_2 \sin\phi_2 = -\frac{ s_{\theta^*} s_{\phi^*} }
   {\gamma (1-\beta c_{\theta^*})},
\end{eqnarray}
\label{conversion}
\esub
which satisfy eq.(\ref{Higgs_xz}). We can now parametrize the 3-body
phase space as
\bsub
\begin{eqnarray}
 d\Phi_3 &\equiv& \left( 2\pi \right)^4 \delta^4 \left(k_1+k_2-p_1-p_2-p_H\right) 
  \left[ \prod_{i=1}^2\, \frac{d^3 p_i} {(2 \pi)^3\, 2 E_i} \right]\,
  \frac{d^3 p_H}{(2 \pi)^3\, 2\EH} \label{3bdyphsp-0}\\
 &=& \frac{1}{ (8 \pi)^2}\, 
  \bar{\beta} \left( \frac{m^2_{f \bar{f}}}{s},\, \frac{\mhsq}{s} \right)\,
  \frac{d\cos\thetaH}{2}\, \frac{d \, m^2_{f \bar{f}}}{2\pi}\,
  \frac{d\cos\theta^* \, d\phi^*}{4\pi} \label{3bdyphsp-1} \\
 &=& \frac{\sqrt{s}}{128 \pi^3}\, 
  \bar{\beta} \left( \frac{m^2_{f \bar{f}}}{s},\, \frac{\mhsq}{s} \right)\,
  d \EH \, d\cos\thetaH \,
  \frac{d \cos\theta^* \, d\phi^*}{4\pi} \label{3bdyphsp-2} \\
 &=& \frac{\sqrt{s}}{128\pi^3}\, 
  \bar{\beta} \left( \frac{m^2_{f \bar{f}}}{s},\, \frac{\mhsq}{s} \right)\,
  d\pTH \, d\yH \,
  \frac{d \cos\theta^* \, d\phi^*}{4\pi} \label{3bdyphsp-3} \\
 &=& \frac{s}{128 \pi^3}
  \frac{x_1(1-x_1-\mhsq/s)}{[1-x_1(1-\cos\theta_{12})/2]^2} dx_1 
  \frac{d\cos\theta_1 \, d\phi_1}{4\pi} 
  \frac{d\cos\theta_2 \, d\phi_2}{4\pi} 
  \label{3bdyphsp-4}
\end{eqnarray}
\esub
where
\bsub
\begin{eqnarray}
 \bar{\beta} (a,b) &=& \sqrt{1-2(a+b)+(a-b)^2}, \label{betafunction} \\
 \cos\theta_{12} &=& \cos\theta_1\cos\theta_2 
  + \sin\theta_1 \sin\theta_2\cos(\phi_2-\phi_1).
\end{eqnarray}
\esub
We use the phase-space parametrization (\ref{3bdyphsp-1}) in the
analysis of the $ZH$ production process, (\ref{3bdyphsp-2}) or
(\ref{3bdyphsp-3}) in the analysis of the $\nu_e \bar{\nu}_e H$ process
where the undetectable $\nu_e$ and $\bar{\nu}_e$ angles
($\cos\theta^*$,$\phi^*$) are integrated out, and (\ref{3bdyphsp-4}) in
the analysis of the double-tag $(eeH)$ events, following the
prescription given in the Appendix A of ref.\cite{hagiphasespace}. In
the analysis of single-tag $(e)eH$ events and no-tag $(ee)H$ events,
the electron mass should be kept in the integration of the forward
scattering angles~\cite{hagiphasespace}.

\section{ Massless Fermion currents}
\label{app:current}
In this appendix,  we show the explicit form of the massless fermion
currents which appear in the $t$-channel and $s$-channel  gauge-boson
exchange processes. The $t$-channel currents of eqs.(\ref{WWcurrent})
and (\ref{tchancurrent}) for $\sigma_1=\lambda_1=\sigma$ and
$\sigma_2=\lambda_2=\overline{\sigma}$ are 
\bsub
\begin{eqnarray}
 j_{e^- (\sigma,\sigma)}^{\mu} &=& \sqrt{s x_1} 
  \left( \cos\frac{\theta_1}{2},
   \sin\frac{\theta_1}{2} e^{i\sigma\phi_1}, 
   -i\sigma\sin\frac{\theta_1}{2} e^{i\sigma\phi_1},
   \cos\frac{\theta_1}{2}\right), \label{app2:current1}\\
 j_{e^+ (\overline{\sigma},\overline{\sigma})}^{\mu} &=& \sqrt{s x_2} 
  \left(\sin \frac{\theta_2}{2}, 
   \cos\frac{\theta_2}{2} e^{-i\overline{\sigma}\phi_2}, 
   i\overline{\sigma}\cos\frac{\theta_2}{2} e^{-i\overline{\sigma}\phi_2},
  -\sin\frac{\theta_2}{2} \right),\label{app2:current2}
\end{eqnarray}
\label{app2:current}
\esub
\hskip -1.4mm
where $\sigma/2$ denotes the $e^-$ helicities in eq.(\ref{app2:current}a),
and $\overline{\sigma}/2$ denotes the $e^+$ helicities in
eq.(\ref{app2:current}b). In the $t$-channel $W$-exchange process,
$e^+e^- \to \nu_e \bar{\nu}_e H$, only the $\sigma=-\,\,\overline{\sigma}=-$
combination contributes, while in $e^+e^- \to e^+ e^- H$ via $t$-channel
$Z$ and $\gamma$ exchange processes, both helicities contribute. The
following 8 combinations of the contractions appear in the cross section
with higher dimensional operators:
\bsub
\begin{eqnarray}
 j_{e^- (-,-)}^{\mu} \, g_{\mu\nu} \, j_{e^+ (+,+)}^{\nu} &=& 
  \biggl[ j_{e^- (+,+)}^{\mu} \,\, g_{\mu\nu} \,\, j_{e^+ (-,-)}^{\nu}
  \biggr]^*  \equiv {\cal F}(\theta_1,\theta_2), \label{wcurrentcontraction1}\\
 j_{e^- (-,-)}^{\mu} \,\, g_{\mu\nu} \,\, j_{e^+ (-,-)}^{\nu} &=&
  \biggl[ j_{e^- (+,+)}^{\mu} \,\, g_{\mu\nu} \,\, j_{e^+ (+,+)}^{\nu}
  \biggr]^*  \equiv 
  {\cal F}(\theta_1,\theta_2) {\cal H}(\theta_1,\theta_2,\phi), \\
  j_{e^- (-,-)}^{\mu} {q_1}_\nu {q_2}_\mu  j_{e^+ (+,+)}^{\nu} &=&
  \biggl[ j_{e^- (+,+)}^{\mu} {q_1}_\nu {q_2}_\mu j_{e^+ (-,-)}^{\nu} 
  \biggr]^* \equiv \frac{s}{8} 
  {\cal F}(\theta_1,\theta_2) {\cal G}(\theta_1,\theta_2,\phi), \\
 j_{e^- (-,-)}^{\mu} {q_1}_\nu {q_2}_\mu j_{e^+ (-,-)}^{\nu} &=&
  \biggl[ j_{e^- (+,+)}^{\mu} {q_1}_\nu {q_2}_\mu j_{e^+ (+,+)}^{\nu} 
  \biggr]^* \equiv \frac{s}{8} 
  {\cal F}(\theta_1,\theta_2) {\cal G'}(\theta_1,\theta_2,\phi).\label{wcurrentcontraction4}
\end{eqnarray}
\label{FHGfuncDef}
\esub
\hskip -.14 cm 
Here, $q_1 =k_1-p_1$ and $q_2 =k_2-p_2$ are the transfer momenta in
the $t$-channel, and $\phi=\phi_1-\phi_2$. The functions 
${\cal F}(\theta_1,\theta_2)$, 
${\cal H}(\theta_1,\theta_2,\phi)$, ${\cal G}(\theta_1,\theta_2,\phi)$
and ${\cal G'}(\theta_1,\theta_2,\phi)$ are
\bsub
\begin{eqnarray}
 {\cal F}(\theta_1,\theta_2) &=& 2s \sqrt{x_1 x_2} \cos\frac{\theta_1}{2}
  \sin\frac{\theta_2}{2} \\
 {\cal H}(\theta_1,\theta_2,\phi) &=& 1 -\tan\frac{\theta_1}{2} 
  \cot\frac{\theta_2}{2}e^{i\phi}  \\
 {\cal G}(\theta_1,\theta_2,\phi) &=& 
  \left[ 2 -x_1 (1 + \cos\theta_1)
   + x_1 \sin\theta_1 \cot\frac{\theta_2}{2} e^{-i\phi}
  \right] \nonumber \\
 & \times &
  \left[ 2 -x_2(1 -\cos\theta_2)
   + x_2 \sin\theta_2 \tan\frac{\theta_1}{2} e^{i\phi} 
  \right] \\
 {\cal G'}(\theta_1,\theta_2,\phi) &=& 
  \left[ 2 -x_1(1 +\cos\theta_1)
   + x_1 \sin\theta_1 \cot\frac{\theta_2}{2} e^{i\phi}
  \right] \nonumber \\
 & \times &
  \left[ 2 -x_2(1 -\cos\theta_2)
   + x_2 \sin\theta_2 \tan\frac{\theta_1}{2} e^{i\phi} 
  \right].
\end{eqnarray}
\label{FHGfunc}
\esub

The currents that appear in the $s$-channel $ZH$ production process are
rather simple. The initial $e^+e^-$ annihilate currents are
\begin{eqnarray}
 j^\mu_{1\sigma} = \overline{v} \left( k_2,-\frac{\sigma}{2} \right)
  \,\, \gamma^\mu \,\, P_{\sigma} \,\, u\left(k_1,\frac{\sigma}{2}\right) 
  = \sqrt{s}(0,-\sigma,-i,0),
\label{schancurrent}
\end{eqnarray}
where the $e^-$ helicity is $\sigma/2$. The final $Z\to f\bar{f}$ decay
currents are
\bsub
\begin{eqnarray}
 j^\mu_{2\sigma'} 
  &=& \overline{u} \left( p_1,\frac{\sigma'}{2} \right) \,\,
  \gamma^\mu \,\, P_{\sigma'}\,\, v 
  \left( p_2,-\frac{\sigma'}{2}\right) \nonumber \\
 &=& \sqrt{s}\,\bigg(0,\,
  -\sigma' \, \cos\theta^* \, \cos\phi^* -i \,\sin\phi^*,\,
  -\sigma' \, \cos\theta^* \, \sin\phi^* +i \,\cos\phi^*,\,
   \sigma' \, \sin\theta^* \bigg), \nonumber\\
\end{eqnarray}
\label{schan1}
\esub
in the $f\bar{f}$ rest frame of eq.(\ref{newlabmom}), where the $f$
helicity is $\sigma'/2$. When contracted with the decaying $Z$-boson
polarization vector
\bsub
\begin{eqnarray}
 \epsilon^{\mu}(\lambda=\pm) &=& \frac{1}{\sqrt{2}}
  (0, \mp 1, -i,0) \\
 \epsilon^{\mu}(\lambda=0) &=& (0,0,0,1)
\end{eqnarray}
\label{schan2}
\esub
we find
\bsub
\begin{eqnarray}
 \epsilon(\pm) \cdot j_{2\sigma'} &=& -\sqrt{\frac{s}{2}} 
  (1 \pm \sigma' c_{\theta^*}) e^{\pm i \phi^*}, \\
  \epsilon(0) \cdot j_{2\sigma'} &=& -\sqrt{s} \sigma' s_{\theta^*}.
\end{eqnarray}
\label{schan3}
\esub
The decay density matrix elements of eq.(\ref{ZH_densityM}) are obtained
from these equations.

\end{document}